\documentclass[aps,twocolumn,superscriptaddress,preprintnumbers,pre]{revtex4}
\usepackage{xr-hyper}
\usepackage{mathrsfs, hyperref}
\usepackage{amssymb, amsbsy, amsmath, latexsym, dsfont, array, layout, graphics,mathrsfs,braket,amsfonts,amsthm,
  amssymb,graphicx,subfigure, youngtab,color,bm,mathtools,braket,verbatim,url,cleveref,natbib,hypernat,extarrows,inputenc,multirow,bbm}
  
\usepackage[usernames,dvipsnames]{xcolor}
\colorlet{RED}{red}
\usepackage{graphicx,psfrag}
\usepackage{amsfonts}
\usepackage[figuresright]{rotating}  
\usepackage{amssymb}
\usepackage{amsmath}
\usepackage{subfigure}
\usepackage{multirow}
\usepackage{tabularx}
\usepackage[resetlabels]{multibib}
\newcites{supp}{Supplemental References}
\usepackage{amssymb, amsbsy, amsmath, latexsym, dsfont, array, layout, graphics,mathrsfs,braket,amsfonts,amsthm,
  amssymb,graphicx,subfigure,dcolumn, youngtab,color,bm,mathtools,braket,verbatim,url,cleveref,natbib,hypernat,extarrows,inputenc,multirow}
\usepackage[usernames,dvipsnames]{xcolor}
\usepackage{mathrsfs, hyperref}
\usepackage{amssymb, amsbsy, amsmath, latexsym, dsfont, array, layout, graphics,mathrsfs,braket,amsfonts,amsthm,
  amssymb,graphicx,subfigure, youngtab,color,bm,mathtools,braket,verbatim,url,cleveref,natbib,hypernat,extarrows,inputenc,multirow,bbm}
\usepackage{graphicx}
\usepackage{dcolumn}
\usepackage{bm}
\usepackage[usernames,dvipsnames]{xcolor}
\usepackage[normalem]{ulem}
\usepackage[T1]{fontenc}
\allowdisplaybreaks

\begin{document}
\title{A New Perspective on Thermally Fluctuating 2D Elastic Membranes: Introducing Odd Elastic Moduli and Non-Equilibrium Effects}

\author{M.E.H. Bahri}
\email{mbahri@princeton.edu}
\affiliation{Department of Mechanical and Aerospace Engineering, Princeton University, Princeton, NJ 08544, USA}
\author{S. Sarkar}
\email{sarkarsi@umich.edu}
\affiliation{Department of Physics, University of Michigan, Ann Arbor, MI 48103, USA}
\author{D.A. Matoz-Fernandez}
\email{dmatoz@ucm.es}
\affiliation{Department of Theoretical Physics, Complutense University of Madrid, 28040 Madrid, Spain}
\author{A. Ko\v{s}mrlj}
\email{andrej@princeton.edu}
\affiliation{Department of Mechanical and Aerospace Engineering, Princeton University, Princeton, NJ 08544, USA}
\affiliation{Princeton Materials Institute, Princeton University, Princeton, NJ 08544, USA}
\date{July 2023}
\begin{abstract}
Non-equilibrium and active effects in mesoscopic scale systems have heralded a new era of scientific inquiries, whether concerning meta-materials or biological systems such as bacteria and cellular components. At mesoscopic scales, experimental and theoretical treatments of membranes, and other quasi-two-dimensional elastic surfaces cannot generically ignore Brownian motion and other thermal effects. In this paper we aim to study the behavior of thermally fluctuating 2-D elastic membranes possessing odd elastic moduli embedded in higher dimensions. We implement an isotropic generalization of the elastic tensor that includes odd elastic moduli, $K_{odd}$ and $A_{odd}$, that break conservation of energy and angular momentum respectively, due to Scheibner et al. \cite{scheibner2020odd}. Naturally this introduces active and non-equilibrium effects. Passive equilibrium thermalized elastic membranes possess effective (renormalized) Lamé coefficients that reduce with increasing system size and a diverging effective bending rigidity \cite{aronovitz1988fluctuations,le1992self}. Introducing two odd elastic moduli means that deformations from a reference state can induce chiral forces that cannot be derived from a Hamiltonian. Thus, the behavior of odd elastic membranes must  instead be investigated with Langevin equations. If fluctuation-dissipation relations hold, we calculate via the renormalization group that at long length scales, active effects due to $K_{odd}$ can be effectively ignored whereas $A_{odd}$ cannot. To validate these findings, we developed an advanced force implementation methodology, inspired by the $(T)$-scheme prevalent in vertex models. This contributed to a new method for the simulation of elastic membranes in higher dimensions, as detailed recently in \cite{matoz2020wrinkle}. The novelty of the simulation method is that microscopic/discrete and continuum in-plane elastic moduli are one-to-one and thus no coarse-graining is needed.
\end{abstract}
\maketitle
\textit{Introduction.} Elasticity has a long history and for much of it, has been typically studied in the presence of general laws and symmetries \cite{landau,timoshenko1959theory}. With the development of the Vicsek model and the Toner-Tu equations, the study of active systems has transformatively shifted the field of non-equilibrium physics to include active systems \cite{vicsek1995novel,toner1995long}. This shift has flung open the door to the burgeoning field of active systems, where long-standing interpretations of laws or symmetries, such as conservation of energy, are being re-examined and cast in a new light. For an interesting discussion of the history of active systems, which does go earlier than the aforementioned papers, see \cite{bowick2022symmetry}. 

Such new considerations have also led to a vast new set of experimental realizations and applications. One way to incorporate active effects is via the inclusion of non-conservative forces. For example, a synthetically engineered active elastic system was constructed out of polycrystals of magnetic colloidal spinners \cite{bililign2022motile}. Within the study, odd elastic forces were found to organize the colloids into a polycrystalline phase with motile dislocations whereby each grain has a tunable rotation rate and size. Odd elastic forces are not conservative but are only present when the system is deformed from a reference configuration \cite{scheibner2020odd}. Another example includes robotic metabeams with piezoelectric properties that can result in odd micropolar elasticity \cite{chen2021realization}. Such a system can convert mechanical and electrical energy into one another via bending/shearing cycles as well as produce direction-dependent bending rigidities. Additionally, work- generating limit cycles via active forces have been exploited to produce robotic elements \cite{brandenbourger2021limit}. Within natural phenomena, active elasticity is also relevant for biological systems. Analysis of already existing data of muscular hydraulics were found to exhibit odd elastic behavior \cite{shankar2022active}. In addition, new phases of active chiral biological matter were obtained in the form of spontaneously formed crystals consisting of starfish embryos \cite{tan2022odd}. By examining the local shear-elongation angle in the vicinity of defects, potential signatures of odd elasticity were also observed. In addition, theoretically, active elastic surfaces have been generalized via symmetry arguments and explored \cite{salbreux2017mechanics,fossati2022odd}, with particular emphasis on wave dynamics. One can also, for example, model the mechanics of the actin cortex, a layer of cross-linked actin that lies beneath the plasma membrane of animal cells \cite{berthoumieux2014active,salbreux2012actin}. The activity of this mesh arises from the myosin motors that exert contractile forces. More specifically amongst animal cells, the mechanical behavior of human red blood cells is a significant topic of interest. At these length scales thermal effects are relevant to the fluctuations and mechanics of membranes \cite{gov2007active,nelson2004statistical}. Active and non-equilibrium effects such as ATP concentration softening the local shear modulus of the cytoskeleton have been confirmed, leading to active height correlations that differ from thermal equilibrium height correlations \cite{gov2007active, park2010metabolic}. This softening effect, due to metabolic activity, can induce interesting curvature-mediated active membrane motions \cite{turlier2016equilibrium}. Given the possibility that odd elastic forces could be present in many biological systems such as the aggregates of star-fish embryos and the cytoskeleton \cite{salbreux2017mechanics, tan2022odd, floyd2022signatures}, it is worthwhile to consider what the behavior of thermalized odd elastic membranes may be. This paper aims to provide a first initial understanding of this topic. 

We use a renormalization-group analysis along with Langevin-dynamic simulations to investigate the long-range equal-time correlations of the displacement fields of 2-D solvent-permeable isotropic odd elastic membranes (which are necessarily chiral) at some non-zero temperature. These membranes are characterized by Lamé coefficients ($\lambda, \mu$), a bending rigidity ($\kappa$) and two odd elastic constants ($A_{odd}, K_{odd}$)\cite{ scheibner2020odd}. Both odd elastic moduli break reflection symmetry, and more specifically $K_{odd}$ breaks conservation of energy and couples pure and simple shears whereas $A_{odd}$ breaks conservation of angular momentum and couples dilation strains to torques. Finally, despite the fact that we performed theoretical calculations for permeable membranes (Rouse dynamics), we expect our scaling results to hold even in the case of impermeable membranes (Zimm dynamics), where the membrane is not completely permeable to the solvent and is modeled with Fourier vector dependent diffusivities. 

\textit{Model and Theoretical Results}
We seek to describe a $D=2$ isotropic odd elastic system embedded in $D+d_c$ dimensions (i.e. with $d_c$ co-dimensions) at a temperature $T$ \cite{scheibner2020odd}. In the absence of thermal fluctuations and deformations, the reference configuration is that of a flat undeformed elastic sheet. A schematic showing the elastic system under non-zero $T$ conditions is shown in Fig.~\ref{fig:Schematic}. Though such an active elastic system cannot be described by a Hamiltonian, the constitutive equation for in-plane stresses ($\sigma_{ij}$) and strains ($u_{ij}$) is still applicable $\sigma_{ij} = \mathcal{C}_{ijkl} u_{kl}$ where 

\begin{equation}
\begin{split}
 \mathcal{C}_{ijkl} =& \lambda \delta_{ij} \delta_{kl} + \mu [\delta_{ik} \delta_{jl} + \delta_{il} \delta_{jk}] \\ & + K_{odd} \mathcal{E}_{ijkl} -A_{odd} \epsilon_{ij} \delta_{kl},
        \label{eq:CModTensor}
\end{split}
\end{equation}

where we have assumed an orthonormal coordinate basis and with $\mathcal{E}_{ijkl} = \frac{1}{2}[\epsilon_{ik}\delta_{jl}+\epsilon_{il}\delta_{jk}+\epsilon_{jk}\delta_{il}+\epsilon_{jl}\delta_{ik}]$. $\epsilon$ is the 2-D Levi-Civita tensor ($\epsilon_{11}=\epsilon_{22}=0,\epsilon_{12}=-\epsilon_{21}=1$) and $\delta$ is the Kronecker delta. In addition, $\lambda, \mu$ are the 2-D Lamé coefficients. We define $u_i(\mathbf{r},t)$ to be the displacement vector along $i$-th axis (in-plane phonon) and $f^{\alpha}(\mathbf{r},t) ; \alpha \in \{1,...,d_c\}$ to be the out-of-plane displacement (flexural phonon). In the Monge gauge the strain tensor takes the form:
\begin{equation}
\begin{split}
    u_{ij}(\mathbf{r},t) = & \frac{1}{2} [\partial_i u_j(\mathbf{r},t) + \partial_j u_i(\mathbf{r},t) \\ & + \partial_i f^{\alpha}(\mathbf{r},t) \partial_j f^{\alpha}(\mathbf{r},t)].
    \label{eq:StrainTensor}
\end{split}
\end{equation}

From these, a system of over-damped Langevin equations can be written down:
\begin{equation}
        \partial_t u_j(\mathbf{r},t) = \mathcal{D} \partial_i\{\mathcal{C}_{ijkl}  u_{kl}(\mathbf{r},t) \}+\eta_j(\mathbf{r},t),
        \label{eq:InPlaneLangevin}
\end{equation}
where the Langevin noise satisfies $\langle \eta_j(\mathbf{r},t) \rangle =0$ and $\langle \eta_j(\mathbf{r},t) \eta_i(\mathbf{r}',t') \rangle = 2 \mathcal{L}_{ij} k_BT \delta(\mathbf{r} - \mathbf{r}') \delta(t-t')$ and we assume $\mathcal{L}_{ij} = \mathcal{L} \delta_{ij}$. $\mathcal{D}$ is the in-plane mobility and is equal to $\mathcal{L}$ if fluctuation-dissipation holds. We may also write a Föppl-von-Kármán equation describing the dynamics of flexural modes:
\begin{equation}
\begin{split}
        \partial_t f^{\alpha}(\mathbf{r},t) = &\mathcal{D}_f [ -  \partial_i \partial_j \{ \mathcal{B}_{ijkl} \partial_k \partial_l f^{\alpha}(\mathbf{r},t)\}\\
        &+ \partial_i\{\sigma_{ij}(\mathbf{r},t) \partial_j f^{\alpha}(\mathbf{r},t)\} ]+\eta_f^{\alpha}(\mathbf{r},t),
        \label{eq:FlexuralLangevin}
\end{split}
\end{equation}
where $\mathcal{B}_{ijkl} = \frac{\kappa}{2}  \delta_{ij} \delta_{kl} + \frac{\kappa}{4}  [\delta_{ik} \delta_{jl}+\delta_{il} \delta_{jk}]$ is the bending rigidity tensor. $\mathcal{D}_f$ is the out-of-plane mobility and is equal to $\mathcal{L}_f$ if fluctuation-dissipation relations hold. One may fairly ask why an odd elastic component has not been incorporated into this bending rigidity tensor but by noticing the symmetrized contraction in $\mathcal{B}_{ijkl}\partial_i \partial_j \partial_k \partial_l$, no anti-symmetric components will remain. Correspondingly the noise term satisfies $\langle \eta_f^{\alpha}(\mathbf{r},t) \rangle =0$ and $\langle \eta_f^{\alpha}(\mathbf{r},t) \eta_f^{\beta}(\mathbf{r}',t') \rangle = 2 \mathcal{L}_f k_BT \delta^{\alpha \beta} \delta(\mathbf{r} - \mathbf{r}') \delta(t-t')  $ where $\delta^{\alpha \beta}$ is the Kronecker delta. We additionally assume that $\langle \eta_f^{\alpha}(\mathbf{r},t) \eta_j(\mathbf{r}',t') \rangle = 0$. To comprehend the effect of thermal fluctuations on the effective elastic moduli, we use the Martin-Siggia-Rose-Janssen-DeDominicis formalism \cite{de1978dynamics,janssen1979field,martin1973statistical, tauber2014critical, frey1991dynamics}. The formalism adopts the path-integral formulation by taking advantage of the form of the distribution of the thermal noises. Thus the transition probability density is of the form:
\begin{equation}
\begin{split}
    \mathcal{W}(\eta_j,\eta_f^{\alpha}) \propto & e^{-\frac{1}{4}\int dt \int d^d \mathbf{r} (k_BT \mathcal{L})^{-1} \eta_i(\mathbf{r},t)^2 } \\
    & e^{-\frac{1}{4}\int dt \int d^d \mathbf{r} (k_BT\mathcal{L}_f)^{-1} \eta_f^{\alpha}(\mathbf{r},t)^2}.
\end{split}
\end{equation}
In this form, inserting in the Langevin equations, Eq.~\eqref{eq:FlexuralLangevin}and Eq.~\eqref{eq:InPlaneLangevin}, renders the expression into a complicated set of terms with high-degree non-linearities. Thus, via an imaginary Hubbard-Stratonovich transformation \cite{altland2010condensed} we introduce the following response non-physical variables and linearize the Langevin noises:
\begin{equation}
    \begin{split}
        \mathcal{W}(\eta_j,\eta_f^{\alpha}) \propto & \int \prod_{i}D[i\Upsilon_i]\prod_{\alpha}D[i\Phi^{\alpha}] \\ & e^{\int dt \int d^d \mathbf{r} [k_BT\mathcal{L} \Upsilon_i(\mathbf{r},t)^2 -\Upsilon_i(\mathbf{r},t)\eta_i(\mathbf{r},t) ]} \\
        &  e^{\int dt \int d^d \mathbf{r}[k_BT\mathcal{L}_f \Phi^{\alpha}(\mathbf{r},t)^2 -\Phi^{\alpha}(\mathbf{r},t)\eta_f^{\alpha}(\mathbf{r},t)]},
    \end{split}
    \label{eq:MSRJDAction}
\end{equation}
where the integral represents a functional integration \cite{tongstatistical}. Assuming periodic boundary conditions, one may then take a Fourier transform of the MSRJD action seen in Eq.~\ref{eq:MSRJDAction}. Using the MSRJD action, and keeping terms only to linear order, we may calculate the equal-time correlations (known as a harmonic/Gaussian average) of our displacement fields to obtain:
\begin{equation}
    \langle f^{\alpha}(\mathbf{q})f^{\beta}(-\mathbf{q}) \rangle = \frac{k_BT \mathcal{L}_f \delta_{\alpha \beta}}{ A  \mathcal{D}_f  \kappa q^4 },
\label{eq:FlexuralPropagatorStaticMT}
\end{equation}
\begin{widetext}
\begin{equation}
\begin{split}
    \langle u_i(\mathbf{q}) u_j(-\mathbf{q})\rangle = &  \frac{ k_BT\mathcal{L}}{A\mathcal{D} [\lambda+3\mu][\mu(\lambda+2\mu)+K_{odd}(K_{odd}-A_{odd})]q^4} \\  & \{ q_i q_j[\mu(\lambda+3\mu)+ K_{odd}(2 K_{odd}-A_{odd})]  \\ & + [\epsilon_{ik}q_jq_k + \epsilon_{jl}q_i q_l][\mu  A_{odd} +(\lambda+\mu) K_{odd}] \\ & + \epsilon_{ik}\epsilon_{jl} q_k q_l [(\lambda+2\mu)(\lambda+3\mu)+ (K_{odd}-A_{odd})(2K_{odd}-A_{odd})] \},
\label{eq:InPlanePropagatorStaticMT}
\end{split}
\end{equation}
\end{widetext}
where $A$ is the area of our system. These calculations are performed in Sec.~III of the supplementary material. Active effects cannot be observed in the out-of-plane correlations Eq.~\ref{eq:FlexuralPropagatorStaticMT}. One may ask if there is a correlation that is otherwise zero in the absence of odd elastic moduli. The solution is in-plane correlations of the type:
$\langle u_1(q_1,0)u_2(-q_1,0) \rangle = k_BT\mathcal{L} [A_{odd} \mu-K_{odd}(\lambda+\mu) ]/\{\mathcal{D}q^2(\lambda+3\mu)(A_{odd}K_{odd}+K_{odd}^2+\mu(\lambda+\mu))\}$. This will be of use when extracting the scaling of the effective odd elastic moduli from our simulation data.

Previous equilibrium studies, utilizing Boltzmann weights, established that due to the geometric non-linearity of the strain tensor, transverse shear modes generate a long-range coupling of Gaussian curvatures and thus the existence of a low-temperature flat phase for such elastic surfaces was established in \cite{nelson1987fluctuations}. The long-range coupling is what permits such 2-D systems to evade the Hohenberg-Mermin-Wagner theorem (which states that continuous symmetries cannot be spontaneously broken in two and less dimensions) \cite{mermin1966absence,mermin1968crystalline,hohenberg1967existence,halperin2019hohenberg}. Furthermore, via thermal fluctuations, the geometric non-linearity of the strain tensor renormalizes the moduli of the theory when fluctuations of the membrane become of the order of its thickness \cite{nelson1987fluctuations}. Further studies followed and established the presence of a globally stable non-Gaussian fixed point named the Aronovitz-Lubensky fixed point, as well as refined anomalous exponents for the moduli via a self-consistent screening analysis (SCSA) \cite{aronovitz1988fluctuations,le1992self}. More specifically it was obtained that beyond the thermal length scale ($\ell_{\text{th}} = \sqrt{16 \pi^3 \kappa^2/3k_BTY}$, where $\kappa$ is the bending rigidity, $k_B$ is the Boltzmann constant, $T$ is the temperature and $Y$ is the 2-D Young's modulus) the effective moduli of the material renormalize with system size, $L$, as $\kappa^R(L) \sim L^{\eta}$ ($\eta \approx 0.8$) whereas the Lamé coefficients scale as $\lambda^R(L),\mu^R(L) \sim L^{-\eta_u}$ ($\eta_u \approx 0.4$). 

One theoretical study utilizing our same model sans odd elastic moduli, has explored the Langevin dynamics of elastic membranes with the addition of hydrodynamic coupling when such membranes may or may not be permeable to the surrounding solvent (Rouse and Zimm dynamics respectively) \cite{frey1991dynamics}. While there are dynamical differences between the two cases, in the study it was found that, regardless of whether the membranes are permeable or not, the static correlations of the displacement fields are in agreement with the equilibrium theory using the Boltzmann measure.

Importantly, however, the MSRJD action has one notable difference from the analogous Boltzmann weight associated with an equilibrium elastic membrane Hamiltonian such as in \cite{aronovitz1988fluctuations,nelson1987fluctuations,nelson2004statistical,guitter1988crumpling}. The MSRJD action does not possess the quasi-rotational symmetry given in \cite{guitter1989thermodynamical}. The absence of this symmetry is significant as this means that the Ward identity, that would be derived from this symmetry, is also absent. Note that this difference holds even in the absence of activity and is thus a property of the Langevin dynamics. Thus, the form of the strain tensor is not necessarily preserved through the renormalization group. As an example, if one inserts Eq.~\ref{eq:StrainTensor} into Eq.~\ref{eq:InPlaneLangevin}, it may be noted that there are the linear and non-linear terms both with coefficient $\mathcal{D} \mathcal{C}_{ijkl}$. Though we may set these two coefficients to be equal at a microscopic length scale, there is no symmetry that says that this necessarily need be the case at larger length scales. In other words, the lack of this Ward identity may lead to the breaking of fluctuation-dissipation relations at larger length scales. Such a scenario, where the symmetry of the strain tensor is broken, has also been treated in \cite{le2021thermal}. To achieve this, the isotropy of the embedding space was broken via an external field to break the form of the strain tensor explicitly. These perturbations will generically result in a renormalization-group flow directed away from the Aronovitz-Lubensky fixed point. More details are expanded upon in Sec.~III of the supplementary material. However, for the purposes of this study, where we performed a performed 1-loop order $1/d_c$ renormalization group expansion, if microscopic fluctuation-dissipation relations hold and the form of the strain tensor is not violated microscopically, then they hold via renormalization as well (one may think of this as an unstable but fixed manifold).

We furthermore mention that by restricting to the Feynman diagrams we used (via the $1/d_c$ analysis expanded upon in the supplementary material) in our theoretical analysis, $A_{odd}$ and $K_{odd}$ do not generate each other. Thus they may be considered as independent parameters when tuning bare moduli such that either of the odd moduli is zero.

\begin{figure*}[t]
\includegraphics[width=\textwidth]{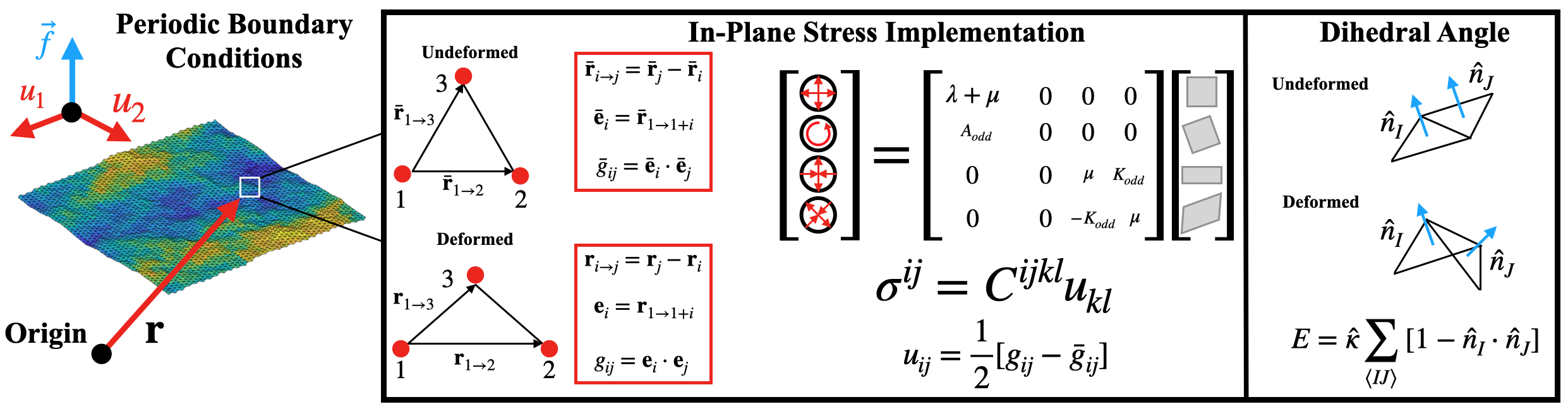}
\caption{A triangular lattice, with periodic boundary conditions, is implemented with fixed connectivity for Langevin simulations performed with a Berendsen barostat. The color field in the schematic indicates the height field with blue indicating lower portions of the elastic sheet and orange indicating higher portions. The vertices associated with each triangular face are assigned an ID $\{1,2,3\}$ in a counter-clockwise fashion. Reference coordinates are referred to as $\{ \bar{\mathbf{r}}_i \}$ where $i$ marks the ID of the vertex for the triangular face. $\{ \mathbf{r}_i \}$ marks the current positions of the vertices. These coordinates may be used to calculate the reference and current metrics and thereby the strain tensor for each triangular face. The dual in-plane stresses for each triangular face were then calculated via the constitutive relation whereas bending forces were implemented using dihedral springs.}
\label{fig:Schematic}
\end{figure*}

With these considerations we numerically integrate our renormalization group equations and performed a stability analysis of the Aronovitz-Lubensky fixed point explained further in the SI. Via our analyses, we find that in the vicinity of the Aronovitz-Lubensky fixed point, $K_{odd}$ is an irrelevant perturbation. Perturbations of non-zero $K_{odd}$, that preserve fluctuation-dissipation relations, converge back to the Aronovitz-Lubensky fixed point with exponent $K_{odd}^R(L) \sim L^{-2\eta_u}$. Thus  $\lambda^R(L),\mu^R(L) \sim L^{-\eta_u}$ and $\kappa^R(L) \sim L^{\eta}$ still hold. On the other hand, perturbations to non-zero $A_{odd}$, that preserve fluctuation-dissipation relations, are marginal and thus expand the Aronovitz-Lubensky fixed point to a higher-dimensional manifold. Thus, $A_{odd}^R(L) \sim L^{-\eta_u}$, $\lambda^R(L),\mu^R(L) \sim L^{-\eta_u}$ and $\kappa^R(L) \sim L^{\eta}$ still hold. In addition, in the supplementary material, we also derive a new form for the thermal length scale, $\ell_{\text{th}}$ that matches with the classical equilibrium form, $\sqrt{16 \pi^3 \kappa^2/3k_BTY}$, in the absence of the odd-elastic moduli and when fluctuation-dissipation relations hold. The new form of $\ell_{\text{th}}$ can be found in Eq.~(S105) in the supplementary material. Thus, for systems sufficiently larger than the thermal length scale and assuming fluctuation-dissipation relations, $K_{odd}^R$ can be ignored but $A_{odd}^R$ cannot, with regard to the effective mechanics of the sheet. Aside from the renormalization of equal-time correlations, from which we derive the scaling of our elastic moduli, dynamic renormalization of the noise variance does not occur \cite{frey1991dynamics}. Thus dynamic critical exponents can be completely determined from our static renormalization results, that is the scaling of our elastic moduli derived from equal-time correlations.

Finally, as far as fluctuation-dissipation relations hold, we make a remark that despite the fact that our results hold for completely permeable membranes, known as Rouse dynamics, these results should hold even in the case of Zimm dynamics where diffusivities are given a power dependence of the Fourier vector magnitude such as in \cite{frey1991dynamics}. This is because inserting a power of the Fourier vector magnitude into the diffusivities will multiply all terms in our theoretical calculation (aside from the temporal frequency, which is not relevant to equal-time correlations) and thus relative ratios of these terms remains the same as in Rouse dynamics. This will just shift the scaling of all the correlation functions by the same factor and thus relative scaling between elastic moduli will not change.

However, the equal-time correlation results change if fluctuation-dissipation relations do not hold microscopically, then we find the Aronovitz-Lubensky fixed point unstable. Thus, the Aronovitz-Lubensky fixed point does not characterize elastic membranes without fluctuation-dissipation relations. In the absence of odd elastic moduli, the governing fixed point was obtained in \cite{le2021thermal}. In the presence of odd elastic moduli, the governing fixed point is yet to be found. For more on the complete theoretical results please refer to Sec.~III of the supplementary material. 


\textit{Simulations Set-Up.}  
To compare and contrast the theoretical results, we also conducted Langevin simulations using following Ref.~\citep{matoz2020wrinkle}. We introduced a new force implementations, explained below, yielding a novel simulation technique for discrete membranes embedded in higher dimensional spaces. Specifically, we simulated a 2-D elastic membrane embedded in 3 dimensions so that $d_c=1$ under the influence of thermal noise. Simulations were performed non-dimensionally so $k_B$, the mass of the vertices and the lattice spacing were set to $1$. A Gaussian random noise associated with a variance of the temperature, $T$, is produced by a random number generator and acts as our Langevin noise. The temperature is a particularly important parameter for our simulations because rather than simulating a large system size, which can be computationally costly, we can fix the system size, $L$, and instead change the temperature, which will change the thermal length scale $\ell_{\text{th}}$ (thus allowing us to tune $L/\ell_{\text{th}}$). Measuring equal-time correlation functions of the displacements, we can then establish anomalous exponents of the effective moduli. This is expanded upon further below. The specific physical system we implemented was a triangular lattice with fixed connectivity under periodic boundary conditions. For the integrator, we implemented a BAOAB-limit method \cite{leimkuhler2013rational}, thus the simulation noise is not completely memory-less, but the memory is short ranged in time. We used a Berendsen barostat to tune the pressure of the system to zero; allowing us to tune the system to its critical point (without a homogeneous stress). We used dihedral springs to implement a bending rigidity for the elastic membrane \cite{seung1988defects}. We set the continuum bending rigidity to $1$ as well. To calculate in-plane forces, for each triangular face, the vertices that make up the corners are assigned an ID $\{1,2,3\}$ with a consistent handedness (counter-clockwise). Thus the same vertex can have a different ID with regards to different faces. $\{ \bar{\mathbf{r}}_i \}$ and $\{ \mathbf{r}_i \}$ mark the reference and current positions of the vertices where $i$ is the ID of the vertex within the triangular face in question. A schematic of this is shown in Fig.~\ref{fig:Schematic}, where the reference basis $\bar{\mathbf{e}}_i$ and current basis $\mathbf{e}_i$ are defined in terms of the coordinates of the vertices. We used the reference metric tensor, $\bar{g}_{ij}(\mathbf{r},t) = \bar{\mathbf{e}}_i(\mathbf{r},t) \cdot \bar{\mathbf{e}}_j(\mathbf{r},t)$, and current metric tensor, $g_{ij}(\mathbf{r},t) = \mathbf{e}_i(\mathbf{r},t) \cdot \mathbf{e}_j(\mathbf{r},t)$,  to calculate the strain tensor $u_{ij}(\mathbf{r},t) = [g_{ij}(\mathbf{r},t)-\bar{g}_{ij}(\mathbf{r},t)]/2$. We will drop the $(\mathbf{r},t)$ argument now as it is implicit. As an aside, we will use notation of differential geometry. For example, the dual of a general tensor, tensor $J_{ij}$ is written down as $J^{ij}$ and is obtained via the following contraction: $J^{ij} = g^{ik}g^{jl}J_{kl}$ where $g^{ij} \equiv (g^{-1})_{ij}$ (meaning that the dual of the metric tensor is its inverse; note that this property does not hold necessarily for other tensors) \cite{do2016differential}. Thus, in general, we use the current local metric tensor to raise and lower indices on a general tensor. Returning to the stress tensor calculation, we used the constitutive relation ($\sigma^{ij}=C^{ijkl}u_{kl}$) via the generalized linear elastic tensor:
\begin{equation}
\begin{split}
 \mathcal{C}^{ijkl} =& \lambda \bar{g}^{ij} \bar{g}^{kl} + \mu [\bar{g}^{ik} \bar{g}^{jl} + \bar{g}^{il} \bar{g}^{jk}] \\&+ \frac{K^{odd}}{\sqrt{\det[\bar{g}]}} \mathcal{E}^{ijkl} -\frac{A_{odd}}{\sqrt{\det[\bar{g}]}} \epsilon^{ij} \bar{g}^{kl},
        \label{eq:CModTensorSim}
\end{split}
\end{equation}
with $\mathcal{E}^{ijkl} = \frac{1}{2}[\epsilon^{ik}\bar{g}^{jl}+\epsilon^{il}\bar{g}^{jk}+\epsilon^{jk}\bar{g}^{il}+\epsilon^{jl}\bar{g}^{ik}]$, where $\epsilon^{ij}$ is again the 2-D Levi-Civita permutation symbol satisfying $\epsilon^{11}=\epsilon^{22}=0, \epsilon^{12} = -\epsilon^{21}=1$.

\begin{figure*}[t]
\includegraphics[width=\textwidth]{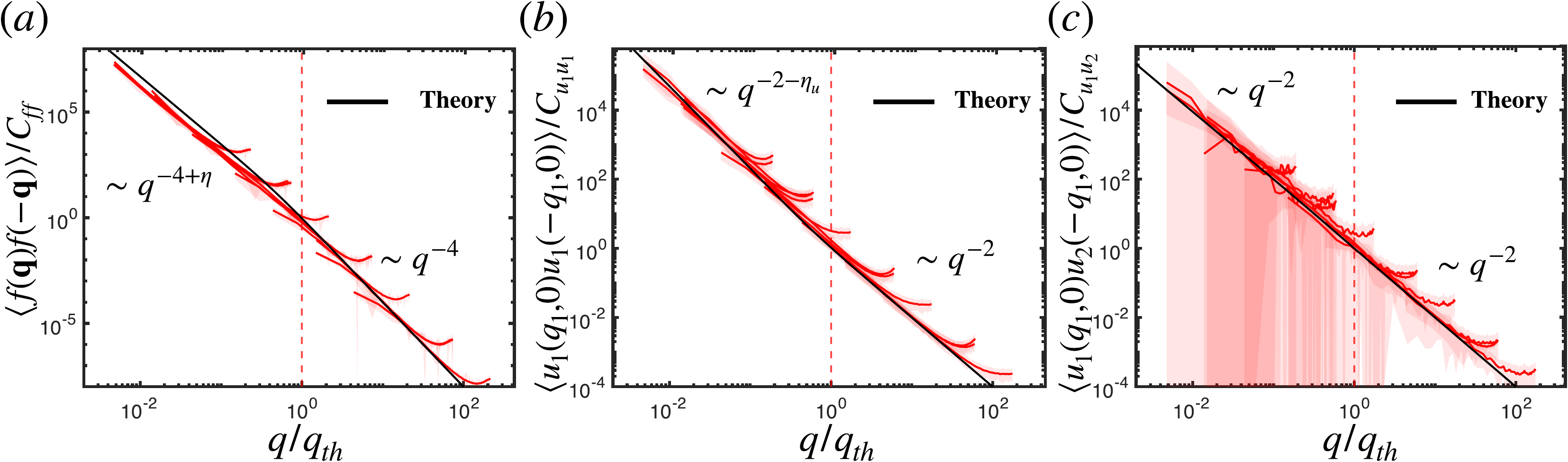}
\caption{Plots of the non-dimensionalized (a) flexural correlation function ($C_{ff} = k_BT/(A \kappa q_{\text{th}}^4)$), (b) longitudinal in-plane phonon correlation function ($C_{u_1u_1} =k_BT(2K_{odd}^2+\mu(\lambda+3\mu))/( A [\lambda +3 \mu][K_{odd}^2+\mu(\lambda+2\mu)] q_{\text{th}}^2 ) $) and (c) semi-transverse in-plane phonon correlation function ($C_{u_1u_2}=k_BT K_{odd}(\lambda+\mu)/(A [\lambda +3 \mu][K_{odd}^2+\mu(\lambda+2\mu)] q_{\text{th}}^2)  $). The shading indicates the error bars showing the standard error of the mean. For $q>q_{\text{th}}$, the harmonic correlations are obtained. The Aronovitz-Lubensky anomalous exponents, $\eta \approx 0.8,\eta_u \approx 0.4$ are observed in (a) and (b) for $q<q_{\text{th}}$. Instead in (c), no anomalous exponents are observed and this is consistent with $K_{odd}^R(q) \sim q^{2 \eta_u}$ being irrelevant. Corresponding data sets can be found in Tab.~\ref{tab:tabdataKplot}.}
\label{fig:KPlot}
\end{figure*}

Once the stress tensor is calculated for a face, we calculated the normal associated with the local basis of the triangular face: $\hat{\mathbf{n}}_{face} = \mathbf{e}_{1} \times \mathbf{e}_{2} / \| \mathbf{e}_{1} \times \mathbf{e}_{2} \|$ and then the normal to an edge of the triangular face via an additional cross product $\mathbf{n}_{i \rightarrow j} = \hat{\mathbf{n}}_{face} \times \mathbf{r}_{i \rightarrow j}$ (note that this normal is not of length $1$; in addition the symbol, $\hat{}$, marks a unit vector). The normal vector can be decomposed in the current basis $\{ \mathbf{e}_{1}, \mathbf{e}_{2} \}$ as follows: $(N_{i \rightarrow j})_k = \mathbf{n}_{i \rightarrow j} \cdot \mathbf{e}_{k} $ for $k=\{1,2\}$. We can then calculate the force on the edge as follows: $(T_{i \rightarrow j})^k = (N_{i \rightarrow j})_l \sigma^{lk}$ and $\mathbf{F}_{ij} = (T_{i \rightarrow j})^k \mathbf{e}_{k}$. To implement the force on the edge, one can halve $\mathbf{F}_{ij}$ and apply each half on each vertex $\{ i,j\}$. Such an implementation is similar to that done in the field of vertex models, the flat analogue bears the name (T)-scheme \cite{lin2022implementation,tlili2015colloquium,tlili2019shaping}. This method bears particular advantages over the use of spring-mass systems. Firstly, via a coarse-graining procedure one obtains that the continuum in-plane moduli, $\mathcal{C}_{ijkl}$, are the exactly the same as those that we use in our force implementation. Secondly, any stable elastic system, anisotropic or not and odd or not, can be simulated very easily via this method. With springs, in order to simulate $A_{odd}$ and $K_{odd}$ independently, for example, one would require a unit cell that is larger than the triangular face and may require a complicated microscopic force-displacement relation. Due to the fact that the continuum description is being directly simulated in this new method, these complications can be avoided. As for the time-step, $\Delta_{\tau}$, we selected a small enough value such that all the inverse of the over-damped frequencies  associated with our system ($a^4/[8 \pi^3 \mathcal{D}_f \kappa], a^2/[2\pi\mathcal{D}\mathcal{C}_{ijkl}]$) were larger. For our simulation procedure we had an equilibration/thermalization period of $2 \times 10^7 \Delta_{\tau}$ after which we began to record realizations of the system for $8 \times 10^8 \Delta_{\tau}$ at an interval of $10^5 \Delta_{\tau}$ time steps. We determined the thermalization length by performing autocorrelations of the displacement fields to know over what period of time the system retains memory of the initial configuration. In certain instances, particularly for simulations at lower temperatures, achieving thermal equilibrium within the system presents a challenge. Detailed elaboration on this topic can be found in Sec.~II of the supplementary material.

Given from our theoretical analysis that $A_{odd}$ and $K_{odd}$ do not generate each other, we simulate odd elastic systems that satisfy $A_{odd}=0$ or $K_{odd}=0$ with regards to the bare moduli. In this way, we may examine the effect of each modulus independently and further avoid potential non-universal characteristics and instabilities \cite{scheibner2020odd}. We then measured the equal-time correlations of our displacement fields by averaging over our snapshots of data. We utilize the form of the equal-time correlations, Eq.~\eqref{eq:FlexuralPropagatorStaticMT} and Eq.~\eqref{eq:InPlanePropagatorStaticMT}, replacing the microscopic values $\kappa,\lambda,\mu,A_{odd},K_{odd}$ with $\kappa^R(q),\lambda^R(q),\mu^R(q),A_{odd}^R(q),K_{odd}^R(q)$ and set $\mathcal{L}=\mathcal{D}, \mathcal{L}_f=\mathcal{D}_f$. Since no spatial symmetries are broken and we do not expect any anisotropies to develop within the system. With these formulas we can then extrapolate the scaling of the parameters $\kappa^R(q),\lambda^R(q),\mu^R(q),A_{odd}^R(q),K_{odd}^R(q)$ from the simulation data.

\textit{Results For $K_{odd}$.}
Correlation functions extracted from simulations with $K_{odd}$ can be found in Fig.~\ref{fig:KPlot}. One can find results consistent with theoretical exponents extracted via the 1-loop $1/d_c$ expansion. That is, on the one hand, for $q>q_{\text{th}}$, equal-time harmonic correlations are obtained, that is, Eq.~\eqref{eq:FlexuralPropagatorStaticMT} and Eq.~\eqref{eq:InPlanePropagatorStaticMT}. Thus below the thermal length scale ($q>q_{\text{th}}$), the mechanical response is described by the bare moduli. On the other hand, for $q<q_{\text{th}}$, $\langle f(\mathbf{q}) f(-\mathbf{q})\rangle \sim q^{-4 + \eta}$ and thus effectively $\kappa^R(q)/\kappa \sim q^{-\eta}$. The equal-time longitudinal-phonon correlation scales as $\langle u_1(q_1,0)u_1(-q_1,0)\rangle \sim q^{-2-\eta_u}$. On the other hand, the semi-transverse in-plane phonon correlations never exhibit anomalous behavior, $\langle u_1(q_1,0)u_2(-q_1,0)\rangle \sim q^{-2}$. With the ansatz that in the vicinity of the Aronovitz-Lubensky fixed point, $\lambda^R(q),\mu^R(q) \sim q^{\eta_u}$ and positing the correlation $K_{odd}^R(q)\sim q^{\alpha}$, then $\langle u_1(q_1,0)u_2(-q_1,0)\rangle \sim q^{-2+\alpha-2\eta_u}$. Thus these in-plane correlations could only be explained by the scaling: $K_{odd}^R(q) \sim q^{2\eta_u}$, which is consistent with our theoretical 1-loop $1/d_c$ expansion calculations. Such a scaling is indicative that small perturbations in $K_{odd}$ to the Aronovitz-Lubensky fixed point are irrelevant and thus $K_{odd}^R(q)/\mu^R(q)$ scales to zero in the thermodynamic limit. The significance of this is that breaking conservation of energy in the manner in which $K_{odd}$ does is irrelevant (though $\mu^R(q),\lambda^R(q)$ do scale to zero in the thermodynamic limit, they are examples of dangerously irrelevant parameters. Instead $K_{odd}^R(q)$ is truly irrelevant).
Correlation functions extracted from simulations with $K_{odd}$ can be found in Fig.~\ref{fig:KPlot}. We remind the readers that these results hold when fluctuation-dissipation relations hold. 

\begin{figure*}[t]
\includegraphics[width=\textwidth]{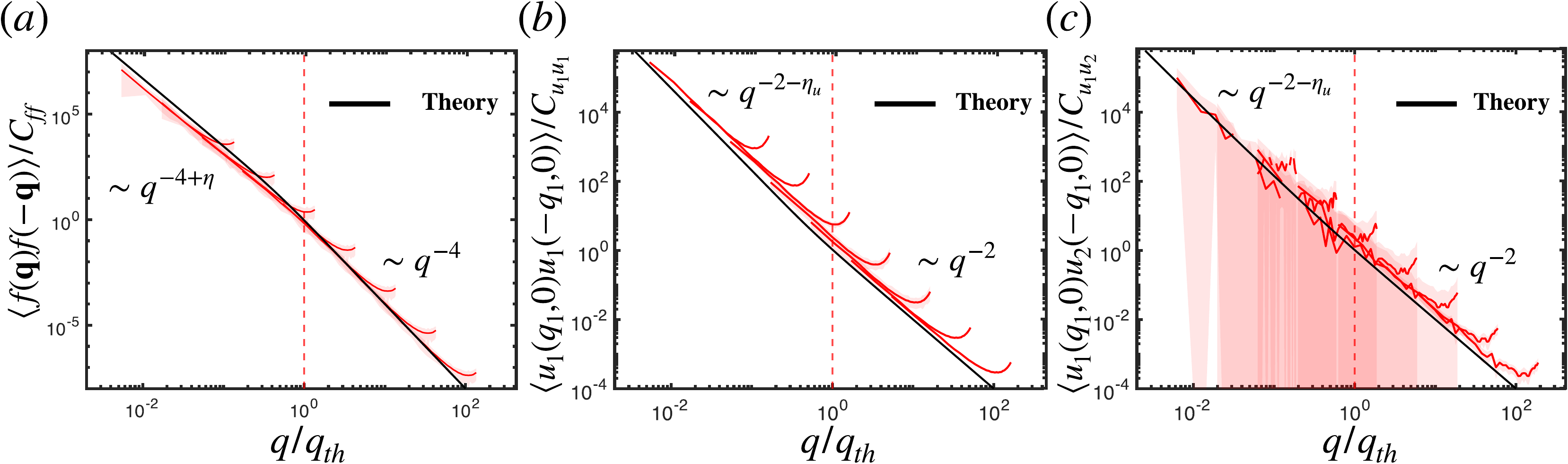}
\caption{Plots of the non-dimensionalized (a) flexural correlation function ($C_{ff} = k_BT/(A \kappa q_{\text{th}}^4)$), (b) longitudinal in-plane phonon correlation function ($C_{u_1u_1} = k_BT/(A (\lambda+2\mu) q_{\text{th}}^2) $) and (c) semi-transverse in-plane phonon correlation function ($C_{u_1u_2}=k_BT A_{odd}/(A [\lambda +3 \mu][\lambda+2\mu] q_{\text{th}}^2  )  $). The shading indicates the error bars showing the standard error of the mean. For $q>q_{\text{th}}$, the harmonic correlations are obtained. The Aronovitz-Lubensky anomalous exponents, $\eta \approx 0.8,\eta_u \approx 0.4$ are observed in (a) and (b) for $q<q_{\text{th}}$. In (c), the $\eta_u$ anomalous exponent is again observed and this is consistent with $A_{odd}^R(q) \sim q^{\eta_u}$ being marginal. Corresponding data sets can be found in Tab.~\ref{tab:tabdataKplot}.}
\label{fig:APlot}
\end{figure*}

\textit{Results For $A_{odd}$.}
Correlation functions extracted from simulations with $A_{odd}$ can be found in Fig.~\ref{fig:APlot}. One can find results consistent with theoretical exponents extracted via the 1-loop $1/d_c$ expansion. That is, on the one hand, for $q>q_{\text{th}}$, harmonic correlations are obtained. Whereas for $q<q_{\text{th}}$, $\langle f(\mathbf{q}) f(-\mathbf{q})\rangle \sim q^{-4 + \eta}$ and $\langle u_1(q_1,0)u_1(-q_1,0)\rangle \sim q^{-2-\eta_u}$ which exhibits the exponents associated with the Aronovitz-Lubensky fixed point once again. The semi-transverse in-plane phonon correlations exhibit the following scaling, $\langle u_1(q_1,0)u_2(-q_1,0)\rangle \sim q^{-2-\eta_u}$, however, some portions of the correlation functions are missing due to large fluctuations. One can make sense of this by understanding that the microscopic value we took for $A_{odd}$ is perturbative to $\lambda$ and $\mu$ and thus $\langle u_1(q_1,0)u_2(-q_1,0)$, which is zero in the absence of odd elastic moduli, experiences a greater amount of noise and error. Returning to the scaling we have obtained: positing the following scaling $A^R_{odd}(q) \sim q^{\alpha}$ then $\langle u_1(q_1,0)u_2(-q_1,0)\rangle \sim q^{-2+\alpha-2\eta_u}$ and can thus only be explained by the scaling: $A_{odd}^R(q) \sim q^{\eta_u}$. Such a scaling is indicative that small perturbations in $A_{odd}$ to the Aronovitz-Lubensky fixed point are marginal and thus $A_{odd}^R(q)/\mu_{odd}^R(q) \sim \text{constant}$ in the thermodynamic limit. Thus breaking conservation of angular momentum, in the manner that $A_{odd}$ does, cannot be ignored at the critical point and thermal fluctuations cannot restore this manner in which the reflection symmetry is broken. Thus the Aronovitz-Lubensky is no longer a fixed point but becomes a higher-dimensional fixed manifold when fluctuation-dissipation relations hold.

\textit{Conclusion.}
We have extracted the behavior of thermally fluctuating elastic membranes in the presence of non-equilibrium odd elastic moduli. For perturbative values of the odd elastic moduli, we observe that $K_{odd}^R(q) \sim q^{2 \eta_u}$ is an irrelevant parameter which can be ignored in the thermodynamic limit whereas $A_{odd}^R(q) \sim q^{\eta_u}$ is marginal and cannot be ignored at any scale. We emphasize once again that this study should be considered as an investigation in the case where these odd elastic moduli are perturbations. Potentially for very large odd elastic moduli, other behaviors and potential non-linear instabilities are possible (for which the stress may not be able to be tuned to zero for given system sizes). In addition, further exploration is merited on the basis that one can also microscopically break the fluctuation-dissipation relations and our eigen-value analysis of the Aronovitz-Lubensky fixed point yields unstable eigen-vectors. This analysis, in the absence of odd elastic moduli, has been obtained in \cite{le2021thermal}, where the same problem with broken fluctuation-dissipation relations can be mapped to an equilibrium problem that breaks the isotropy of the embedding space.

Obtaining a globally stable fixed point in the full phase space where fluctuation-dissipation relations do not hold and in the presence of odd elastic moduli is unresolved at the current time. Despite this, an important point to note is that active components making up an elastic network, such as actin or myosin in biological systems, are not to be necessarily associated with the breaking of fluctuation-dissipation relations. The origin of the non-equilibrium nature of biological systems, such as the cytoskeleton, would not originate from odd active behavior necessarily. Instead, it may come from non-Gaussian noise distributions as well as chemical reactions. For example, in the context of the KPZ equation, multiplicative noise was even considered to probe non-equilibrium properties \cite{tu1997systems}. On the other hand, in the context of chemical reactions, \cite{park2010metabolic} found that metabolic activity of ATP can help facilitate non-equilibrium fluctuations of membranes. To model this, one may conduct simulations and repeat a similar analysis with a concentration parameter analagous to what is done in \cite{turlier2016equilibrium}. In addition, our results are valid for permeable membranes, further can be done to explore membrane-solvent interactions as was done in \cite{frey1991dynamics}. We expect that our equal-time correlation results should hold even in the case of Zimm dynamics. Purely hydrodynamic solvent interactions are expected to modify the dynamical scaling but not the static~\cite{frey1991dynamics}, though this remains to be verified via simulations. Lastly, many more open directions can be investigated by incorporating other forms of non-equilibrium features and activity such as non-Gaussian noises and odd visco-elastic components for example \cite{tu1997systems, floyd2022signatures}. Thus, various open problems remain to fully explore the non-equilibrium inducing aspects of our system to properly model biological and potential synthetic systems at mesoscopic scales. 

\textit{Acknowledgements.}
The authors thanks Uwe Taüber for a useful discussion in development of the ideas with regards to fluctuation-dissipation relations, and John Toner for a discussion on mass renormalization. The authors would also like to acknowledge the following funding source: NSF Career award DMR-1752100. DMF was supported by the Comunidad de Madrid and the Complutense University of Madrid (Spain) through the Atracción de Talento program 2022-T1/TIC-24007.

\let\oldaddcontentsline\addcontentsline
\renewcommand{\addcontentsline}[3]{}
\bibliographystyle{apsrev4-1}
\bibliography{refer}

\begin{thebibliography}{53}%
\makeatletter
\providecommand \@ifxundefined [1]{%
 \@ifx{#1\undefined}
}%
\providecommand \@ifnum [1]{%
 \ifnum #1\expandafter \@firstoftwo
 \else \expandafter \@secondoftwo
 \fi
}%
\providecommand \@ifx [1]{%
 \ifx #1\expandafter \@firstoftwo
 \else \expandafter \@secondoftwo
 \fi
}%
\providecommand \natexlab [1]{#1}%
\providecommand \enquote  [1]{``#1''}%
\providecommand \bibnamefont  [1]{#1}%
\providecommand \bibfnamefont [1]{#1}%
\providecommand \citenamefont [1]{#1}%
\providecommand \href@noop [0]{\@secondoftwo}%
\providecommand \href [0]{\begingroup \@sanitize@url \@href}%
\providecommand \@href[1]{\@@startlink{#1}\@@href}%
\providecommand \@@href[1]{\endgroup#1\@@endlink}%
\providecommand \@sanitize@url [0]{\catcode `\\12\catcode `\$12\catcode
  `\&12\catcode `\#12\catcode `\^12\catcode `\_12\catcode `\%12\relax}%
\providecommand \@@startlink[1]{}%
\providecommand \@@endlink[0]{}%
\providecommand \url  [0]{\begingroup\@sanitize@url \@url }%
\providecommand \@url [1]{\endgroup\@href {#1}{\urlprefix }}%
\providecommand \urlprefix  [0]{URL }%
\providecommand \Eprint [0]{\href }%
\providecommand \doibase [0]{http://dx.doi.org/}%
\providecommand \selectlanguage [0]{\@gobble}%
\providecommand \bibinfo  [0]{\@secondoftwo}%
\providecommand \bibfield  [0]{\@secondoftwo}%
\providecommand \translation [1]{[#1]}%
\providecommand \BibitemOpen [0]{}%
\providecommand \bibitemStop [0]{}%
\providecommand \bibitemNoStop [0]{.\EOS\space}%
\providecommand \EOS [0]{\spacefactor3000\relax}%
\providecommand \BibitemShut  [1]{\csname bibitem#1\endcsname}%
\let\auto@bib@innerbib\@empty
\bibitem [{\citenamefont {Scheibner}\ \emph {et~al.}(2020)\citenamefont
  {Scheibner}, \citenamefont {Souslov}, \citenamefont {Banerjee}, \citenamefont
  {Sur{\'o}wka}, \citenamefont {Irvine},\ and\ \citenamefont
  {Vitelli}}]{scheibner2020odd}%
  \BibitemOpen
  \bibfield  {author} {\bibinfo {author} {\bibfnamefont {C.}~\bibnamefont
  {Scheibner}}, \bibinfo {author} {\bibfnamefont {A.}~\bibnamefont {Souslov}},
  \bibinfo {author} {\bibfnamefont {D.}~\bibnamefont {Banerjee}}, \bibinfo
  {author} {\bibfnamefont {P.}~\bibnamefont {Sur{\'o}wka}}, \bibinfo {author}
  {\bibfnamefont {W.~T.}\ \bibnamefont {Irvine}}, \ and\ \bibinfo {author}
  {\bibfnamefont {V.}~\bibnamefont {Vitelli}},\ }\href@noop {} {\bibfield
  {journal} {\bibinfo  {journal} {Nature Physics}\ }\textbf {\bibinfo {volume}
  {16}},\ \bibinfo {pages} {475} (\bibinfo {year} {2020})}\BibitemShut
  {NoStop}%
\bibitem [{\citenamefont {Aronovitz}\ and\ \citenamefont
  {Lubensky}(1988)}]{aronovitz1988fluctuations}%
  \BibitemOpen
  \bibfield  {author} {\bibinfo {author} {\bibfnamefont {J.~A.}\ \bibnamefont
  {Aronovitz}}\ and\ \bibinfo {author} {\bibfnamefont {T.~C.}\ \bibnamefont
  {Lubensky}},\ }\href@noop {} {\bibfield  {journal} {\bibinfo  {journal}
  {Physical review letters}\ }\textbf {\bibinfo {volume} {60}},\ \bibinfo
  {pages} {2634} (\bibinfo {year} {1988})}\BibitemShut {NoStop}%
\bibitem [{\citenamefont {Le~Doussal}\ and\ \citenamefont
  {Radzihovsky}(1992)}]{le1992self}%
  \BibitemOpen
  \bibfield  {author} {\bibinfo {author} {\bibfnamefont {P.}~\bibnamefont
  {Le~Doussal}}\ and\ \bibinfo {author} {\bibfnamefont {L.}~\bibnamefont
  {Radzihovsky}},\ }\href@noop {} {\bibfield  {journal} {\bibinfo  {journal}
  {Physical review letters}\ }\textbf {\bibinfo {volume} {69}},\ \bibinfo
  {pages} {1209} (\bibinfo {year} {1992})}\BibitemShut {NoStop}%
\bibitem [{\citenamefont {Matoz-Fernandez}\ \emph {et~al.}(2020)\citenamefont
  {Matoz-Fernandez}, \citenamefont {Davidson}, \citenamefont {Stanley-Wall},\
  and\ \citenamefont {Sknepnek}}]{matoz2020wrinkle}%
  \BibitemOpen
  \bibfield  {author} {\bibinfo {author} {\bibfnamefont {D.}~\bibnamefont
  {Matoz-Fernandez}}, \bibinfo {author} {\bibfnamefont {F.~A.}\ \bibnamefont
  {Davidson}}, \bibinfo {author} {\bibfnamefont {N.~R.}\ \bibnamefont
  {Stanley-Wall}}, \ and\ \bibinfo {author} {\bibfnamefont {R.}~\bibnamefont
  {Sknepnek}},\ }\href@noop {} {\bibfield  {journal} {\bibinfo  {journal}
  {Physical Review Research}\ }\textbf {\bibinfo {volume} {2}},\ \bibinfo
  {pages} {013165} (\bibinfo {year} {2020})}\BibitemShut {NoStop}%
\bibitem [{\citenamefont {Landau}\ \emph {et~al.}(2012)\citenamefont {Landau},
  \citenamefont {Pitaevskii}, \citenamefont {Kosevich},\ and\ \citenamefont
  {Lifshitz}}]{landau}%
  \BibitemOpen
  \bibfield  {author} {\bibinfo {author} {\bibfnamefont {L.~D.}\ \bibnamefont
  {Landau}}, \bibinfo {author} {\bibfnamefont {L.~P.}\ \bibnamefont
  {Pitaevskii}}, \bibinfo {author} {\bibfnamefont {A.~M.}\ \bibnamefont
  {Kosevich}}, \ and\ \bibinfo {author} {\bibfnamefont {E.~M.}\ \bibnamefont
  {Lifshitz}},\ }\href@noop {} {\emph {\bibinfo {title} {Theory of
  Elasticity}}},\ \bibinfo {edition} {3rd}\ ed.\ (\bibinfo  {publisher}
  {Butterworth-Heinemann},\ \bibinfo {year} {2012})\BibitemShut {NoStop}%
\bibitem [{\citenamefont {Timoshenko}\ and\ \citenamefont
  {Woinowsky-Krieger}(1959)}]{timoshenko1959theory}%
  \BibitemOpen
  \bibfield  {author} {\bibinfo {author} {\bibfnamefont {S.~P.}\ \bibnamefont
  {Timoshenko}}\ and\ \bibinfo {author} {\bibfnamefont {S.}~\bibnamefont
  {Woinowsky-Krieger}},\ }\href@noop {} {\emph {\bibinfo {title} {Theory of
  plates and shells}}}\ (\bibinfo  {publisher} {McGraw-hill},\ \bibinfo {year}
  {1959})\BibitemShut {NoStop}%
\bibitem [{\citenamefont {Vicsek}\ \emph {et~al.}(1995)\citenamefont {Vicsek},
  \citenamefont {Czir{\'o}k}, \citenamefont {Ben-Jacob}, \citenamefont
  {Cohen},\ and\ \citenamefont {Shochet}}]{vicsek1995novel}%
  \BibitemOpen
  \bibfield  {author} {\bibinfo {author} {\bibfnamefont {T.}~\bibnamefont
  {Vicsek}}, \bibinfo {author} {\bibfnamefont {A.}~\bibnamefont {Czir{\'o}k}},
  \bibinfo {author} {\bibfnamefont {E.}~\bibnamefont {Ben-Jacob}}, \bibinfo
  {author} {\bibfnamefont {I.}~\bibnamefont {Cohen}}, \ and\ \bibinfo {author}
  {\bibfnamefont {O.}~\bibnamefont {Shochet}},\ }\href@noop {} {\bibfield
  {journal} {\bibinfo  {journal} {Physical review letters}\ }\textbf {\bibinfo
  {volume} {75}},\ \bibinfo {pages} {1226} (\bibinfo {year}
  {1995})}\BibitemShut {NoStop}%
\bibitem [{\citenamefont {Toner}\ and\ \citenamefont
  {Tu}(1995)}]{toner1995long}%
  \BibitemOpen
  \bibfield  {author} {\bibinfo {author} {\bibfnamefont {J.}~\bibnamefont
  {Toner}}\ and\ \bibinfo {author} {\bibfnamefont {Y.}~\bibnamefont {Tu}},\
  }\href@noop {} {\bibfield  {journal} {\bibinfo  {journal} {Physical review
  letters}\ }\textbf {\bibinfo {volume} {75}},\ \bibinfo {pages} {4326}
  (\bibinfo {year} {1995})}\BibitemShut {NoStop}%
\bibitem [{\citenamefont {Bowick}\ \emph {et~al.}(2022)\citenamefont {Bowick},
  \citenamefont {Fakhri}, \citenamefont {Marchetti},\ and\ \citenamefont
  {Ramaswamy}}]{bowick2022symmetry}%
  \BibitemOpen
  \bibfield  {author} {\bibinfo {author} {\bibfnamefont {M.~J.}\ \bibnamefont
  {Bowick}}, \bibinfo {author} {\bibfnamefont {N.}~\bibnamefont {Fakhri}},
  \bibinfo {author} {\bibfnamefont {M.~C.}\ \bibnamefont {Marchetti}}, \ and\
  \bibinfo {author} {\bibfnamefont {S.}~\bibnamefont {Ramaswamy}},\ }\href@noop
  {} {\bibfield  {journal} {\bibinfo  {journal} {Physical Review X}\ }\textbf
  {\bibinfo {volume} {12}},\ \bibinfo {pages} {010501} (\bibinfo {year}
  {2022})}\BibitemShut {NoStop}%
\bibitem [{\citenamefont {Bililign}\ \emph {et~al.}(2022)\citenamefont
  {Bililign}, \citenamefont {Balboa~Usabiaga}, \citenamefont {Ganan},
  \citenamefont {Poncet}, \citenamefont {Soni}, \citenamefont {Magkiriadou},
  \citenamefont {Shelley}, \citenamefont {Bartolo},\ and\ \citenamefont
  {Irvine}}]{bililign2022motile}%
  \BibitemOpen
  \bibfield  {author} {\bibinfo {author} {\bibfnamefont {E.~S.}\ \bibnamefont
  {Bililign}}, \bibinfo {author} {\bibfnamefont {F.}~\bibnamefont
  {Balboa~Usabiaga}}, \bibinfo {author} {\bibfnamefont {Y.~A.}\ \bibnamefont
  {Ganan}}, \bibinfo {author} {\bibfnamefont {A.}~\bibnamefont {Poncet}},
  \bibinfo {author} {\bibfnamefont {V.}~\bibnamefont {Soni}}, \bibinfo {author}
  {\bibfnamefont {S.}~\bibnamefont {Magkiriadou}}, \bibinfo {author}
  {\bibfnamefont {M.~J.}\ \bibnamefont {Shelley}}, \bibinfo {author}
  {\bibfnamefont {D.}~\bibnamefont {Bartolo}}, \ and\ \bibinfo {author}
  {\bibfnamefont {W.~T.}\ \bibnamefont {Irvine}},\ }\href@noop {} {\bibfield
  {journal} {\bibinfo  {journal} {Nature Physics}\ }\textbf {\bibinfo {volume}
  {18}},\ \bibinfo {pages} {212} (\bibinfo {year} {2022})}\BibitemShut
  {NoStop}%
\bibitem [{\citenamefont {Chen}\ \emph {et~al.}(2021)\citenamefont {Chen},
  \citenamefont {Li}, \citenamefont {Scheibner}, \citenamefont {Vitelli},\ and\
  \citenamefont {Huang}}]{chen2021realization}%
  \BibitemOpen
  \bibfield  {author} {\bibinfo {author} {\bibfnamefont {Y.}~\bibnamefont
  {Chen}}, \bibinfo {author} {\bibfnamefont {X.}~\bibnamefont {Li}}, \bibinfo
  {author} {\bibfnamefont {C.}~\bibnamefont {Scheibner}}, \bibinfo {author}
  {\bibfnamefont {V.}~\bibnamefont {Vitelli}}, \ and\ \bibinfo {author}
  {\bibfnamefont {G.}~\bibnamefont {Huang}},\ }\href@noop {} {\bibfield
  {journal} {\bibinfo  {journal} {Nature communications}\ }\textbf {\bibinfo
  {volume} {12}},\ \bibinfo {pages} {5935} (\bibinfo {year}
  {2021})}\BibitemShut {NoStop}%
\bibitem [{\citenamefont {Brandenbourger}\ \emph {et~al.}(2021)\citenamefont
  {Brandenbourger}, \citenamefont {Scheibner}, \citenamefont {Veenstra},
  \citenamefont {Vitelli},\ and\ \citenamefont
  {Coulais}}]{brandenbourger2021limit}%
  \BibitemOpen
  \bibfield  {author} {\bibinfo {author} {\bibfnamefont {M.}~\bibnamefont
  {Brandenbourger}}, \bibinfo {author} {\bibfnamefont {C.}~\bibnamefont
  {Scheibner}}, \bibinfo {author} {\bibfnamefont {J.}~\bibnamefont {Veenstra}},
  \bibinfo {author} {\bibfnamefont {V.}~\bibnamefont {Vitelli}}, \ and\
  \bibinfo {author} {\bibfnamefont {C.}~\bibnamefont {Coulais}},\ }\href@noop
  {} {\bibfield  {journal} {\bibinfo  {journal} {arXiv preprint
  arXiv:2108.08837}\ } (\bibinfo {year} {2021})}\BibitemShut {NoStop}%
\bibitem [{\citenamefont {Shankar}\ and\ \citenamefont
  {Mahadevan}(2022)}]{shankar2022active}%
  \BibitemOpen
  \bibfield  {author} {\bibinfo {author} {\bibfnamefont {S.}~\bibnamefont
  {Shankar}}\ and\ \bibinfo {author} {\bibfnamefont {L.}~\bibnamefont
  {Mahadevan}},\ }\href@noop {} {\bibfield  {journal} {\bibinfo  {journal}
  {bioRxiv}\ ,\ \bibinfo {pages} {2022}} (\bibinfo {year} {2022})}\BibitemShut
  {NoStop}%
\bibitem [{\citenamefont {Tan}\ \emph {et~al.}(2022)\citenamefont {Tan},
  \citenamefont {Mietke}, \citenamefont {Li}, \citenamefont {Chen},
  \citenamefont {Higinbotham}, \citenamefont {Foster}, \citenamefont {Gokhale},
  \citenamefont {Dunkel},\ and\ \citenamefont {Fakhri}}]{tan2022odd}%
  \BibitemOpen
  \bibfield  {author} {\bibinfo {author} {\bibfnamefont {T.~H.}\ \bibnamefont
  {Tan}}, \bibinfo {author} {\bibfnamefont {A.}~\bibnamefont {Mietke}},
  \bibinfo {author} {\bibfnamefont {J.}~\bibnamefont {Li}}, \bibinfo {author}
  {\bibfnamefont {Y.}~\bibnamefont {Chen}}, \bibinfo {author} {\bibfnamefont
  {H.}~\bibnamefont {Higinbotham}}, \bibinfo {author} {\bibfnamefont {P.~J.}\
  \bibnamefont {Foster}}, \bibinfo {author} {\bibfnamefont {S.}~\bibnamefont
  {Gokhale}}, \bibinfo {author} {\bibfnamefont {J.}~\bibnamefont {Dunkel}}, \
  and\ \bibinfo {author} {\bibfnamefont {N.}~\bibnamefont {Fakhri}},\
  }\href@noop {} {\bibfield  {journal} {\bibinfo  {journal} {Nature}\ }\textbf
  {\bibinfo {volume} {607}},\ \bibinfo {pages} {287} (\bibinfo {year}
  {2022})}\BibitemShut {NoStop}%
\bibitem [{\citenamefont {Salbreux}\ and\ \citenamefont
  {J{\"u}licher}(2017)}]{salbreux2017mechanics}%
  \BibitemOpen
  \bibfield  {author} {\bibinfo {author} {\bibfnamefont {G.}~\bibnamefont
  {Salbreux}}\ and\ \bibinfo {author} {\bibfnamefont {F.}~\bibnamefont
  {J{\"u}licher}},\ }\href@noop {} {\bibfield  {journal} {\bibinfo  {journal}
  {Physical Review E}\ }\textbf {\bibinfo {volume} {96}},\ \bibinfo {pages}
  {032404} (\bibinfo {year} {2017})}\BibitemShut {NoStop}%
\bibitem [{\citenamefont {Fossati}\ \emph {et~al.}(2022)\citenamefont
  {Fossati}, \citenamefont {Scheibner}, \citenamefont {Fruchart},\ and\
  \citenamefont {Vitelli}}]{fossati2022odd}%
  \BibitemOpen
  \bibfield  {author} {\bibinfo {author} {\bibfnamefont {M.}~\bibnamefont
  {Fossati}}, \bibinfo {author} {\bibfnamefont {C.}~\bibnamefont {Scheibner}},
  \bibinfo {author} {\bibfnamefont {M.}~\bibnamefont {Fruchart}}, \ and\
  \bibinfo {author} {\bibfnamefont {V.}~\bibnamefont {Vitelli}},\ }\href@noop
  {} {\bibfield  {journal} {\bibinfo  {journal} {arXiv preprint
  arXiv:2210.03669}\ } (\bibinfo {year} {2022})}\BibitemShut {NoStop}%
\bibitem [{\citenamefont {Berthoumieux}\ \emph {et~al.}(2014)\citenamefont
  {Berthoumieux}, \citenamefont {Ma{\^\i}tre}, \citenamefont {Heisenberg},
  \citenamefont {Paluch}, \citenamefont {J{\"u}licher},\ and\ \citenamefont
  {Salbreux}}]{berthoumieux2014active}%
  \BibitemOpen
  \bibfield  {author} {\bibinfo {author} {\bibfnamefont {H.}~\bibnamefont
  {Berthoumieux}}, \bibinfo {author} {\bibfnamefont {J.-L.}\ \bibnamefont
  {Ma{\^\i}tre}}, \bibinfo {author} {\bibfnamefont {C.-P.}\ \bibnamefont
  {Heisenberg}}, \bibinfo {author} {\bibfnamefont {E.~K.}\ \bibnamefont
  {Paluch}}, \bibinfo {author} {\bibfnamefont {F.}~\bibnamefont
  {J{\"u}licher}}, \ and\ \bibinfo {author} {\bibfnamefont {G.}~\bibnamefont
  {Salbreux}},\ }\href@noop {} {\bibfield  {journal} {\bibinfo  {journal} {New
  Journal of Physics}\ }\textbf {\bibinfo {volume} {16}},\ \bibinfo {pages}
  {065005} (\bibinfo {year} {2014})}\BibitemShut {NoStop}%
\bibitem [{\citenamefont {Salbreux}\ \emph {et~al.}(2012)\citenamefont
  {Salbreux}, \citenamefont {Charras},\ and\ \citenamefont
  {Paluch}}]{salbreux2012actin}%
  \BibitemOpen
  \bibfield  {author} {\bibinfo {author} {\bibfnamefont {G.}~\bibnamefont
  {Salbreux}}, \bibinfo {author} {\bibfnamefont {G.}~\bibnamefont {Charras}}, \
  and\ \bibinfo {author} {\bibfnamefont {E.}~\bibnamefont {Paluch}},\
  }\href@noop {} {\bibfield  {journal} {\bibinfo  {journal} {Trends in cell
  biology}\ }\textbf {\bibinfo {volume} {22}},\ \bibinfo {pages} {536}
  (\bibinfo {year} {2012})}\BibitemShut {NoStop}%
\bibitem [{\citenamefont {Gov}(2007)}]{gov2007active}%
  \BibitemOpen
  \bibfield  {author} {\bibinfo {author} {\bibfnamefont {N.~S.}\ \bibnamefont
  {Gov}},\ }\href@noop {} {\bibfield  {journal} {\bibinfo  {journal} {Physical
  Review E}\ }\textbf {\bibinfo {volume} {75}},\ \bibinfo {pages} {011921}
  (\bibinfo {year} {2007})}\BibitemShut {NoStop}%
\bibitem [{\citenamefont {Nelson}\ \emph {et~al.}(2004)\citenamefont {Nelson},
  \citenamefont {Piran},\ and\ \citenamefont
  {Weinberg}}]{nelson2004statistical}%
  \BibitemOpen
  \bibfield  {author} {\bibinfo {author} {\bibfnamefont {D.}~\bibnamefont
  {Nelson}}, \bibinfo {author} {\bibfnamefont {T.}~\bibnamefont {Piran}}, \
  and\ \bibinfo {author} {\bibfnamefont {S.}~\bibnamefont {Weinberg}},\
  }\href@noop {} {\emph {\bibinfo {title} {Statistical mechanics of membranes
  and surfaces}}}\ (\bibinfo  {publisher} {World Scientific},\ \bibinfo {year}
  {2004})\BibitemShut {NoStop}%
\bibitem [{\citenamefont {Park}\ \emph {et~al.}(2010)\citenamefont {Park},
  \citenamefont {Best}, \citenamefont {Auth}, \citenamefont {Gov},
  \citenamefont {Safran}, \citenamefont {Popescu}, \citenamefont {Suresh},\
  and\ \citenamefont {Feld}}]{park2010metabolic}%
  \BibitemOpen
  \bibfield  {author} {\bibinfo {author} {\bibfnamefont {Y.}~\bibnamefont
  {Park}}, \bibinfo {author} {\bibfnamefont {C.~A.}\ \bibnamefont {Best}},
  \bibinfo {author} {\bibfnamefont {T.}~\bibnamefont {Auth}}, \bibinfo {author}
  {\bibfnamefont {N.~S.}\ \bibnamefont {Gov}}, \bibinfo {author} {\bibfnamefont
  {S.~A.}\ \bibnamefont {Safran}}, \bibinfo {author} {\bibfnamefont
  {G.}~\bibnamefont {Popescu}}, \bibinfo {author} {\bibfnamefont
  {S.}~\bibnamefont {Suresh}}, \ and\ \bibinfo {author} {\bibfnamefont {M.~S.}\
  \bibnamefont {Feld}},\ }\href@noop {} {\bibfield  {journal} {\bibinfo
  {journal} {Proceedings of the National Academy of Sciences}\ }\textbf
  {\bibinfo {volume} {107}},\ \bibinfo {pages} {1289} (\bibinfo {year}
  {2010})}\BibitemShut {NoStop}%
\bibitem [{\citenamefont {Turlier}\ \emph {et~al.}(2016)\citenamefont
  {Turlier}, \citenamefont {Fedosov}, \citenamefont {Audoly}, \citenamefont
  {Auth}, \citenamefont {Gov}, \citenamefont {Sykes}, \citenamefont {Joanny},
  \citenamefont {Gompper},\ and\ \citenamefont
  {Betz}}]{turlier2016equilibrium}%
  \BibitemOpen
  \bibfield  {author} {\bibinfo {author} {\bibfnamefont {H.}~\bibnamefont
  {Turlier}}, \bibinfo {author} {\bibfnamefont {D.~A.}\ \bibnamefont
  {Fedosov}}, \bibinfo {author} {\bibfnamefont {B.}~\bibnamefont {Audoly}},
  \bibinfo {author} {\bibfnamefont {T.}~\bibnamefont {Auth}}, \bibinfo {author}
  {\bibfnamefont {N.~S.}\ \bibnamefont {Gov}}, \bibinfo {author} {\bibfnamefont
  {C.}~\bibnamefont {Sykes}}, \bibinfo {author} {\bibfnamefont {J.-F.}\
  \bibnamefont {Joanny}}, \bibinfo {author} {\bibfnamefont {G.}~\bibnamefont
  {Gompper}}, \ and\ \bibinfo {author} {\bibfnamefont {T.}~\bibnamefont
  {Betz}},\ }\href@noop {} {\bibfield  {journal} {\bibinfo  {journal} {Nature
  physics}\ }\textbf {\bibinfo {volume} {12}},\ \bibinfo {pages} {513}
  (\bibinfo {year} {2016})}\BibitemShut {NoStop}%
\bibitem [{\citenamefont {Floyd}\ \emph {et~al.}(2022)\citenamefont {Floyd},
  \citenamefont {Dinner},\ and\ \citenamefont
  {Vaikuntanathan}}]{floyd2022signatures}%
  \BibitemOpen
  \bibfield  {author} {\bibinfo {author} {\bibfnamefont {C.}~\bibnamefont
  {Floyd}}, \bibinfo {author} {\bibfnamefont {A.~R.}\ \bibnamefont {Dinner}}, \
  and\ \bibinfo {author} {\bibfnamefont {S.}~\bibnamefont {Vaikuntanathan}},\
  }\href@noop {} {\bibfield  {journal} {\bibinfo  {journal} {arXiv preprint
  arXiv:2210.01159}\ } (\bibinfo {year} {2022})}\BibitemShut {NoStop}%
\bibitem [{\citenamefont {De~Dominicis}(1978)}]{de1978dynamics}%
  \BibitemOpen
  \bibfield  {author} {\bibinfo {author} {\bibfnamefont {C.}~\bibnamefont
  {De~Dominicis}},\ }\href@noop {} {\bibfield  {journal} {\bibinfo  {journal}
  {Physical Review B}\ }\textbf {\bibinfo {volume} {18}},\ \bibinfo {pages}
  {4913} (\bibinfo {year} {1978})}\BibitemShut {NoStop}%
\bibitem [{\citenamefont {Janssen}(1979)}]{janssen1979field}%
  \BibitemOpen
  \bibfield  {author} {\bibinfo {author} {\bibfnamefont {H.}~\bibnamefont
  {Janssen}},\ }in\ \href@noop {} {\emph {\bibinfo {booktitle} {Dynamical
  critical phenomena and related topics}}}\ (\bibinfo  {publisher} {Springer},\
  \bibinfo {year} {1979})\ pp.\ \bibinfo {pages} {25--47}\BibitemShut {NoStop}%
\bibitem [{\citenamefont {Martin}\ \emph {et~al.}(1973)\citenamefont {Martin},
  \citenamefont {Siggia},\ and\ \citenamefont {Rose}}]{martin1973statistical}%
  \BibitemOpen
  \bibfield  {author} {\bibinfo {author} {\bibfnamefont {P.~C.}\ \bibnamefont
  {Martin}}, \bibinfo {author} {\bibfnamefont {E.}~\bibnamefont {Siggia}}, \
  and\ \bibinfo {author} {\bibfnamefont {H.}~\bibnamefont {Rose}},\ }\href@noop
  {} {\bibfield  {journal} {\bibinfo  {journal} {Physical Review A}\ }\textbf
  {\bibinfo {volume} {8}},\ \bibinfo {pages} {423} (\bibinfo {year}
  {1973})}\BibitemShut {NoStop}%
\bibitem [{\citenamefont {T{\"a}uber}(2014)}]{tauber2014critical}%
  \BibitemOpen
  \bibfield  {author} {\bibinfo {author} {\bibfnamefont {U.~C.}\ \bibnamefont
  {T{\"a}uber}},\ }\href@noop {} {\emph {\bibinfo {title} {Critical dynamics: a
  field theory approach to equilibrium and non-equilibrium scaling behavior}}}\
  (\bibinfo  {publisher} {Cambridge University Press},\ \bibinfo {year}
  {2014})\BibitemShut {NoStop}%
\bibitem [{\citenamefont {Frey}\ and\ \citenamefont
  {Nelson}(1991)}]{frey1991dynamics}%
  \BibitemOpen
  \bibfield  {author} {\bibinfo {author} {\bibfnamefont {E.}~\bibnamefont
  {Frey}}\ and\ \bibinfo {author} {\bibfnamefont {D.~R.}\ \bibnamefont
  {Nelson}},\ }\href@noop {} {\bibfield  {journal} {\bibinfo  {journal}
  {Journal de Physique I}\ }\textbf {\bibinfo {volume} {1}},\ \bibinfo {pages}
  {1715} (\bibinfo {year} {1991})}\BibitemShut {NoStop}%
\bibitem [{\citenamefont {Altland}\ and\ \citenamefont
  {Simons}(2010)}]{altland2010condensed}%
  \BibitemOpen
  \bibfield  {author} {\bibinfo {author} {\bibfnamefont {A.}~\bibnamefont
  {Altland}}\ and\ \bibinfo {author} {\bibfnamefont {B.~D.}\ \bibnamefont
  {Simons}},\ }\href@noop {} {\emph {\bibinfo {title} {Condensed matter field
  theory}}}\ (\bibinfo  {publisher} {Cambridge university press},\ \bibinfo
  {year} {2010})\BibitemShut {NoStop}%
\bibitem [{\citenamefont {Tong}()}]{tongstatistical}%
  \BibitemOpen
  \bibfield  {author} {\bibinfo {author} {\bibfnamefont {D.}~\bibnamefont
  {Tong}},\ }\href@noop {} {\bibinfo  {journal} {Department of Applied
  Mathematics and Theoretical Physics, Centre for Mathematical Sciences,
  University of Cambridge}\ }\BibitemShut {NoStop}%
\bibitem [{\citenamefont {Nelson}\ and\ \citenamefont
  {Peliti}(1987)}]{nelson1987fluctuations}%
  \BibitemOpen
\bibfield  {journal} {  }\bibfield  {author} {\bibinfo {author} {\bibfnamefont
  {D.}~\bibnamefont {Nelson}}\ and\ \bibinfo {author} {\bibfnamefont
  {L.}~\bibnamefont {Peliti}},\ }\href@noop {} {\bibfield  {journal} {\bibinfo
  {journal} {Journal de physique}\ }\textbf {\bibinfo {volume} {48}},\ \bibinfo
  {pages} {1085} (\bibinfo {year} {1987})}\BibitemShut {NoStop}%
\bibitem [{\citenamefont {Mermin}\ and\ \citenamefont
  {Wagner}(1966)}]{mermin1966absence}%
  \BibitemOpen
  \bibfield  {author} {\bibinfo {author} {\bibfnamefont {N.~D.}\ \bibnamefont
  {Mermin}}\ and\ \bibinfo {author} {\bibfnamefont {H.}~\bibnamefont
  {Wagner}},\ }\href@noop {} {\bibfield  {journal} {\bibinfo  {journal}
  {Physical Review Letters}\ }\textbf {\bibinfo {volume} {17}},\ \bibinfo
  {pages} {1133} (\bibinfo {year} {1966})}\BibitemShut {NoStop}%
\bibitem [{\citenamefont {Mermin}(1968)}]{mermin1968crystalline}%
  \BibitemOpen
  \bibfield  {author} {\bibinfo {author} {\bibfnamefont {N.~D.}\ \bibnamefont
  {Mermin}},\ }\href@noop {} {\bibfield  {journal} {\bibinfo  {journal}
  {Physical Review}\ }\textbf {\bibinfo {volume} {176}},\ \bibinfo {pages}
  {250} (\bibinfo {year} {1968})}\BibitemShut {NoStop}%
\bibitem [{\citenamefont {Hohenberg}(1967)}]{hohenberg1967existence}%
  \BibitemOpen
  \bibfield  {author} {\bibinfo {author} {\bibfnamefont {P.~C.}\ \bibnamefont
  {Hohenberg}},\ }\href@noop {} {\bibfield  {journal} {\bibinfo  {journal}
  {Physical Review}\ }\textbf {\bibinfo {volume} {158}},\ \bibinfo {pages}
  {383} (\bibinfo {year} {1967})}\BibitemShut {NoStop}%
\bibitem [{\citenamefont {Halperin}(2019)}]{halperin2019hohenberg}%
  \BibitemOpen
  \bibfield  {author} {\bibinfo {author} {\bibfnamefont {B.~I.}\ \bibnamefont
  {Halperin}},\ }\href@noop {} {\bibfield  {journal} {\bibinfo  {journal}
  {Journal of Statistical Physics}\ }\textbf {\bibinfo {volume} {175}},\
  \bibinfo {pages} {521} (\bibinfo {year} {2019})}\BibitemShut {NoStop}%
\bibitem [{\citenamefont {Guitter}\ \emph {et~al.}(1988)\citenamefont
  {Guitter}, \citenamefont {David}, \citenamefont {Leibler},\ and\
  \citenamefont {Peliti}}]{guitter1988crumpling}%
  \BibitemOpen
  \bibfield  {author} {\bibinfo {author} {\bibfnamefont {E.}~\bibnamefont
  {Guitter}}, \bibinfo {author} {\bibfnamefont {F.}~\bibnamefont {David}},
  \bibinfo {author} {\bibfnamefont {S.}~\bibnamefont {Leibler}}, \ and\
  \bibinfo {author} {\bibfnamefont {L.}~\bibnamefont {Peliti}},\ }\href@noop {}
  {\bibfield  {journal} {\bibinfo  {journal} {Physical review letters}\
  }\textbf {\bibinfo {volume} {61}},\ \bibinfo {pages} {2949} (\bibinfo {year}
  {1988})}\BibitemShut {NoStop}%
\bibitem [{\citenamefont {Guitter}\ \emph {et~al.}(1989)\citenamefont
  {Guitter}, \citenamefont {David}, \citenamefont {Leibler},\ and\
  \citenamefont {Peliti}}]{guitter1989thermodynamical}%
  \BibitemOpen
  \bibfield  {author} {\bibinfo {author} {\bibfnamefont {E.}~\bibnamefont
  {Guitter}}, \bibinfo {author} {\bibfnamefont {F.}~\bibnamefont {David}},
  \bibinfo {author} {\bibfnamefont {S.}~\bibnamefont {Leibler}}, \ and\
  \bibinfo {author} {\bibfnamefont {L.}~\bibnamefont {Peliti}},\ }\href@noop {}
  {\bibfield  {journal} {\bibinfo  {journal} {Journal de Physique}\ }\textbf
  {\bibinfo {volume} {50}},\ \bibinfo {pages} {1787} (\bibinfo {year}
  {1989})}\BibitemShut {NoStop}%
\bibitem [{\citenamefont {Le~Doussal}\ and\ \citenamefont
  {Radzihovsky}(2021)}]{le2021thermal}%
  \BibitemOpen
  \bibfield  {author} {\bibinfo {author} {\bibfnamefont {P.}~\bibnamefont
  {Le~Doussal}}\ and\ \bibinfo {author} {\bibfnamefont {L.}~\bibnamefont
  {Radzihovsky}},\ }\href@noop {} {\bibfield  {journal} {\bibinfo  {journal}
  {Physical Review Letters}\ }\textbf {\bibinfo {volume} {127}},\ \bibinfo
  {pages} {015702} (\bibinfo {year} {2021})}\BibitemShut {NoStop}%
\bibitem [{\citenamefont {Leimkuhler}\ and\ \citenamefont
  {Matthews}(2013)}]{leimkuhler2013rational}%
  \BibitemOpen
  \bibfield  {author} {\bibinfo {author} {\bibfnamefont {B.}~\bibnamefont
  {Leimkuhler}}\ and\ \bibinfo {author} {\bibfnamefont {C.}~\bibnamefont
  {Matthews}},\ }\href@noop {} {\bibfield  {journal} {\bibinfo  {journal}
  {Applied Mathematics Research eXpress}\ }\textbf {\bibinfo {volume} {2013}},\
  \bibinfo {pages} {34} (\bibinfo {year} {2013})}\BibitemShut {NoStop}%
\bibitem [{\citenamefont {Seung}\ and\ \citenamefont
  {Nelson}(1988)}]{seung1988defects}%
  \BibitemOpen
  \bibfield  {author} {\bibinfo {author} {\bibfnamefont {H.~S.}\ \bibnamefont
  {Seung}}\ and\ \bibinfo {author} {\bibfnamefont {D.~R.}\ \bibnamefont
  {Nelson}},\ }\href@noop {} {\bibfield  {journal} {\bibinfo  {journal}
  {Physical Review A}\ }\textbf {\bibinfo {volume} {38}},\ \bibinfo {pages}
  {1005} (\bibinfo {year} {1988})}\BibitemShut {NoStop}%
\bibitem [{\citenamefont {Do~Carmo}(2016)}]{do2016differential}%
  \BibitemOpen
  \bibfield  {author} {\bibinfo {author} {\bibfnamefont {M.~P.}\ \bibnamefont
  {Do~Carmo}},\ }\href@noop {} {\emph {\bibinfo {title} {Differential geometry
  of curves and surfaces: revised and updated second edition}}}\ (\bibinfo
  {publisher} {Courier Dover Publications},\ \bibinfo {year}
  {2016})\BibitemShut {NoStop}%
\bibitem [{\citenamefont {Lin}\ \emph {et~al.}(2022)\citenamefont {Lin},
  \citenamefont {Merkel},\ and\ \citenamefont
  {Rupprecht}}]{lin2022implementation}%
  \BibitemOpen
  \bibfield  {author} {\bibinfo {author} {\bibfnamefont {S.-Z.}\ \bibnamefont
  {Lin}}, \bibinfo {author} {\bibfnamefont {M.}~\bibnamefont {Merkel}}, \ and\
  \bibinfo {author} {\bibfnamefont {J.-F.}\ \bibnamefont {Rupprecht}},\
  }\href@noop {} {\bibfield  {journal} {\bibinfo  {journal} {The European
  Physical Journal E}\ }\textbf {\bibinfo {volume} {45}},\ \bibinfo {pages} {4}
  (\bibinfo {year} {2022})}\BibitemShut {NoStop}%
\bibitem [{\citenamefont {Tlili}\ \emph {et~al.}(2015)\citenamefont {Tlili},
  \citenamefont {Gay}, \citenamefont {Graner}, \citenamefont {Marcq},
  \citenamefont {Molino},\ and\ \citenamefont
  {Saramito}}]{tlili2015colloquium}%
  \BibitemOpen
  \bibfield  {author} {\bibinfo {author} {\bibfnamefont {S.}~\bibnamefont
  {Tlili}}, \bibinfo {author} {\bibfnamefont {C.}~\bibnamefont {Gay}}, \bibinfo
  {author} {\bibfnamefont {F.}~\bibnamefont {Graner}}, \bibinfo {author}
  {\bibfnamefont {P.}~\bibnamefont {Marcq}}, \bibinfo {author} {\bibfnamefont
  {F.}~\bibnamefont {Molino}}, \ and\ \bibinfo {author} {\bibfnamefont
  {P.}~\bibnamefont {Saramito}},\ }\href@noop {} {\bibfield  {journal}
  {\bibinfo  {journal} {The European Physical Journal E}\ }\textbf {\bibinfo
  {volume} {38}},\ \bibinfo {pages} {1} (\bibinfo {year} {2015})}\BibitemShut
  {NoStop}%
\bibitem [{\citenamefont {Tlili}\ \emph {et~al.}(2019)\citenamefont {Tlili},
  \citenamefont {Yin}, \citenamefont {Rupprecht}, \citenamefont
  {Mendieta-Serrano}, \citenamefont {Weissbart}, \citenamefont {Verma},
  \citenamefont {Teng}, \citenamefont {Toyama}, \citenamefont {Prost},\ and\
  \citenamefont {Saunders}}]{tlili2019shaping}%
  \BibitemOpen
  \bibfield  {author} {\bibinfo {author} {\bibfnamefont {S.}~\bibnamefont
  {Tlili}}, \bibinfo {author} {\bibfnamefont {J.}~\bibnamefont {Yin}}, \bibinfo
  {author} {\bibfnamefont {J.-F.}\ \bibnamefont {Rupprecht}}, \bibinfo {author}
  {\bibfnamefont {M.}~\bibnamefont {Mendieta-Serrano}}, \bibinfo {author}
  {\bibfnamefont {G.}~\bibnamefont {Weissbart}}, \bibinfo {author}
  {\bibfnamefont {N.}~\bibnamefont {Verma}}, \bibinfo {author} {\bibfnamefont
  {X.}~\bibnamefont {Teng}}, \bibinfo {author} {\bibfnamefont {Y.}~\bibnamefont
  {Toyama}}, \bibinfo {author} {\bibfnamefont {J.}~\bibnamefont {Prost}}, \
  and\ \bibinfo {author} {\bibfnamefont {T.}~\bibnamefont {Saunders}},\
  }\href@noop {} {\bibfield  {journal} {\bibinfo  {journal} {Proceedings of the
  National Academy of Sciences}\ }\textbf {\bibinfo {volume} {116}},\ \bibinfo
  {pages} {25430} (\bibinfo {year} {2019})}\BibitemShut {NoStop}%
\bibitem [{\citenamefont {Tu}\ \emph {et~al.}(1997)\citenamefont {Tu},
  \citenamefont {Grinstein},\ and\ \citenamefont {Munoz}}]{tu1997systems}%
  \BibitemOpen
  \bibfield  {author} {\bibinfo {author} {\bibfnamefont {Y.}~\bibnamefont
  {Tu}}, \bibinfo {author} {\bibfnamefont {G.}~\bibnamefont {Grinstein}}, \
  and\ \bibinfo {author} {\bibfnamefont {M.}~\bibnamefont {Munoz}},\
  }\href@noop {} {\bibfield  {journal} {\bibinfo  {journal} {Physical review
  letters}\ }\textbf {\bibinfo {volume} {78}},\ \bibinfo {pages} {274}
  (\bibinfo {year} {1997})}\BibitemShut {NoStop}%
\bibitem [{\citenamefont {Falasco}\ \emph {et~al.}(2016)\citenamefont
  {Falasco}, \citenamefont {Baldovin}, \citenamefont {Kroy},\ and\
  \citenamefont {Baiesi}}]{falasco2016mesoscopic}%
  \BibitemOpen
  \bibfield  {author} {\bibinfo {author} {\bibfnamefont {G.}~\bibnamefont
  {Falasco}}, \bibinfo {author} {\bibfnamefont {F.}~\bibnamefont {Baldovin}},
  \bibinfo {author} {\bibfnamefont {K.}~\bibnamefont {Kroy}}, \ and\ \bibinfo
  {author} {\bibfnamefont {M.}~\bibnamefont {Baiesi}},\ }\href@noop {}
  {\bibfield  {journal} {\bibinfo  {journal} {New Journal of Physics}\ }\textbf
  {\bibinfo {volume} {18}},\ \bibinfo {pages} {093043} (\bibinfo {year}
  {2016})}\BibitemShut {NoStop}%
\bibitem [{\citenamefont {L{\"o}wen}(2020)}]{lowen2020inertial}%
  \BibitemOpen
  \bibfield  {author} {\bibinfo {author} {\bibfnamefont {H.}~\bibnamefont
  {L{\"o}wen}},\ }\href@noop {} {\bibfield  {journal} {\bibinfo  {journal} {The
  Journal of chemical physics}\ }\textbf {\bibinfo {volume} {152}},\ \bibinfo
  {pages} {040901} (\bibinfo {year} {2020})}\BibitemShut {NoStop}%
\bibitem [{\citenamefont {Solon}\ \emph {et~al.}(2015)\citenamefont {Solon},
  \citenamefont {Fily}, \citenamefont {Baskaran}, \citenamefont {Cates},
  \citenamefont {Kafri}, \citenamefont {Kardar},\ and\ \citenamefont
  {Tailleur}}]{solon2015pressure}%
  \BibitemOpen
  \bibfield  {author} {\bibinfo {author} {\bibfnamefont {A.~P.}\ \bibnamefont
  {Solon}}, \bibinfo {author} {\bibfnamefont {Y.}~\bibnamefont {Fily}},
  \bibinfo {author} {\bibfnamefont {A.}~\bibnamefont {Baskaran}}, \bibinfo
  {author} {\bibfnamefont {M.~E.}\ \bibnamefont {Cates}}, \bibinfo {author}
  {\bibfnamefont {Y.}~\bibnamefont {Kafri}}, \bibinfo {author} {\bibfnamefont
  {M.}~\bibnamefont {Kardar}}, \ and\ \bibinfo {author} {\bibfnamefont
  {J.}~\bibnamefont {Tailleur}},\ }\href@noop {} {\bibfield  {journal}
  {\bibinfo  {journal} {Nature Physics}\ }\textbf {\bibinfo {volume} {11}},\
  \bibinfo {pages} {673} (\bibinfo {year} {2015})}\BibitemShut {NoStop}%
\bibitem [{\citenamefont {Barab{\'a}si}\ \emph {et~al.}(1995)\citenamefont
  {Barab{\'a}si}, \citenamefont {Stanley} \emph
  {et~al.}}]{barabasi1995fractal}%
  \BibitemOpen
  \bibfield  {author} {\bibinfo {author} {\bibfnamefont {A.-L.}\ \bibnamefont
  {Barab{\'a}si}}, \bibinfo {author} {\bibfnamefont {H.~E.}\ \bibnamefont
  {Stanley}},  \emph {et~al.},\ }\href@noop {} {\emph {\bibinfo {title}
  {Fractal concepts in surface growth}}}\ (\bibinfo  {publisher} {Cambridge
  university press},\ \bibinfo {year} {1995})\BibitemShut {NoStop}%
\bibitem [{\citenamefont {Radzihovsky}\ and\ \citenamefont
  {Toner}(1995)}]{radzihovsky1995new}%
  \BibitemOpen
  \bibfield  {author} {\bibinfo {author} {\bibfnamefont {L.}~\bibnamefont
  {Radzihovsky}}\ and\ \bibinfo {author} {\bibfnamefont {J.}~\bibnamefont
  {Toner}},\ }\href@noop {} {\bibfield  {journal} {\bibinfo  {journal}
  {Physical review letters}\ }\textbf {\bibinfo {volume} {75}},\ \bibinfo
  {pages} {4752} (\bibinfo {year} {1995})}\BibitemShut {NoStop}%
\bibitem [{\citenamefont {Kardar}(2007)}]{kardar2007statistical}%
  \BibitemOpen
  \bibfield  {author} {\bibinfo {author} {\bibfnamefont {M.}~\bibnamefont
  {Kardar}},\ }\href@noop {} {\emph {\bibinfo {title} {Statistical physics of
  fields}}}\ (\bibinfo  {publisher} {Cambridge University Press},\ \bibinfo
  {year} {2007})\BibitemShut {NoStop}%
\bibitem [{\citenamefont {Peskin}(2018)}]{peskin2018introduction}%
  \BibitemOpen
  \bibfield  {author} {\bibinfo {author} {\bibfnamefont {M.~E.}\ \bibnamefont
  {Peskin}},\ }\href@noop {} {\emph {\bibinfo {title} {An introduction to
  quantum field theory}}}\ (\bibinfo  {publisher} {CRC press},\ \bibinfo {year}
  {2018})\BibitemShut {NoStop}%
\bibitem [{\citenamefont {Ko{\v{s}}mrlj}\ and\ \citenamefont
  {Nelson}(2016)}]{kovsmrlj2016response}%
  \BibitemOpen
  \bibfield  {author} {\bibinfo {author} {\bibfnamefont {A.}~\bibnamefont
  {Ko{\v{s}}mrlj}}\ and\ \bibinfo {author} {\bibfnamefont {D.~R.}\ \bibnamefont
  {Nelson}},\ }\href@noop {} {\bibfield  {journal} {\bibinfo  {journal}
  {Physical Review B}\ }\textbf {\bibinfo {volume} {93}},\ \bibinfo {pages}
  {125431} (\bibinfo {year} {2016})}\BibitemShut {NoStop}%
\end{thebibliography}%
\pagebreak
\let\addcontentsline\oldaddcontentsline
\onecolumngrid

\makeatletter
\renewcommand \thesection{S-\@arabic\c@section}
\renewcommand\thetable{S\@arabic\c@table}
\renewcommand \thefigure{S\@arabic\c@figure}
\renewcommand \theequation{S\@arabic\c@equation}
\makeatother
\setcounter{equation}{0}  
\setcounter{figure}{0}  
\setcounter{section}{0}  

{
    \center \bf \large 
    Supplemental Material\vspace*{0.1cm}\\ 
    \vspace*{0.0cm}
}
\maketitle
\tableofcontents
\let\oldsection\section
\renewcommand\section{\clearpage\oldsection}

\section{Choice Of Over-Damped Equations}
We will illustrate why we opt for over-damped Langevin equations for examining our model system, as described in sub-sections~\ref{sec:instability}, \ref{sec:intertialirrelevant}, and \ref{sec:steadystate}. The choice is driven by three key reasons:
\begin{enumerate}
    \item The presence of the inertial term does not cause instability that might lead to substantial differences from the over-damped scenario.
    \item At large length scales, the inertial term becomes irrelevant in the context of field theory physics.
    \item Adopting the over-damped condition simplifies simulations, as it enables us to circumvent potential non-steady state behaviors.
\end{enumerate}
Regarding our theoretical analysis, it's crucial to note that we will conduct all calculations based on an orthonormal basis, contrary to the non-orthogonal basis employed for the simulations. This approach will streamline the application of the Fourier transformation, a critical step in our forthcoming analysis.

\subsection{Stability of Fully Inertial System \label{sec:instability}}
We first show that inclusion of the intertial term will not provide us with a qualitative difference from the over-damped version of the equations via some instability. Thus, we begin with the fully inertial athermal (sans thermal noise) equations, which are the following:
\begin{equation}
\begin{split}
\label{eq:LangevinUNN}
     \rho \partial_t^2 u_j(\mathbf{r},t) +\partial_t u_j(\mathbf{r},t) &= \mathcal{D} \partial_i \sigma_{ij}(\mathbf{r},t)  = \mathcal{D}  \mathcal{C}_{ijkl} \partial_i u_{kl}(\mathbf{r},t)
    \\ 
     \rho \partial_t^2 f^{\alpha}(\mathbf{r},t) +\partial_t f^{\alpha}(\mathbf{r},t)  &= \mathcal{D}_f[-\partial^i \partial^j M_{ij}^{\alpha}(\mathbf{r},t)
    +\partial_i[\sigma_{ij}(\mathbf{r},t) \partial_j f^{\alpha}(\mathbf{r},t)]] \\ & = \mathcal{D}_f[-\partial_i \partial_j\{ \mathcal{B}_{ijkl}  \partial_k \partial_l f^{\alpha}(\mathbf{r},t) \}
    +\partial_i \{\sigma_{ij}(\mathbf{r},t) \partial_j f^{\alpha}(\mathbf{r},t)\}]
\end{split}
\end{equation}
where $\rho$ is the density. Roman indices vary through the number of in-plane dimensions, which in this case $D=2$, whereas Greek indices vary through the co-dimension $d_c=d-D$ of the membrane (with $d$ being the total dimension of the embedding space). $\mathcal{D}$ and $\mathcal{D}_f$ are the in-plane and out-of-plane mobilities of the membrane respectively.
The strain tensor $u_{ij}$, moment tensor $M_{ij}$, and stress tensor $\sigma_{ij}$ are the following:
\begin{equation}
\begin{split}
    & u_{ij}(\mathbf{r},t) = \frac{1}{2} [\partial_i u_j(\mathbf{r},t) + \partial_j u_i(\mathbf{r},t) + \partial_i f^{\alpha}(\mathbf{r},t) \partial_j f^{\alpha}(\mathbf{r},t)] \\ &
    M_{ij}^{\alpha}(\mathbf{r},t) = \mathcal{B}_{ijkl} \partial_k \partial_l f^{\alpha}(\mathbf{r},t)\\ &
    \sigma_{ij}(\mathbf{r},t) = \mathcal{C}_{ijkl}u_{kl}(\mathbf{r},t)
\end{split}
\end{equation}
We shall also use an isotropic bending rigidity $\mathcal{B}_{ijkl} = \frac{\kappa}{2}  \delta_{ij} \delta_{kl} + \frac{\kappa}{4}  [\delta_{ik} \delta_{jl}+\delta_{il} \delta_{jk}]$ so that $\mathcal{B}_{ijkl} \partial_i \partial_j \partial_k \partial_l = \kappa \Delta^2$. Under periodic boundary conditions, the Gaussian bending rigidity may be ignored by means of the Gauss-Bonnet theorem. To perform a stability analysis we neglect non-linear terms and arrive at the linearized form in Fourier space (via the defined Fourier transform $g(\mathbf{r},t) = \frac{1}{A \tau} \sum_{\mathbf{q},\omega} g(\mathbf{q},\omega) e^{i(\mathbf{q} \cdot \mathbf{r} -\omega t)}$ with $A$ being the area of the system and $\tau$ the length of time):
\begin{equation}
\begin{split}
\label{eq:LangevinLinear}
    & - \rho \omega^2 u_j(\mathbf{q},\omega) -i \omega  u_j(\mathbf{q},\omega) = - \mathcal{D} \mathcal{C}_{ijkl} q_i q_k u_{l}(\mathbf{q},\omega)
    \\ & 
     - \rho \omega^2 f^{\alpha}(\mathbf{q},\omega) -i \omega f^{\alpha}(\mathbf{q},\omega)   =  \mathcal{D}_f[-\kappa  q^4 f^{\alpha}(\mathbf{q},\omega)
    - \mathcal{C}_{ijkl} u^{o}_{kl} q_i q_j f^{\alpha}(\mathbf{q},\omega)]
\end{split}
\end{equation}
where $u^{o}_{kl}$ is the homogeneous strain tensor defined via the Fourier transform of the strain tensor:
\begin{equation}
\begin{split}
    u_{ij}(\mathbf{r},t) =& u_{ij}^{o} + \frac{i}{2} \sum_{\mathbf{q},\omega} [q_i u_j(\mathbf{q},\omega) +q_j u_i(\mathbf{q},\omega)] e^{i (\mathbf{q}\cdot \mathbf{r} - \omega t)}  \\ & -\frac{1}{2} \sum_{\mathbf{q}_1,\mathbf{q}_2,\omega_1,
    \omega_2} \mathbf{q}_{1,i} \mathbf{q}_{2,j} f^{\alpha}(\mathbf{q}_1,\omega_1)  f^{\alpha}(\mathbf{q}_2,\omega_2) e^{i(\{\mathbf{q}_1 +\mathbf{q}_2\} \cdot \mathbf{r} - \{ \omega_1 + \omega_2\} t)}
\end{split}
\end{equation}
Note that in addition via the constitutive relation: $\sigma_{ij}^o = \mathcal{C}_{ijkl}u^{o}_{kl}$ can be reformulated as the homogeneous stress tensor. Thus we may rewritten  Eq.~\ref{eq:LangevinLinear} we may rewrite the equations as:
\begin{equation}
\begin{split}
\label{eq:LangevinLinearInterim}
    & - \rho \omega^2 u_j(\mathbf{q},\omega) -i \omega  u_j(\mathbf{q},\omega) = - \mathcal{D} \mathcal{C}_{ijkl} q_i q_k u_{l}(\mathbf{q},\omega)
    \\ & 
     - \rho \omega^2 f^{\alpha}(\mathbf{q},\omega) -i \omega f^{\alpha}(\mathbf{q},\omega)   =  \mathcal{D}_f[-\kappa  q^4 
    - \sigma_{ij}^o q^i q^j ]f^{\alpha}(\mathbf{q},\omega)
\end{split}
\end{equation}
In addition, assuming isotropy we can further rewrite $\sigma_{ij}^o q_i q_j = \sigma^o q^2$ and so one obtains:
\begin{equation}
\begin{split}
\label{eq:LangevinLinearFinal}
    & - \rho \omega^2 u_j(\mathbf{q},\omega) -i \omega  u_j(\mathbf{q},\omega) = - \mathcal{D} \mathcal{C}_{ijkl} q_i q_k u_{l}(\mathbf{q},\omega)
    \\ & 
     - \rho \omega^2 f^{\alpha}(\mathbf{q},\omega) -i \omega f^{\alpha}(\mathbf{q},\omega)   =  \mathcal{D}_f[-\kappa  q^4 
    - \sigma^o q^2 ]f^{\alpha}(\mathbf{q},\omega)
\end{split}
\end{equation}
From here the stability condition for the flexural field, $f^{\alpha}$, is clear. We require simply that the system have a positive bending rigidity ($\kappa>0$) and that the elastic sheet not be buckled, i.e. $\kappa q^4 + \sigma^o q^2 > 0$. This leaves us with the stability analysis of the linearized in-plane phonon equation. To do this we can follow an analysis similar to that given by \cite{scheibner2020odd}. The strain tensor is taken to be the same as Eq.~(1) in the main text:
\begin{equation}
\begin{split}
 \mathcal{C}_{ijkl} = \lambda \delta_{ij} \delta_{kl} + \mu [\delta_{ik} \delta_{jl} + \delta_{il} \delta_{jk}] + K_{odd} \mathcal{E}_{ijkl} -A_{odd}\epsilon_{ij} \delta_{kl}
\end{split}
\end{equation}
with with $\mathcal{E}_{ijkl} = \frac{1}{2}[\epsilon_{ik}\delta_{jl}+\epsilon_{il}\delta_{jk}+\epsilon_{jk}\delta_{il}+\epsilon_{jl}\delta_{ik}]$. The symbol $\epsilon_{il}$ is the permutation Levi-Civita tensor ($\epsilon_{11}=\epsilon_{22}=0,\epsilon_{12}=-\epsilon_{21}=1$). 
We begin by reformulating the equations:
\begin{equation}
    (-\rho \omega^2 - i \omega ) \begin{bmatrix}
u_1 \\ u_2
\end{bmatrix} = \mathcal{D} q^2 \left[
 \begin{array}{ccccc}
 B+ \mu & K_{odd} \\
 -K_{odd}-A_{odd} & \mu 
   \\
\end{array}  
\right]\begin{bmatrix}
u_{\parallel} \\ u_{\perp}
\end{bmatrix}
\end{equation}
where $B= \lambda+\mu$ and $u_{\parallel} = \hat{q}_i u_i$ and $u_{\perp} = \epsilon_{ij} \hat{q}_i u_j$ (where the symbol $\hat{}$ marks a unit vector). By doing the eigenvalue analysis of the matrix we obtain:
\begin{equation}
    (- \rho \omega^2 - i  \omega ) = - \mathcal{D} \bigg[ \frac{B}{2} + \mu \pm \sqrt{\bigg(\frac{B}{2}\bigg)^2 -A_{odd}K_{odd}-K_{odd}^2}\bigg]q^2
\end{equation}
For simplicity let us redefine $J=A_{odd}K_{odd}+K_{odd}^2$. We can then solve this quadratic equation to obtain:
\begin{equation}
    \omega = - \frac{i }{2 \rho} \pm \frac{1}{2 \rho}\sqrt{-1 +  \rho \mathcal{D} \bigg[ \frac{B}{2} + \mu \pm \sqrt{\bigg(\frac{B}{2}\bigg)^2 -J}\bigg]q^2}
\end{equation}
We do not want $\omega$ to have a positive imaginary branch, otherwise this leads to a real positive exponential solution which leads to a blow-up of the equations. Thus when the second term in the above equation becomes more "positive" than the first, then we have obtained an instability. We find 3 regimes which are as follows:
\subsubsection{ $J<-\mu(B+\mu)$}
If this condition is satisfied then the following inequality is implied:
\begin{equation}
    \sqrt{\bigg(\frac{B}{2}\bigg)^2 -J} >\frac{B}{2}+\mu
\end{equation}
and thus the following eigen-value:
\begin{equation}
    i \omega =  \frac{ 1}{2 \rho} \pm \frac{i}{2 \rho}\sqrt{-1 + \rho \mathcal{D} \bigg[ \frac{B}{2} + \mu - \sqrt{\bigg(\frac{B}{2}\bigg)^2 -J}\bigg]q^2}  
\end{equation}
will induce a solution that blows up for all $q$. Hence, one must require that $J>-\mu(B+\mu)$ in order to have a stable odd elastic system.

\subsubsection{ $-\mu(B+\mu)<J<(B/2)^2$}
In this case a careful analysis of all 4 eigenvalues shows that there is no possibility for $\omega$ to have a positive imaginary branch since:
\begin{equation}
    \vert \mathrm{Im} \bigg( \sqrt{-1 + \rho \mathcal{D} \bigg[ \frac{B}{2} + \mu \pm \sqrt{\bigg(\frac{B}{2}\bigg)^2 -J}\bigg]q^2} \bigg) \vert <1
\end{equation}
Thus we have a stable system for all $q$.
\subsubsection{$J>(B/2)^2$} 
In this case we can rewrite the eigenvalues as:
\begin{equation}
    \omega = - \frac{i }{2 \rho} \pm \frac{1}{2 \rho}\sqrt{-1 + \rho \mathcal{D} \bigg[ \frac{B}{2} + \mu \pm i\sqrt{J - \bigg(\frac{B}{2}\bigg)^2 }\bigg]q^2}
\end{equation}
By rewriting:
\begin{equation}
    R e^{- i \theta_{\pm }} \equiv 1 - \rho \mathcal{D} \bigg[ \frac{B}{2} + \mu \pm i\sqrt{J - \bigg(\frac{B}{2}\bigg)^2 }\bigg]q^2
\end{equation}
And taking the principal branch cut in the complex plane to be the non-positive x-axis (so that $-\pi<\theta_{\pm}<\pi$), we can take the square root and write:
\begin{equation}
    \omega = - \frac{i }{2 \rho} \pm \frac{i}{2 \rho}\sqrt{R}e^{-i \theta_{\pm}/2} 
\end{equation}
And thus:
\begin{equation}
    \mathrm{Im}(\omega) = - \frac{ 1}{2 \rho} \pm \frac{1}{2 \rho}\sqrt{R}\cos{ \theta_{\pm}/2} 
\end{equation}
Since $-\pi<\theta_{\pm}<\pi$, then $\cos{ \frac{\theta_{\pm}}{2}}>0$. Therefore we need only examine:
\begin{equation}
    \mathrm{Im}(\omega) = - \frac{ 1}{2 \rho} + \frac{1}{2 \rho}\sqrt{R}\cos{ \theta_{\pm}/2} 
\end{equation}
Once $\sqrt{R}\cos{ (\theta_{\pm}/2) } > 1$ we have an instability. This is equivalent to checking when $R(\cos{ ( \theta_{\pm}/2)})^2 > 1$. Via the half-angle formula this becomes:
\begin{equation}
    R (1+\cos{\theta_{\pm}}) > 2
\end{equation}
Thus we obtain:
\begin{equation}
    \sqrt{(1- \rho q^2 \mathcal{D} \bigg[ \frac{B}{2} +\mu\bigg])^2 + \rho^2 \mathcal{D} q^4 \bigg[ J- \bigg(\frac{B}{2}\bigg)^2\bigg]} + 1- \rho \mathcal{D} q^2 \bigg[ \frac{B}{2} +\mu\bigg] > 2 
\end{equation}
which leads us to:
\begin{equation}
    \sqrt{(1- \rho \mathcal{D} q^2 \bigg[ \frac{B}{2} +\mu\bigg])^2 + \rho^2 \mathcal{D}^2 q^4 \bigg[ J- \bigg(\frac{B}{2}\bigg)^2\bigg]}  > 1 + \rho \mathcal{D} q^2 \bigg[ \frac{B}{2} +\mu\bigg]>0
\end{equation}
By squaring the inequality once more and solving we see that we obtain instabilities when:
\begin{equation}
    \rho^2 \mathcal{D}^2 q^4 \bigg[ J- \bigg(\frac{B}{2}\bigg)^2\bigg] > 4 \rho \mathcal{D} q^2 \bigg[ \frac{B}{2} +\mu\bigg]
\end{equation}
resulting in:
\begin{equation}
    q> \sqrt{\frac{4 }{\rho \mathcal{D}} \frac{\bigg[ \frac{B}{2} +\mu\bigg]}{\bigg[ J- \bigg(\frac{B}{2}\bigg)^2\bigg]}} \equiv q_c
\end{equation}
Note the difference in the direction of this inequality is different from that obtained in the stability analysis of the visco-elastic equations in \cite{scheibner2020odd}, where the dissipative term has two extra spatial derivatives, $\eta^{ijkl} \partial_t \partial_i \partial_k u_l$. Hence we see that in this regime, if a system has a small enough physical length scale (i.e. the lattice spacing of discrete model is small enough) then we have instabilities. If we were to explore over-damping via the zero-inertia Brownian method ($\rho = 0$), we see that $q_c = \infty$ and thus we should not be able to observe this instability. In the Langevin method where we have under-damping, if $q_c< 2 \pi/a$ where $a$ is the lattice spacing of our discrete system, then we should once again not see this instability. Otherwise, the under-damped Langevin case should show this instability. 

\subsection{Inertial Term Irrelevance \label{sec:intertialirrelevant}}
Further justifying the over-damped case, we show the irrelevance of the inertial term near the Gaussian fixed point of the corresponding MSRJD action. To do this a power-counting analysis \cite{tongstatistical} is done in Section~\ref{sec:powercount}. From this one can see that $\partial_t^2 f, \partial_t^2 u_j$ scale to zero in the low frequency limit (since $\zeta_t=4$). Thus high-frequency phenomena will be unimportant to the long-wavelength limiting behavior of the theory.

\subsection{Potential Non-Steady State Behaviors \label{sec:steadystate}}
As a final note, in the case where over-damping is not assumed, it is well known that one must consider an active heat flow \cite{falasco2016mesoscopic, lowen2020inertial}. This active heat flow is a form of work done by non-equilibrium or active forces that are not derived from a free energy. However, in the over-damped limit, one may disregard such terms. This will help us to simplify simulations that we performed where we use barostats and thermostats which in general should contain an active heat flow term.

In addition, other variables such as the pressure may also not be treated as state functions \cite{solon2015pressure}. Thus we also consider Rouse dynamics so that we do not couple hydrodynamics to the membrane fluctuations \cite{frey1991dynamics}.  

\section{Details of Simulations \label{SimulationDetailsSI}}

\subsection{Implementation of Dihedral Forces}

The bending forces were determined directly from a set of dihedral springs between the faces of the triangles. The elastic bending energy of such a system can be formulated as:
\begin{equation}
    \text{E}_{\text{bend}} =  \hat{\kappa} \sum_{\langle I J \rangle} [1-\hat{n}_I\cdot \hat{n}_J]
\end{equation}
where $\hat{\kappa}$ is the discrete dihedral spring stiffness and $\hat{n}_I$ is the normal to face $I$. We can relate the microscopic spring stiffness, $\hat{\kappa}$, to the coarse-grained bending rigidity, $\kappa$ \cite{seung1988defects}:
\begin{equation}
    \kappa = \frac{\sqrt{3}}{2} \hat{\kappa} 
\end{equation}

\subsection{Explicit In-Plane Constitutive Relations}
For our simulations we used the implementation of the constitutive relation to enact forces explained in the main body of the text. We also added in a non-linear area potential (which we call a bulk strain term) which helped to prevent the areas of faces of the triangles from becoming too small and thus aids with the stabilization of our numerical simulations. We show here the derivation of explicit constitutive relations with respect to each of these contributions to the in-plane stresses:

We begin by defining the strain tensor for the each triangular face as follows:
\begin{equation}
u_{ij} = \frac{1}{2}[g_{ij}-\bar{g}_{ij}]
\end{equation}
where $\{ g, \bar{g}\}$ are the current and reference metric tensors respectively. We use the following elastic modulus tensor in our constitutive relation:
\begin{equation}
\begin{split}
 \mathcal{C}^{ijkl} =& \lambda \bar{g}^{ij} \bar{g}^{kl} + \mu [\bar{g}^{ik} \bar{g}^{jl} + \bar{g}^{il} \bar{g}^{jk}] + \frac{K^{odd}}{\sqrt{\det[\bar{g}]}} \mathcal{E}^{ijkl} -\frac{A_{odd}}{\sqrt{\det[\bar{g}]}} \epsilon^{ij} \bar{g}^{kl}
        \label{eq:CModTensorSimSI}
\end{split}
\end{equation}
with $\mathcal{E}^{ijkl} = \frac{1}{2}[\epsilon^{ik}\bar{g}^{jl}+\epsilon^{il}\bar{g}^{jk}+\epsilon^{jk}\bar{g}^{il}+\epsilon^{jl}\bar{g}^{ik}]$, where $\epsilon^{ij}$ is again the 2-D Levi-Civita permutation symbol satisfying $\epsilon^{11}=\epsilon^{22}=0, \epsilon^{12} = -\epsilon^{21}=1$.

Then, in the linear elastic limit where deformations are small, one can write down the contribution of $A_{odd}$ to the constitutive relation:
\begin{equation}
\begin{split}
    \sigma_A^{ij}  = -\frac{A_{odd}}{\sqrt{\vert\bar{g}\vert}} \epsilon^{ij} \mathrm{Tr}(u)
    \end{split}
\end{equation}
where $\mathrm{Tr}$ is the trace and $\epsilon$ is again the 2-D Levi-Civita tensor ($\epsilon^{11}=\epsilon^{22}=0,\epsilon^{12}=-\epsilon^{21}=1$). So now:
\begin{equation}
    \begin{split}
        \sigma_A^{11} &= 0\\
        \sigma_A^{12} &= -\frac{A_{odd}}{\sqrt{\vert\bar{g}\vert}} \mathrm{Tr}(u)\\
        \sigma_A^{22} &= 0\\
        \sigma_A^{21} &= -\sigma_A^{12}\\
    \end{split}
\end{equation}

Next, we move the contribution to $K$,

\begin{equation}
\begin{split}
\sigma_K^{ij} = \frac{K_{odd}}{2\sqrt{\vert\bar{g}\vert}} \left(2\epsilon^{ik}u_{k}^j +2 \epsilon^{jk} u_{k}^i \right)\\
    \end{split}
\end{equation}

Thus we obtain:
\begin{equation}
\begin{split}
    \sigma_K^{11}
&= \frac{K_{odd}}{\sqrt{\vert\bar{g}\vert}} \left( 2 u_{2}^1 \right) \\
    \sigma_K^{12}
&= \frac{K_{odd}}{\sqrt{\vert\bar{g}\vert}}  \left(u_{2}^2-u_{1}^1 \right) \\
    \sigma_K^{21}
&= \sigma_K^{12} \\
    \sigma_K^{22}
&= 
-\frac{K_{odd}}{\sqrt{\vert\bar{g}\vert}} \left( 2 u_{1}^2 \right) \\
    \end{split}
\end{equation}

We can write the same for the normal elastic parameters $\lambda, \mu$:

\begin{equation}
\begin{split}
 \sigma_{\lambda,\mu}^{ij} = \lambda \bar{g}^{ij} \mathrm{Tr} (u)+ 2 \mu \bar{g}^{ik} u_{k}^{j} \\
\end{split}
\end{equation}

\begin{equation}
\begin{split}
\sigma_{\lambda,\mu}^{11} &= \lambda \bar{g}^{11} \mathrm{Tr}(u) + 2 \mu (\bar{g}^{11} u_{1}^{1} + \bar{g}^{12} u_{2}^{1})\\
\sigma_{\lambda,\mu}^{12} &= \lambda\bar{g}^{12} \mathrm{Tr}(u) + 2 \mu (\bar{g}^{11} u_{1}^{2} + \bar{g}^{12} u_{2}^{2})\\
\sigma_{\lambda,\mu}^{21} &= \lambda\bar{g}^{21} \mathrm{Tr}(u) + 2 \mu (\bar{g}^{21} u_{1}^{2} + \bar{g}^{22} u_{2}^{1})\\
\sigma_{\lambda,\mu}^{22} &=\lambda \bar{g}^{22} \mathrm{Tr}(u) + 2 \mu (\bar{g}^{21} u_{1}^{2} + \bar{g}^{22} u_{2}^{2})\\
\end{split}
\end{equation}
 Thus we have established stress-strain relations in terms of the metric tensor with respect to $A_{odd},K_{odd},\lambda,\mu$. We now move on, to showing the implementation of the area potential. 
 
 \paragraph{Area Potential}
For reasons concerned with the stability of the triangular lattice under large deformations in the presence of odd elastic parameters, we also added in an area potential (to prevent small areas of the triangular faces and ultimately two edges crossing each other; which can cause numerical divergences by means of the calculation of our normal to the face: $\hat{n}_{face} = \mathbf{e}_1 \times \mathbf{e}_2 / \| \mathbf{e}_1 \times \mathbf{e}_2 \|$ where $\{ \mathbf{e}_i\}$ is the current basis derived from the edges of a triangular face as was discussed in the main text). To add such a term we have to potentially implement the following energy term:
\begin{equation}
    \mathcal{F} = C_1 \log \frac{\det g}{ \det \bar{g}}+C_2 \bigg(\log \frac{\det g}{ \det \bar{g}} \bigg)^2
\end{equation}
where $g$ and $\bar{g}$ are the current and reference metrics respectively. $\det g$ and $\det \bar{g}$ are intuitively allowing us to compare the local change in area. Thus the idea of this potential is to create a restoring force that does not allow $\det g$ to become too small. Of course we want to avoid using the energy term as we don't have any conservation of energy in our system. Thus we are instead interested in deriving a stress strain relation:
\begin{equation}
    \sigma^{ij} = \frac{\delta \mathcal{F}}{\delta u_{ij}}
\end{equation}
In order to do this we must express the current metric tensor in terms of the reference basis and the strains. The strain tensor takes the following form:
\begin{equation}
    u_{ij} = \frac{1}{2} [g_{ij}-\bar{g}_{ij}]
\end{equation}
We use the same notation as in the main text: $\{ \mathbf{e}_i\}$ is our current basis and $\{ \bar{\mathbf{e}}_i\}$ is our reference basis. Meanwhile the current and reference metric tensors are expressed respectively as $g_{ij} = \mathbf{e}_i \cdot \mathbf{e}_j$ and $\bar{g}_{ij} = \bar{\mathbf{e}}_i \cdot \bar{\mathbf{e}}_j$. With this, we can write:
\begin{equation}
\begin{split}
    \det g = & (\mathbf{e}_1 \cdot \mathbf{e}_1)(\mathbf{e}_2 \cdot \mathbf{e}_2) - (\mathbf{e}_1 \cdot \mathbf{e}_2)^2 \\
    = & [2u_{11} +\bar{\mathbf{e}}_1\cdot \bar{\mathbf{e}}_1][2u_{22} +\bar{\mathbf{e}}_2\cdot \bar{\mathbf{e}}_2]-[2u_{12} +\bar{\mathbf{e}}_1\cdot \bar{\mathbf{e}}_2]^2
\end{split}
\end{equation}
Using this we can rewrite and define:
\begin{equation}
    \mathcal{G} = \log \frac{\det g}{ \det \bar{g}} = \log \frac{[2u_{11} +\bar{\mathbf{e}}_1\cdot \bar{\mathbf{e}}_1][2u_{22} +\bar{\mathbf{e}}_2\cdot \bar{\mathbf{e}}_2]-[2u_{12} +\bar{\mathbf{e}}_1\cdot \bar{\mathbf{e}}_2]^2}{(\bar{\mathbf{e}}_1 \cdot \bar{\mathbf{e}}_1)(\bar{\mathbf{e}}_2 \cdot \bar{\mathbf{e}}_2) - (\bar{\mathbf{e}}_1 \cdot \bar{\mathbf{e}}_2)^2}
\end{equation}
Differentiating with respect to the strain we obtain:
\begin{equation}
    \frac{\delta \mathcal{G}}{\delta u_{11}} = \frac{2[2u_{22} +\bar{\mathbf{e}}_2\cdot \bar{\mathbf{e}}_2]}{\det g}
\end{equation}
\begin{equation}
    \frac{\delta \mathcal{G}}{\delta u_{22}} = \frac{2[2u_{11} +\bar{\mathbf{e}}_1\cdot \bar{\mathbf{e}}_1]}{\det g}
\end{equation}
\begin{equation}
    \frac{\delta \mathcal{G}}{\delta u_{12}} = -\frac{2[2u_{12} +\bar{\mathbf{e}}_1\cdot \bar{\mathbf{e}}_2]}{\det g}
\end{equation}
This gives us:
\begin{equation}
    \sigma^{ij}=\frac{\delta \mathcal{F}}{\delta u_{ij}} = C_1 \frac{\delta \mathcal{G}}{\delta u_{ij}}+2C_2 \mathcal{G} \frac{\delta \mathcal{G}}{\delta u_{ij}}
\end{equation}
And thus, having obtained the stress tensor we can use the rest of the implementation in the main text to determine the necessary forces to apply on the vertices. By evaluating these expressions at $u_{ij}=0$, we immediately obtain that $C_1=0$ so that the stress tensor is zero in the absence of strain. Thus we are left with:
\begin{equation}
    \sigma^{ij}=\frac{\delta \mathcal{F}}{\delta u_{ij}} = 2C_2 \mathcal{G} \frac{\delta \mathcal{G}}{\delta u_{ij}}
\end{equation}
To obtain the linear elastic response from the area potential in terms of $C_2$, we can differentiate again with respect to $u_{kl}$ and then set $u_{ij}=0$. This in turn gives us the respective contribution, $\mathcal{C}^{ijkl}_{C_2}$, to the elastic modulus tensor:
\begin{equation}
    \mathcal{C}^{ijkl}_{C_2} = \frac{\delta \sigma^{ij}}{\delta u_{kl}} = 2C_2 \bigg[ \mathcal{G} \frac{\delta^2 \mathcal{G}}{\delta u_{ij}\delta u_{kl}} + \frac{\delta \mathcal{G}}{\delta u_{ij}}\frac{\delta \mathcal{G}}{\delta u_{kl}}\bigg] \vert_{u=0} = 2C_2 \frac{\delta \mathcal{G}}{\delta u_{ij}}\frac{\delta \mathcal{G}}{\delta u_{kl}}\vert_{u=0}
\end{equation}
We thus obtain that:
\begin{equation}
    \mathcal{C}^{1111}_{C_2} = 2C_2 \bigg[  \frac{2 \bar{\mathbf{e}}_2 \cdot \bar{\mathbf{e}}_2}{\det \bar{g}} \bigg]^2  
\end{equation}
\begin{equation}
    \mathcal{C}^{2222}_{C_2} = 2C_2 \bigg[  \frac{2 \bar{\mathbf{e}}_1 \cdot \bar{\mathbf{e}}_1}{\det \bar{g}} \bigg]^2  
\end{equation}
\begin{equation}
    \mathcal{C}^{1122}_{C_2} = \mathcal{C}^{2211}_{C_2} = 2C_2\bigg[  \frac{2 \bar{\mathbf{e}}_2 \cdot \bar{\mathbf{e}}_2}{\det \bar{g}} \bigg]  \bigg[  \frac{2 \bar{\mathbf{e}}_1 \cdot \bar{\mathbf{e}}_1}{\det \bar{g}} \bigg]  
\end{equation}
\begin{equation}
    \mathcal{C}^{1212}_{C_2} = \mathcal{C}^{2112}_{C_2}= \mathcal{C}^{2121}_{C_2}=\mathcal{C}^{1221}_{C_2}= 2C_2 \bigg[  \frac{2 \bar{\mathbf{e}}_1 \cdot \bar{\mathbf{e}}_2}{\det \bar{g}} \bigg]^2  
\end{equation}
\begin{equation}
\begin{split}
    & \mathcal{C}^{1112}_{C_2}= \mathcal{C}^{1121}_{C_2}=\mathcal{C}^{1211}_{C_2}=\mathcal{C}^{2111}_{C_2} =   2C_2 \bigg[  \frac{2 \bar{\mathbf{e}}_1 \cdot \bar{\mathbf{e}}_2}{\det \bar{g}} \bigg]   \bigg[  \frac{2 \bar{\mathbf{e}}_2 \cdot \bar{\mathbf{e}}_2}{\det \bar{g}} \bigg]
\end{split}
\end{equation}
\begin{equation}
\begin{split}
    &\mathcal{C}^{2221}_{C_2}= \mathcal{C}^{2212}_{C_2}=\mathcal{C}^{2122}_{C_2}=\mathcal{C}^{1222}_{C_2} =    2C_2 \bigg[  \frac{2 \bar{\mathbf{e}}_1 \cdot \bar{\mathbf{e}}_2}{\det \bar{g}} \bigg]   \bigg[  \frac{2 \bar{\mathbf{e}}_1 \cdot \bar{\mathbf{e}}_1}{\det \bar{g}} \bigg]
\end{split}
\end{equation}

\subsection{Barostat, Integrators and Simulation Procedure}

It is known that there are certain cases, such as active brownian fluids with alignment interactions, for which pressure is not an equation of state \cite{solon2015pressure,falasco2016mesoscopic}. Thus, we begin by remarking on the fact that active systems have to be simulated differently from equilibrium systems due to active heat flow as seen in \cite{falasco2016mesoscopic}. For a generic active matter system with $N$ particles in $d$ dimensions experiencing a damping $\gamma$, the equation of state takes the form:
\begin{equation}
     \frac{1}{\gamma}\sum_{i=1}^{Nd} \langle F^{nc}_i \frac{dr_i }{dt} \rangle +dNk_BT = \int_V d\mathbf{r} Tr(\sigma)-\sum_{i=1}^{Nd} \langle F^{c}_i r_i \rangle-\sum_{i=1}^{Nd} \langle F^{nc}_i r_i \rangle 
\end{equation}
where $F^{nc}$ are the non-conservative forces, $F^c$ are conservative forces and $r_i$ is the position of the i-th degree of freedom. The brackets $\langle \cdot \rangle $ represent ensemble averages. The first term on the right-hand side of the equation is the pressure (this formula is how we calculate the pressure in our simulations as well). The second and third terms on the right-hand side represent the work being done by conservative and non-conservative forces respectively. In the absence of non-conservative forces, this equation simplifies to the usual equilibrium equation of state. The left most term is the active heat flow and represents the energy change by means of non-conservative forces \cite{falasco2016mesoscopic}. However, with over-damped systems ($\gamma \rightarrow \infty$), this term can be ignored and velocity distributions take the values in thermal equilibrium (i.e. $p_i^2/2m = k_BT$ as by the virial theorem). This enables simpler equations of motion and allows us to treat the pressure as a steady-state variable.

Thus, for the simulations we used a Berendsen barostat to tune the box pressure, $\int_V d\mathbf{r} Tr(\sigma)/[dV]$, to zero and to tune the temperature of the system, we utilized a Gaussian distributed noise-force whose variance dictated the magnitude of the temperature. For the integrator, we implemented a BAOAB-limit method \cite{leimkuhler2013rational}, thus the noise is not completely memory-less, but the memory is short ranged in time. We selected the BAOAB-limit method as it reduces error while being very efficient in terms of the descriptive equations of motion. Specifically the BAOAB-limit differential equation is of the form:
\begin{equation}
\label{eq:BAOAB}
    \mathbf{r}_i([n+1]\Delta_{\tau}) = \mathbf{r}_i(n\Delta_{\tau}) + \frac{\Delta_{\tau} \mathbf{\Gamma}}{2} [ \mathbf{F}_{face}+ \mathbf{F}_{dih}]+\sqrt{\frac{k_BT \Delta_{\tau }\mathbf{\Gamma}}{4}} (\mathbf{R}_{n,i}+\mathbf{R}_{n+1,i})
\end{equation}
where $\mathbf{\Gamma}$ is the a diagonal diffusivity matrix with components $(\mathcal{D},\mathcal{D},\mathcal{D}_f)$, $\Delta_{\tau}$ is the time step, $r_i(n\Delta_{\tau})$ is the position of the $i$-th atom at time-step n, $k_BT$ sets the temperature (we take $k_B=1$), $\mathbf{R}_{n,i}$ is a random number generator of variance 1 and mean 0 for time step $n$ operating on vertex $i$. $\mathbf{F}_{dih}$ is the force originating from dihedral degrees of freedom whereas $\mathbf{F}_{face}$ are the forces from the in-plane degrees of freedom of the faces. These results are merely an implementation of forces discussed in the main text. As one can see from this equation, the noise that is used is not independent between adjacent time steps but the correlation is short-ranged nonetheless.

In our simulations we took the matricial diffusivity matrix, $\mathbf{\Gamma}$ in Eq.~\eqref{eq:BAOAB} to be the identity matrix and thus the fluctuation-dissipation is assumed at the microscopic scale.
\subsection{Procedure} We simulated $A_{odd}$ and $K_{odd}$ separately in an otherwise normal elastic system. We took a variety of values of each of the parameters while varying the temperature. Varying the temperature allows us to access different effective length scales (by effectively changing $L/\ell_{\text{th}}$ without changing the linear dimension of the system $L$; look further below for our derivation of the new form of $\ell_{\text{th}}$), rather than doing a computationally costly simulation with a large system size. We allowed each elastic system to ``thermally equilibrate'' for around $2 \cdot 10^7$ time steps, after which we would begin recording instantaneous snapshots of the configurations of the system. The time step was determined by the limiting factors given by the over-damped frequencies of the system:
\begin{equation}
\begin{split}
     \Delta_{\tau,\kappa} = a^4/[8 \pi^3 \mathcal{D}_f\kappa], \Delta_{\tau,\mathcal{C}_{ijkl}} = a^2/[2 \pi \mathcal{D}\mathcal{C}_{ijkl}]
\end{split}
\end{equation}
where $a$ is the lattice spacing. Thus $\Delta_{\tau} \leq \text{Min} \{ \Delta_{\tau,\kappa}, \Delta_{\tau,\mathcal{C}_{ijkl}} \}$.

Each simulation was thermally equilibrated with $2 \times 10^{7}$ time steps. This length of simulation steps was determined by performing auto-correlations of the displacement fields to know the amount of time it takes for the system to lose memor of the initial configuration. For low temperatures, however, thermal equilibration is very difficult as the auto-correlation times grow. After thermal equilibration steps were taken, the simulation ran through $8 \times 10^{8}$ time steps, recording snapshots every $10^5$ time steps. This gave us a total of $8\times10^3$ snapshots of data. For each simulation we measured the auto-correlation time for the amplitude peak to decay by an order of magnitude and used this time-interval to sample fully independent realizations of membrane configurations out of the total data set. With this reduced independent data set, we calculated averaged equal-time correlations. In computation time, this amounted to about 80 hours.

\subsection{Data Tables \label{datatables}}
Here we show the data tables for the plots we generated in Fig.~2 and Fig.~3 of the main text.

\begin{table}[h!]
\centering
\caption{\label{tab:tabdataKplot}}
\begin{tabular}{ |p{1cm}|p{1.5cm}|p{3cm}|p{3cm}|  }
 \hline
 \multicolumn{4}{|c|}{Data Sets for Fig.~2} \\
 \hline
 $L/a$ &  $\kappa/k_BT$& $K_{odd}/(\lambda+2\mu)$& $C_2/(\lambda+2\mu)$\\
 \hline
 $35 $ &   $1 $  & $4.76$ & $.048$ \\
 $35 $&     $10$ & $4.76$ & $.048$ \\
 $35 $&   $1e3$  & $4.76$ & $.048$ \\
 $35$&   $1e5$  & $4.76$& $.048$ \\
 $35$ &   $1$  & $.73$& $.007$ \\
 $35$&   $10$  & $.73$& $.007$ \\
 $35$&   $1e3$  & $.73$& $.007$ \\
 $35$&   $1e5$  & $.73$&$.007$ \\
 $35$&   $1$  & $.076$ &$8e-4$\\
 $35$&   $10$  & $.076$&$8e-4$ \\
 $35$&   $1e3$  & $.076$ &$8e-4$\\
 $35$&   $1e5$  & $.076$ &$8e-4$\\
 \hline
\end{tabular} 

\end{table}

\begin{table}[h!]
\centering
\caption{\label{tab:tabdataAplot}}
\begin{tabular}{ |p{1cm}|p{1.5cm}|p{3cm}|p{3cm}|  }
 \hline
 \multicolumn{4}{|c|}{Data Sets for Fig.~3} \\
 \hline
 $L/a$ &  $\kappa/k_BT$& $A_{odd}/(\lambda+2\mu)$& $C_2/(\lambda+2\mu)$\\
 \hline
 $35 $ &   $1 $  & $.1$ & $.048$ \\
 $35 $&     $10$ & $.1$ & $.048$ \\
 $35 $&   $100$  & $.1$ & $.048$ \\
 $35$&   $1e3$  & $.1$& $.048$ \\
 $35$ &   $1e4$  & $.1$& $.048$ \\
 $35$&   $1e5$  & $.1$& $.048$ \\
 $35$&   $1e6$  & $.1$& $.048$ \\
 \hline
\end{tabular} 
\end{table}

\section{Field Theoretic and Renormalization Group Analysis \label{sec:FieldTheorySection}}

\subsection{Fourier Transform of MSRJD Action and Propagators}
We can take Eq.~(6) of the main text and begin by taking the Fourier transform of the MSRJD action (assuming once again that we are operating in an orthonormal basis), $\mathcal{A}_{MSRJD}$:
\begin{equation}
\begin{split}
         & \mathcal{W}(\eta_j,\eta_f^{\alpha},\Upsilon_i,\Phi^{\alpha}) \propto e^{-\mathcal{A}_{MSRJD}} \equiv  e^{\int d\omega A \cdot \tau \sum_{\mathbf{q}} [k_BT\mathcal{L} \vert \Upsilon_i(\mathbf{q},\omega)\vert^2 -\Upsilon_i(\mathbf{q},\omega)\eta_i(-\mathbf{q},-\omega) +k_BT\mathcal{L}_f \vert \Phi^{\alpha}(\mathbf{q},\omega)\vert^2 -\Phi^{\alpha}(\mathbf{q},\omega)\eta_f^{\alpha}(-\mathbf{q},-\omega)]}
\end{split}
\end{equation}
where $A$ is the area of the system and $\tau$ is the length of time. For completeness, the corresponding Fourier equations of the noises are the following:
\begin{equation}
    \eta_j(\mathbf{q},\omega) = \mathcal{D} \mathcal{C}_{ijkl}q_i[ q_k u_l(\mathbf{q},\omega) -\frac{i}{2} \sum_{\mathbf{p},\gamma} p_k(p_l-q_l)f^{\alpha}(\mathbf{p},\gamma)f^{\alpha}(\mathbf{q}-\mathbf{p},\omega-\gamma)]-i\omega u_j(\mathbf{q},\omega)
\end{equation}
\begin{equation}
\begin{split}
    &\eta_f^{\alpha}(\mathbf{q},\omega)   = (\mathcal{D}_f[\kappa q^4 +\mathcal{C}_{ijkl}u^{o}_{kl}q_iq_j]-i\omega) f^{\alpha}(\mathbf{q},\omega)\\ &+\mathcal{D}_f \mathcal{C}_{ijkl} q_i \bigg[\sum_{(\mathbf{p},\gamma) \neq \mathbf{0}} \bigg(ip_k u_l( \mathbf{p},\gamma) - \frac{1}{2} \sum_{\mathbf{z},\xi} (p_k-z_k)z_l f^{\beta}(\mathbf{p}-\mathbf{z},\gamma-\xi)f^{\beta}(\mathbf{z},\xi) \bigg) (q_j-p_j)f^{\alpha}(\mathbf{q}-\mathbf{p},\omega-\gamma) \bigg]
\end{split}
\end{equation}
where we have used the isotropic bending rigidity, $\kappa$ in place of the bending rigidity tensor and we can further assume isotropy of the homogeneous strain term: $\mathcal{C}_{ijkl}u^{o}_{kl}q_iq_j = \sigma^o q^2$. The MSRJD action, $\mathcal{A}_{MSRJD}$, possesses the following harmonic terms in matrix form:

\begin{equation}
    \begin{split}
     & \mathcal{A}_{MSRJD}^{harm.}=   \frac{1}{2} \begin{bmatrix}\Upsilon_j(\mathbf{q},\omega) \\ u_j(\mathbf{q},\omega)\end{bmatrix}^T \begin{bmatrix}
         -2k_BT\mathcal{L} \delta_{jl}  & i \omega \delta_{jl} + \mathcal{D} \mathcal{C}_{ijkl} q_i q_k \\ -i \omega \delta_{jl} +\mathcal{D} \mathcal{C}_{ilkj} q_i q_k & 0
         \end{bmatrix}  \begin{bmatrix}\Upsilon_l(-\mathbf{q},-\omega) \\ u_l(-\mathbf{q},-\omega)\end{bmatrix} \\
             &+\frac{\delta_{\alpha \beta}}{2} \begin{bmatrix}\Phi^{\alpha}(\mathbf{q},\omega) \\ f^{\alpha}(\mathbf{q},\omega)\end{bmatrix}^T \begin{bmatrix}
    -2k_BT \mathcal{L}_f & i \omega  + \mathcal{D}_f \kappa q^4 + \mathcal{D}_f\sigma^o q^2 \\ -i \omega + \mathcal{D}_f \kappa q^4 + \mathcal{D}_f\sigma^o q^2 & 0
    \end{bmatrix}  \begin{bmatrix} \Phi^{\beta}(-\mathbf{q},-\omega) \\ f^{\beta}(-\mathbf{q},-\omega) \end{bmatrix}
    \end{split}
\end{equation}
From here we can obtain the propagators of the linear theory:
\begin{equation}
    \langle f^{\alpha}(\mathbf{q},\omega)f^{\beta}(-\mathbf{q},-\omega) \rangle = \frac{2k_BT\mathcal{L}_f \delta_{\alpha \beta}}{ A \tau\{ \mathcal{D}_f^2 [ \kappa q^4 + \sigma^o q^2]^2+\omega^2\}}
\label{eq:FlexuralPropagatorDynamic}
\end{equation}
This corresponds to a static propagator, which is obtained by integrating over all $\omega$, thus applying the Cauchy residue theorem \cite{barabasi1995fractal,tauber2014critical}:
\begin{equation}
    \langle f^{\alpha}(\mathbf{q})f^{\beta}(-\mathbf{q}) \rangle = \frac{k_BT \mathcal{L}_f \delta_{\alpha \beta}}{ A  \mathcal{D}_f [ \kappa q^4 + \sigma^o q^2]}
\label{eq:FlexuralPropagatorStatic}
\end{equation}

Analogously we can write the in-plane propagator explicitly as:
\begin{equation}
\begin{split}
    &\langle u_h(\mathbf{q},\omega) u_v(-\mathbf{q},-\omega)\rangle =  \frac{2 k_BT \mathcal{L} \epsilon_{hl}\epsilon_{vp}}{A \tau \text{Det}} [-i \omega \delta_{jl} +q^2 \mathcal{D} \mathcal{C}_{ijkl} P^L_{ik}(\mathbf{q})][i \omega \delta_{jp} +q^2 \mathcal{D} \mathcal{C}_{mjrp} P^L_{rp}(\mathbf{q})]
\label{eq:InPlanePropagatorDynamic}
\end{split}
\end{equation}
where $\text{Det}$ is the determinant of the matrix that couples the harmonic terms $u_j$ of the action (the determinant of $[-i \omega \delta_{jl} +q^2 \mathcal{D} \mathcal{C}_{ijkl} P^L_{ik}(\mathbf{q})][i \omega \delta_{jp} +q^2 \mathcal{D} \mathcal{C}_{mjrp} P^L_{rp}(\mathbf{q})]$). In addition $P^L_{ij}(\mathbf{q}) = q_i q_j /q^2$ is the linear projection operator and $\epsilon$ is the 2-D Levi-Civita tensor. To obtain the static propagators one must apply the Cauchy residue formula once again. The lower-half-plane residues of the determinant denominator can be found as well in \cite{scheibner2020odd} and are:
\begin{equation}
    \omega_{res} = -i\frac{\mathcal{D} q^2}{2}[\lambda+3\mu \pm  \sqrt{(\lambda+\mu)^2 -4 K_{odd}(A_{odd}+K_{odd})}]
\end{equation}
Applying the residue formula results in the following static propagator:
\begin{equation}
\begin{split}
    &\langle u_h(\mathbf{q}) u_v(-\mathbf{q})\rangle = \frac{k_BT\mathcal{L}}{A\mathcal{D} [\lambda+3\mu][\mu(\lambda+2\mu)+K_{odd}(K_{odd}-A_{odd})]q^4} \\ & \{ [\mu(\lambda+3\mu)+K_{odd}(2 K_{odd}-A_{odd})]q_h q_v  \\ &+ [\epsilon_{hk}q_vq_k + \epsilon_{vl}q_h q_l][A_{odd}\mu + K_{odd}(\lambda+\mu)] \\ & + \epsilon_{hk}\epsilon_{vl} q_k q_l [(K_{odd}-A_{odd})(2K_{odd}-A_{odd})+(\lambda+2\mu)(\lambda+3\mu)] \}
\label{eq:InPlanePropagatorStatic}
\end{split}
\end{equation}
\subsection{Reformulating MSRJD Action \label{sec:refMSRJD}}
We can further simplify these equations via two observations. Firstly, we can scale out $\kappa$ via the following transformation:
\begin{equation}
\begin{split}
    &\{\Phi^{\alpha},f^{\alpha},\mathcal{C}_{ijkl},u_j,\Upsilon_j,\mathcal{D},\mathcal{L},\mathcal{D}_{f},\mathcal{L}_f,\sigma^o\} \rightarrow  \{ \sqrt{\kappa} \Phi^{\alpha}, \frac{1}{\sqrt{\kappa}} f^{\alpha} ,  \mathcal{C}_{ijkl} \kappa^2, u_j/\kappa, \kappa \Upsilon_j, \frac{1}{\kappa^2} \mathcal{D}, \frac{1}{\kappa^2} \mathcal{L}, \mathcal{D}_{f}/ \kappa,\mathcal{L}_f/\kappa,\sigma^o\kappa \}
\end{split}
\end{equation}
The second simplifying step can be taken by recognizing that the absolute value of the critical temperature of the theory, $\mathcal{L}_f/\mathcal{D}_{f}$, is not important \cite{tauber2014critical}. Thus one can rescale all the order parameters and diffusivities such that $\mathcal{L}_f$ disappears from the equations:
\begin{equation}
\begin{split}
\label{eq:DfScaleTransformation}
    & \{\Phi^{\alpha},f^{\alpha},\mathcal{C}_{ijkl},u_j,\Upsilon_j,\mathcal{D},\mathcal{L} \} =  \{ \sqrt{\frac{\mathcal{D}_{f}}{\mathcal{L}_f}} \tilde{\Phi}^{\alpha}, \sqrt{\frac{\mathcal{L}_f}{\mathcal{D}_{f}}} \tilde{f}^{\alpha} , \frac{\mathcal{D}_{f}}{\mathcal{L}_f} \tilde{\mathcal{C}}_{ijkl}, \frac{\mathcal{L}_f}{\mathcal{D}_{f}} \tilde{u}_j, \frac{\mathcal{D}_{f}}{\mathcal{L}_f} \tilde{\Upsilon}_j, \frac{\mathcal{L}_f}{\mathcal{D}_{f}} \tilde{\mathcal{D}}, \bigg( \frac{\mathcal{L}_f}{\mathcal{D}_{f}} \bigg)^2 \tilde{\mathcal{L}}\}
\end{split}
\end{equation}
Hence, the harmonic theory adjusts as follows:
\begin{equation}
    \begin{split}
     & \mathcal{A}_{MSRJD}^{harm.}=   \frac{1}{2} \begin{bmatrix}\tilde{\Upsilon}_j(\mathbf{q},\omega) \\ \tilde{u}_j(\mathbf{q},\omega)\end{bmatrix}^T \begin{bmatrix}
         -2 k_BT\tilde{\mathcal{L}} \delta_{jl}  & i \omega \delta_{jl} + \tilde{\mathcal{D}} \tilde{\mathcal{C}}_{ijkl} q_i q_k \\ -i \omega \delta_{jl} + \tilde{\mathcal{D}} \tilde{\mathcal{C}}_{ilkj} q_i q_k & 0
         \end{bmatrix}  \begin{bmatrix}\tilde{\Upsilon}_l(-\mathbf{q},-\omega) \\ \tilde{u}_l(-\mathbf{q},-\omega)\end{bmatrix} \\
             &+\frac{\delta_{\alpha \beta}}{2} \begin{bmatrix}\tilde{\Phi}^{\alpha}(\mathbf{q},\omega) \\ \tilde{f}^{\alpha}(\mathbf{q},\omega)\end{bmatrix}^T \begin{bmatrix}
    -2k_BT\mathcal{D}_{f}  & i \omega  + \mathcal{D}_{f} q^4 + \mathcal{D}_{f}\sigma^o q^2 \\ -i \omega + \mathcal{D}_{f} q^4 + \mathcal{D}_{f} \sigma^o q^2 & 0
    \end{bmatrix}  \begin{bmatrix} \tilde{\Phi}^{\beta}(-\mathbf{q},-\omega) \\ \tilde{f}^{\beta}(-\mathbf{q},-\omega) \end{bmatrix}
    \end{split}
\end{equation}

Despite having introduced these extra Hubbard-Stratonovich variables, a key simplifying feature of the MSRJD approach can be seen from the form of these matrices: Feynman diagrams which contain the contraction of two response variables together will be null \cite{tauber2014critical, altland2010condensed}. Thus, such Feynman diagrams need not be considered. But before obtaining information about the renormalization factors and the Feynman diagrams, we first review the scaling of the harmonic theory.

\subsection{Scaling of Harmonic Theory \label{sec:powercount}}
We briefly discuss the scaling of the harmonic theory. Since we have two order parameters with different dispersion relations, we are first confronted with how to rescale frequencies. To resolve this we must go to the propagators in the linear theory:
\begin{equation}
    \langle \tilde{\Phi}^{\alpha}(\mathbf{q},\omega) \tilde{f}^{\alpha}(-\mathbf{q},-\omega) \rangle \sim \frac{k_BT \mathcal{D}_f}{A\tau[-i \omega +\mathcal{D}_f( q^4 +\sigma^o q^2)]}
\end{equation}
where we will make the assumptions of vanishing stress, $\sigma^o q^2<<\kappa q^4$, and
\begin{equation}
    \langle \tilde{\Upsilon}_j(\mathbf{q},\omega) \tilde{u}_l(-\mathbf{q},-\omega) \rangle \sim \frac{k_BT\tilde{\mathcal{L}}}{A\tau[-i \omega \delta_{jl} +\tilde{\mathcal{D}} \tilde{\mathcal{C}}_{ijkl} q_iq_k]}
\end{equation}
From this we see that in the small stress limit, the residues scale as $\omega \sim q^4$ in the flexural response propagator and $\omega \sim q^2$ for the in-plane phonon response propagator \cite{frey1991dynamics,radzihovsky1995new,tauber2014critical}. Thus in the small frequency limit, the flexural modes are much slower and thus in-plane phonons should be considered as ``fast'' variables in the field theoretic sense. Thus, the terms that establish the harmonic theory in $\mathcal{A}_{MSRJD}$ are:
\begin{equation}
    \tilde{\Phi}^{\alpha} \partial_t \tilde{f}^{\alpha}, \tilde{\Phi}^{\alpha} \nabla^4 \tilde{f}^{\alpha}, \tilde{\Phi}^{\alpha 2}
\end{equation}
This makes the coefficients of these terms automatically scale-invariant. We now perform a power counting procedure to obtain the scaling of the theory and render the action, $\mathcal{A}_{MSRJD}$, massless \cite{kardar2007statistical}. Thus if we assign scale powers in a momentum-shell sense \cite{kardar2007statistical}:
\begin{equation}
    \{ \mathbf{r} \rightarrow b \mathbf{r} \}, \{ t \rightarrow b^{\zeta_t} t \}, \{ \tilde{f}^{\alpha} \rightarrow b^{\zeta_f} \tilde{f}^{\alpha} \}, \{ \tilde{\Phi}^{\alpha} \rightarrow b^{\zeta_{\Phi}} \tilde{\Phi}^{\alpha} \}
\end{equation}
then we obtain that:
\begin{equation}
    \zeta_t=4, \zeta_f =  \frac{-D+\zeta_t}{2}, \zeta_{\Phi} = \frac{-D-\zeta_t}{2}
\end{equation}
Though we are taking $D=2$ (the number of in-plane dimensions), we leave $D$ un-inserted to show the general scaling. If we also assign the following rescaling factors:
\begin{equation}
    \{\tilde{u}_i \rightarrow b^{\zeta_u} \tilde{u}_i \},\{ \tilde{\Upsilon}_i \rightarrow b^{\zeta_{\Upsilon}} \tilde{\Upsilon}_i\}
\end{equation}
then via the form of the strain tensor we also obtain:
\begin{equation}
    \zeta_u = 2 \zeta_f -1 = -D+\zeta_t-1 
\end{equation}
and thus via observing that $\tilde{\Upsilon}_i \partial_t \tilde{u}_i$ must also be scale invariant we obtain:
\begin{equation}
    \zeta_{\Upsilon}  = D+1 
\end{equation}
The linear scaling of the theory thus results in the following mass dimensions:
\begin{equation}
    \{ \tilde{\mathcal{C}}_{ijkl} \rightarrow b^{4-D}\tilde{\mathcal{C}}_{ijkl} \} , \{ \tilde{\mathcal{L}}, \tilde{\mathcal{D}} \rightarrow b^{D-2} \tilde{\mathcal{L}} , \tilde{\mathcal{D}} \}, \{\sigma^o \rightarrow b^2 \sigma^o \}
\end{equation}
This establishes that the upper critical dimension for $\tilde{\mathcal{C}}_{ijkl}$ is $4$. With this scaling, one can see that the terms $\tilde{\Upsilon}_i \partial_t^2 \tilde{u}_i$ and $\tilde{\Phi}^{\alpha} \partial_t^2 \tilde{f}^{\alpha}$ would be irrelevant, which both scale with exponent $-4$.

\subsection{Absence of Ward Identity}
An important tool in the renormalization group are the use of symmetries of the action. In the case of the free energy associated with equilibrium fluctuating elastic membranes, Guitter et al. \cite{guitter1989thermodynamical} established a symmetry of both the strain tensor:
\begin{equation}
\label{eq:GuitterSymmetry}
\begin{split}
    & f^{\alpha}(\mathbf{r}) \rightarrow  f^{\alpha}(\mathbf{r}) + A_i^{\alpha} r_i \\ 
    & u_i(\mathbf{r}) \rightarrow  u_i(\mathbf{r}) - A_i^{\alpha} f^{\alpha}(\mathbf{r}) -\frac{1}{2} A_i^{\alpha} A_j^{\alpha} r_j
\end{split}
\end{equation}
which helps to establish the Ward identity associated with the effective action, $\Gamma$ \cite{peskin2018introduction}:
\begin{equation}
    \int d^D\mathbf{r} \bigg[ r_i \frac{\delta \Gamma}{\delta f^{\alpha}} - f^{\alpha} \frac{\delta \Gamma}{\delta u_i}\bigg] = 0
\end{equation}
This Ward identity spares extra calculations as it enforces that the coefficient of the $\partial_i u_j \partial_k u_l$ vertex will renormalize exactly as $\partial_i u_j \partial_k f^{\beta} \partial_l f^{\beta}$ and $ \partial_i f^{\alpha} \partial_j f^{\alpha} \partial_k f^{\beta} \partial_l f^{\beta}$. Thus one need only renormalize one of these vertices to obtain the renormalization of the in-plane elastic constants, $C_{ijkl}$. 

Because the symmetry \cite{guitter1989thermodynamical} obtained is a symmetry of the strain tensor, as soon as one incorporates components that prevent an elastic action from being entirely formulated in terms of $u_{ij}$ and $\nabla^2 f$, the symmetry and its associated Ward identity no longer hold. Indeed the only dynamic term that could be added to a free energy should be of the form $(\partial_t u_{ij})^2$, which would lead to non-physical forces and equations that we wouldn't derive kinematically. One need not the formalism of the Ward identity to observe this either. From Eq.~(3) and Eq.~(4) of the main text, since the kinematic condition is not necessarily satisfied, $\partial_i \sigma_{ij} = \mathcal{C}_{ijkl}\partial_i u_{kl} \neq 0$, Eq.~\ref{eq:GuitterSymmetry} is no longer a symmetry of the dynamic equations. This holds whether we are working with the odd elastic equations or not. Thus \cite{frey1991dynamics} is also missing the Ward identity. In the case of equilibrium Langevin elastic membranes, Feynman diagrams will retain the same effective structure as in the case of \cite{guitter1989thermodynamical} and thus they derive the matching results. However, for odd elastic systems where there is no analogue and the structure of elastic tensors has been generalized, this is no longer a guarantee. Thus, without a Ward identity, further care will be required because many more Feynman diagrams must be calculated.  

One immediate consequence of the lack of the Ward identity means that renormalization may not preserve equality between the coefficients of vertices in Eq.~(3) and Eq.~(4) of the main text, potentially resulting in breaking any microscopic fluctuation-dissipation. Thus these equations must be generalized into the following form:
\begin{equation}
\label{eq:LangevinUDampedGeneralized}
   \partial_t \tilde{u}_j(\mathbf{r},t) =  \tilde{\mathcal{C}}_{ijkl}^{u,\mathcal{D}} \partial_i \partial_k \tilde{u}_{l}(\mathbf{r},t) +\frac{1}{2} \tilde{\mathcal{C}}_{ijkl}^{f,\mathcal{D}} \partial_i (\partial_k \tilde{f}^{\alpha}(\mathbf{r},t)\partial_l \tilde{f}^{\alpha}(\mathbf{r},t))+\eta_j(\mathbf{r},t)
\end{equation}
\begin{equation}
\label{eq:LangevinFDampedGeneralized}
\begin{split}
    \partial_t \tilde{f}^{\alpha}(\mathbf{r},t) & = [- \mathcal{D}_f \Delta^2 +\tilde{\sigma}^o \Delta]  \tilde{f}^{\alpha}(\mathbf{r},t)+ \tilde{\mathcal{C}}_{ijkl}^{u,\mathcal{D}_f}\partial_i[\partial_k \tilde{u}_{l}(\mathbf{r},t)\partial_j \tilde{f}^{\alpha}(\mathbf{r},t)]+ \tilde{\mathcal{C}}_{ijkl}^{f,\mathcal{D}_f}\partial_i[\frac{1}{2}\partial_k \tilde{f}^{\beta}(\mathbf{r},t)\partial_l \tilde{f}^{\beta}(\mathbf{r},t)\partial_j \tilde{f}^{\alpha}(\mathbf{r},t)]  +\eta_f^{\alpha}(\mathbf{r},t)
\end{split}
\end{equation}
where one can observe that the diffusivities and elastic tensor have been combined and $\tilde{\sigma}^o = \mathcal{D}_f \sigma^o$. Consequently the MSRJD action must also be generalized accordingly. We re-state them in their final form before commencing the renormalization scheme:

\begin{equation}
\begin{split}
\label{eq:FourierULangevinFinal}
    \eta_j(\mathbf{q},\omega) =&   q_i[\tilde{\mathcal{C}}_{ijkl}^{u,\mathcal{D}} q_k \tilde{u}_l(\mathbf{q},\omega) -\frac{i}{2} \tilde{\mathcal{C}}_{ijkl}^{f,\mathcal{D}} \sum_{\mathbf{p},\gamma} p_k(p_l-q_l)\tilde{f}^{\alpha}(\mathbf{p},\gamma)\tilde{f}^{\alpha}(\mathbf{q}-\mathbf{p},\omega-\gamma)]-i\omega \tilde{u}_j(\mathbf{q},\omega)
\end{split}
\end{equation}

\begin{equation}
\label{eq:FourierFLangevinFinal}
\begin{split}
 &\eta_f^{\alpha}(\mathbf{q},\omega)  = ([\mathcal{D}_f q^4 +\tilde{\sigma}^o q^2]-i\omega) \tilde{f}^{\alpha}(\mathbf{q},\omega)\\ & +  q_i \bigg[\sum_{(\mathbf{p},\gamma) \neq \mathbf{0}} \bigg(i \tilde{\mathcal{C}}_{ijkl}^{u,\mathcal{D}_f} p_k u_l( \mathbf{p},\gamma) - \frac{\tilde{\mathcal{C}}_{ijkl}^{f,\mathcal{D}_f}}{2} \sum_{\mathbf{z},\xi} (p_k-z_k)z_l \tilde{f}^{\beta}(\mathbf{p}-\mathbf{z},\gamma-\xi)\tilde{f}^{\beta}(\mathbf{z},\xi) \bigg) (q_j-p_j)\tilde{f}^{\alpha}(\mathbf{q}-\mathbf{p},\omega-\gamma) \bigg]
\end{split}
\end{equation}

\begin{equation}
\begin{split}
\label{eq:MeasureOddFinal}
         & \mathcal{W}(\eta_j,\eta_f^{\alpha},\Upsilon_i,\tilde{\Phi}^{\alpha}) \propto e^{-\mathcal{A}_{MSRJD}} =   e^{\int d\omega A \cdot \tau \sum_{\mathbf{q}} [k_BT\tilde{\mathcal{L}} \vert \tilde{\Upsilon}_i(\mathbf{q},\omega)\vert^2 -\tilde{\Upsilon}_i(\mathbf{q},\omega)\eta_i(-\mathbf{q},-\omega) +k_BT\mathcal{D}_f \vert\tilde{\Phi}^{\alpha}(\mathbf{q},\omega)\vert^2 -\tilde{\Phi}^{\alpha}(\mathbf{q},\omega)\eta_f^{\alpha}(-\mathbf{q},-\omega)]}
\end{split}
\end{equation}

with the harmonic portion of the action taking the form:

\begin{equation}
\label{eq:ActionOddFinal}
    \begin{split}
     &\mathcal{A}_{MSRJD}^{harm.}=   \frac{1}{2} \begin{bmatrix}\tilde{\Upsilon}_j(\mathbf{q},\omega) \\ \tilde{u}_j(\mathbf{q},\omega)\end{bmatrix}^T \begin{bmatrix}
         -2 k_BT\tilde{\mathcal{L}} \delta_{jl}  & i \omega \delta_{jl} + \tilde{\mathcal{C}}_{ijkl}^{u,\mathcal{D}} q_i q_k \\ -i \omega \delta_{jl} + \tilde{\mathcal{C}}_{ijkl}^{u,\mathcal{D}} q_i q_k & 0
         \end{bmatrix}  \begin{bmatrix}\tilde{\Upsilon}_l(-\mathbf{q},-\omega) \\ \tilde{u}_l(-\mathbf{q},-\omega)\end{bmatrix} \\
             &+\frac{\delta_{\alpha \beta}}{2} \begin{bmatrix}\tilde{\Phi}^{\alpha}(\mathbf{q},\omega) \\ \tilde{f}^{\alpha}(\mathbf{q},\omega)\end{bmatrix}^T \begin{bmatrix}
    -2 k_BT \mathcal{D}_f  & i \omega  + \mathcal{D}_f q^4 + \tilde{\sigma}^o q^2 \\ -i \omega + \mathcal{D}_f q^4 + \tilde{\sigma}^o q^2 & 0
    \end{bmatrix}  \begin{bmatrix} \tilde{\Phi}^{\beta}(-\mathbf{q},-\omega) \\ \tilde{f}^{\beta}(-\mathbf{q},-\omega) \end{bmatrix}
    \end{split}
\end{equation}

with additional anharmonic terms:
\begin{equation}
\label{eq:ActionOddAnharmonic}
    \begin{split}
     &\mathcal{A}_{MSRJD}^{anharm.} = - \tilde{\Upsilon}_i (\mathbf{q},\omega) \bigg[ \frac{iq_i}{2} \tilde{\mathcal{C}}^{f,\mathcal{D}}_{ijkl} \sum_{\mathbf{p},\gamma} p_k (p_l+q_l) \tilde{f}^{\alpha}(\mathbf{p},\gamma) \tilde{f}^{\alpha} (-\mathbf{q}-\mathbf{p},-\omega-\gamma)  \bigg] \\ & + \tilde{\Phi}^{\alpha}(\mathbf{q},\omega) q_i \bigg[\sum_{(\mathbf{p},\gamma) \neq \mathbf{0}} \bigg(i \tilde{\mathcal{C}}_{ijkl}^{u,\mathcal{D}_f} p_k u_l( \mathbf{p},\gamma)  - \frac{\tilde{\mathcal{C}}_{ijkl}^{f,\mathcal{D}_f}}{2} \sum_{\mathbf{z},\xi} (p_k-z_k)z_l \tilde{f}^{\beta}(\mathbf{p}-\mathbf{z},\gamma-\xi)\tilde{f}^{\beta}(\mathbf{z},\xi) \bigg) (-q_j-p_j)\tilde{f}^{\alpha}(-\mathbf{q}-\mathbf{p},-\omega-\gamma) \bigg]
    \end{split}
\end{equation}
These non-linear/anharmonic terms can be found in the next section in Fig.~\ref{fig:FeynmanIsolatedOdd} as diagrams $(a), (b),(c)$. From here forwards we will take $k_BT=1$ as was done in \cite{frey1991dynamics}, due to the fact that there are many parameters that are present in the theory. Setting $k_BT=1$ will also not change the asymptotic results of the scaling of the moduli.
\subsection{\label{sec:RenormalizationOdd} Renormalization of Over-Damped Odd Elastic Membranes}

\subsubsection{\label{sec:RenormalizationDiagramsODE} Feynman Diagrams and Renormalization Group Equations}
\begin{figure}[h!]
\includegraphics[width=\textwidth]{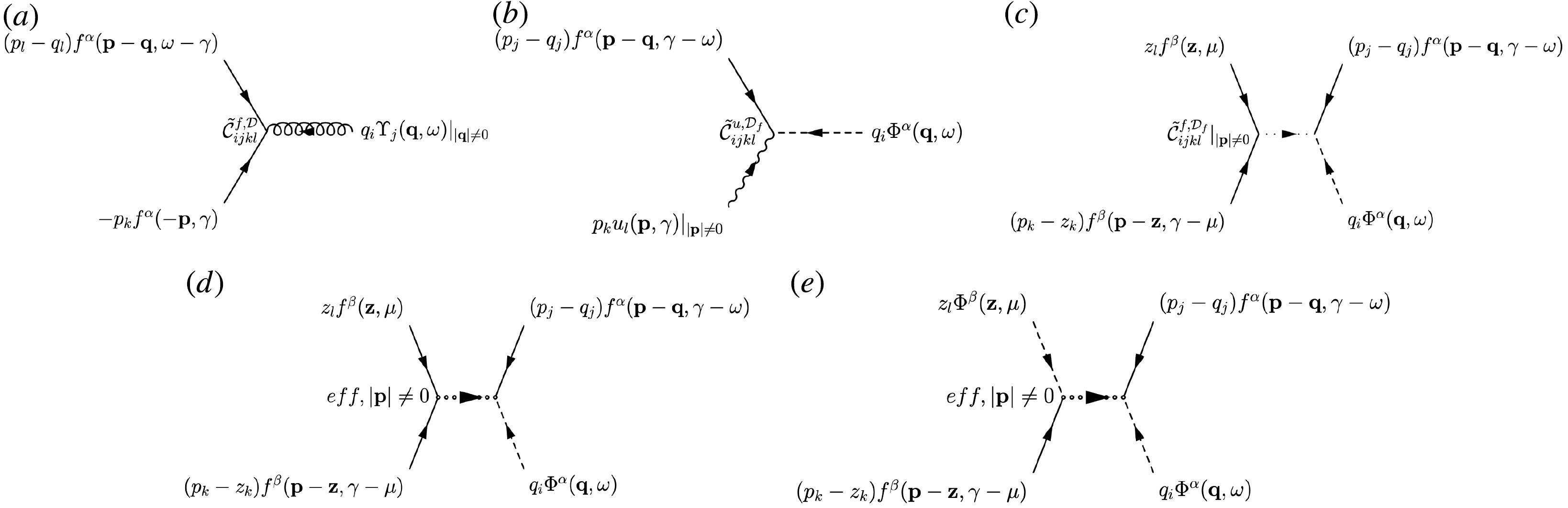}
\caption{The Feynman diagrams corresponding to the linear terms in equations Eq.~\eqref{eq:FourierULangevinFinal} and Eq.~\eqref{eq:FourierFLangevinFinal} are shown in (a), (b) and (c). If one integrates out the in-plane order parameters $\tilde{u}_j,\tilde{\Upsilon}_j$, then one obtains in effective flexural vertices shown in (d) and (e).}
\label{fig:FeynmanIsolatedOdd}
\end{figure}

With the set up of the equations complete, we can commence the process of renormalization. Each term can be renormalized by the contraction of an-harmonicities Taylor-expanded from the MSRJD action. We intend to show the calculations diagrammatically with an attached Mathematica code which performs the accompanying analytic calculations. We begin by representing an-harmonic terms diagrammatically in isolation. This is done in Fig.~\ref{fig:FeynmanIsolatedOdd} (a),(b) and (c). By further integrating out the harmonic in-plane field using Eq.~\eqref{eq:MeasureOddFinal} and Eq.~\eqref{eq:ActionOddFinal}, one can also obtain effective vertices of the flexural field shown in (d) and (e). This is analytically done in the attached Mathematica code, but is too complicated to put in closed form in text. The effective flexural vertices will aid us by allowing us to calculate less Feynman diagrams in total.

\begin{figure}[h!]
\includegraphics[width=\textwidth]{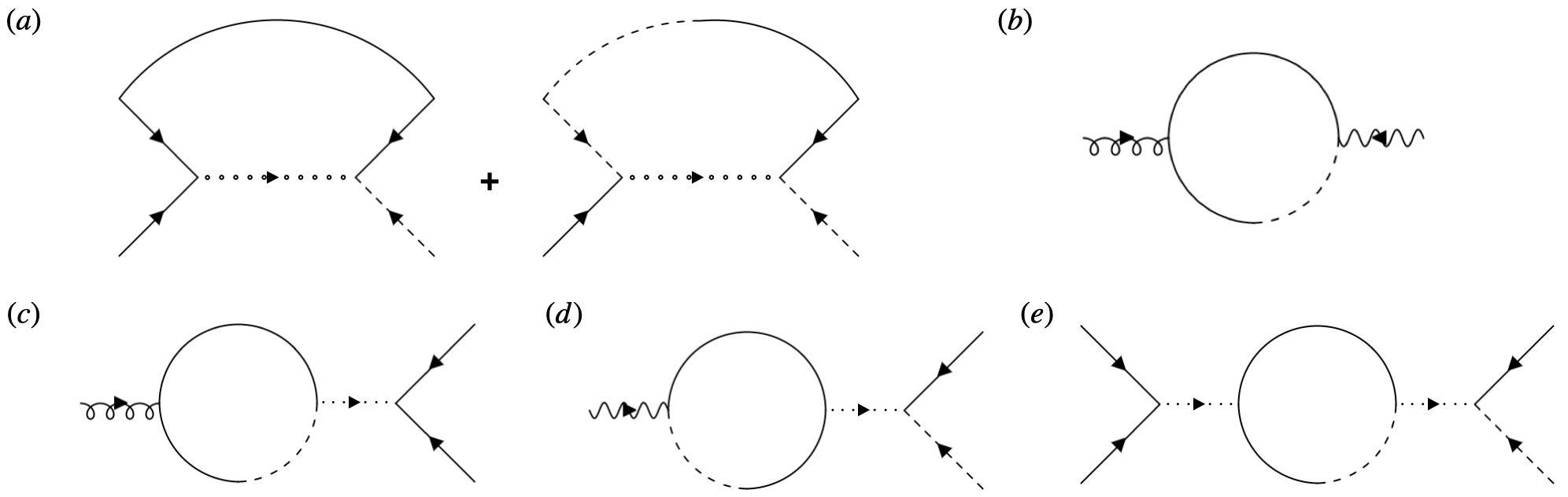}
\caption{The Feynman diagrams corresponding to a 1-loop renormalization group to leading order in $d_c$ are shown. Diagrams in (a) renormalize $\mathcal{D}_f$ and $\tilde{\sigma}^o$, (b) $\tilde{\mathcal{C}}_{ijkl}^{u,\mathcal{D}}$, (c) $\tilde{\mathcal{C}}_{ijkl}^{f,\mathcal{D}}$, (d) $\tilde{\mathcal{C}}_{ijkl}^{u,\mathcal{D}_f}$, (e) $\tilde{\mathcal{C}}_{ijkl}^{f,\mathcal{D}_f}$. }
\label{fig:FeynmanContractOdd}
\end{figure}

\begin{figure}[h!]
\includegraphics[width=\textwidth]{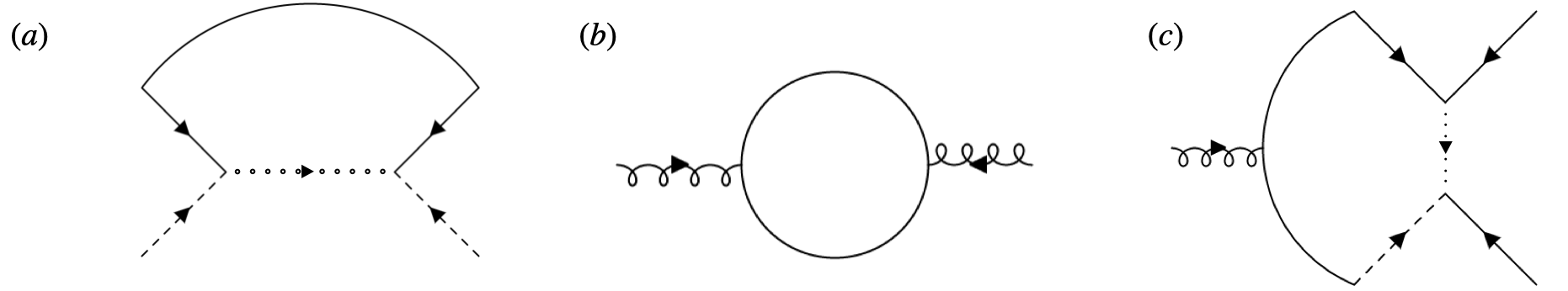}
\caption{Examples of numerous Feynman diagrams that have been ignored are shown. (a) renormalizes $(\tilde{\Phi}^{\alpha})^2$ to non-zero orders of the wave-vector and thus would be irrelevant with respect to its scalar coefficient. (b) is ignored on basis of the same argument with respect to $\tilde{\Upsilon}_j^2$. (c) Diagrams of this type have been ignored as they contribute to $\tilde{\mathcal{C}}_{ijkl}^{f,\mathcal{D}}$ to a lower order in $d_c$ than the diagram found in Fig.~\ref{fig:FeynmanContractOdd}(c).}
\label{fig:FeynmanIgnoredOdd}
\end{figure}

We perform a 1-loop perturbative momentum-shell renormalization group scheme to leading order in $d_c$. The 1-particle-irreducible diagrams included in such a scheme are given in Fig.~\ref{fig:FeynmanContractOdd} \cite{peskin2018introduction,tauber2014critical,barabasi1995fractal,frey1991dynamics}. Other diagrams in a 1-loop scheme are ignored as they are lower in order $d_c$, such as that found in Fig.~\ref{fig:FeynmanIgnoredOdd}(c). However, such diagrams would have to be hypothetically included in an $\epsilon$-expansion analysis, as all 1-loop diagrams are of order $\epsilon$. In the case when fluctuation-dissipation relations hold and in the absence of odd elasticity, these extra diagrams vanish (thus leading to the exact same analysis as presented by \cite{frey1991dynamics}). This is not the case if those two conditions do not hold. 
The contributions of these extra diagrams in the non-odd-elastic case has been recently done by \cite{le2021thermal}, where breaking the isotropy of the embedding space is effectively equivalent to our case in which fluctuation-dissipation relations do not hold. We also note here that these other ignored Feynman diagrams can also allow $A_{odd}$ and $K_{odd}$ to generate one another.
Thus further investigation is merited in exploring these other potential Feynman diagrams that we have ignored here. Other diagrams such as Fig.~\ref{fig:FeynmanIgnoredOdd}(a,b) can be ignored on the grounds that they produce only higher-order-wavelength contributions to the original vertices and thus are respectively irrelevant.

Considering these Feynman diagrams, we can write down the renormalization factors in order to calculate our renormalization group equations. To remove the $UV$ divergences due to an-harmonic Feynman diagrams and reformulate the theory in terms of renormalized variables, we make the following ansatz of how to rescale the theory:

\begin{equation}
\begin{split}
       & \tilde{u}_i^R =Z^{-1}\tilde{u}_i, \tilde{f}^{\alpha,R} =Z_f^{-1/2}\tilde{f}^{\alpha},\\ &\tilde{\Upsilon}^R=Z_{\Upsilon}^{-1}\tilde{\Upsilon}, \tilde{\Phi}^{\alpha, R} = Z_{\Phi}^{-1/2} \tilde{\Phi}^{\alpha}
\end{split}
\end{equation}
\begin{equation}
    \mathcal{D}_f^R=Z_{\mathcal{D}_f} D_{f}
\end{equation}
\begin{equation}
    \tilde{\mathcal{L}}^R=Z_{L} \tilde{\mathcal{L}}
\end{equation}
\begin{equation}
    \tilde{\mathcal{C}}_{ijkl}^{u,\mathcal{D},R} = Z_{ijkl}^{u,\mathcal{D}}\tilde{\mathcal{C}}_{ijkl}^{u,\mathcal{D}}
\end{equation}
\begin{equation}
    \tilde{\mathcal{C}}_{ijkl}^{f,\mathcal{D},R} = Z_{ijkl}^{f,\mathcal{D}}\tilde{\mathcal{C}}_{ijkl}^{f,\mathcal{D}}
\end{equation}
\begin{equation}
    \tilde{\mathcal{C}}_{ijkl}^{u,\mathcal{D}_f,R} = Z_{\mathcal{D}_f}^{-1} Z_{ijkl}^{u,\mathcal{D}_f} \tilde{\mathcal{C}}_{ijkl}^{u,\mathcal{D}_f}
\end{equation}
\begin{equation}
    \tilde{\mathcal{C}}_{ijkl}^{f,\mathcal{D}_f,R} = Z_{\mathcal{D}_f}^{-1}Z_{ijkl}^{f,\mathcal{D}_f}\tilde{\mathcal{C}}_{ijkl}^{f,\mathcal{D}_f}
\end{equation}
\begin{equation}
    \tilde{\sigma}^{R} = Z_{\sigma}\tilde{\sigma}^o
\end{equation}
where we have used our scale transformations in Eq.~\ref{eq:DfScaleTransformation} to account for how the diffusivity, $\mathcal{D}_f$, has been absorbed into the tensors $\tilde{\mathcal{C}}_{ijkl}^{u,\mathcal{D}_f},\tilde{\mathcal{C}}_{ijkl}^{f,\mathcal{D}_f}$ and the stress $\tilde{\sigma}^o$. From this ansatz, similar as to \cite{frey1991dynamics}, we can obtain certain reductions in the number of independent renormalization factors. We obtain symmetries of $\Gamma_{M,\Tilde{M},N,\Tilde{N}}$ which is the effective vertex function with $M$ $f$ fields, $\Tilde{M}$ $\Phi$ fields, $N$ $u$ fields and $\Tilde{N}$ $\Upsilon$ fields as in \cite{frey1991dynamics,peskin2018introduction}. Since:
\begin{equation}
\begin{split}
   & \partial_{\omega} \Gamma_{1100}(\mathbf{q}=0,\omega) \sim i \sqrt{Z_fZ_{\Phi}}
   \\ & \partial_{\omega} \Gamma_{0011}(\mathbf{q}=0,\omega) \sim iZ_uZ_{\Upsilon}
\end{split}
\end{equation}
and since all an-harmonic vertices vanish when $\mathbf{q}=0$, we obtain that $Z=1/Z_{\Upsilon}$ and $Z_f = 1/Z_{\Phi}$. Furthermore:
\begin{equation}
    \Gamma_{0200}(\mathbf{q}=0,\omega) \sim Z_{\mathcal{D}_f} Z_{\Phi}
\end{equation}
which also implies that $Z_{\mathcal{D}_f} = 1/Z_{\Phi}$. Thus we have established that $Z_f = Z_{\mathcal{D}_f}$. However, unlike \cite{frey1991dynamics}, we cannot use any Ward identity as we have shown that it is not valid for a dynamical action and thus we cannot establish that $Z=Z_f$. This is an important point because without the Ward identity, we cannot establish further simplifications of the following renormalization factors $Z_{ijkl}^{u,\mathcal{D}}, Z_{ijkl}^{f,\mathcal{D}},Z_{ijkl}^{u,\mathcal{D}_f},Z_{ijkl}^{f,\mathcal{D}_f}$. Thus we must treat these renormalization factors as independent. With renormalization factors established, we can write down the following renormalization group ODEs for the following parameters with the use of our evaluated Feynman diagrams:
\begin{subequations}
\label{eq:OddElasticRGEquations}
\begin{align}
& \beta_{\tilde{\mu}^{u,\mathcal{D},R}} = 2 \tilde{\mu}^{u,\mathcal{D},R} -d_c\mathcal{D}_f^R \Lambda^2 \frac{\tilde{\mu}^{f,\mathcal{D},R} \tilde{\mu}^{u,\mathcal{D}_f,R} - \tilde{K}_{odd}^{f,\mathcal{D},R} \tilde{K}_{odd}^{u,\mathcal{D}_f,R} }{8 \pi [\mathcal{D}_f^R \Lambda^2 +\tilde{\sigma}^{R}]^2}  \\
&  \beta_{\tilde{\lambda}^{u,\mathcal{D},R}} = 2 \tilde{\lambda}^{u,\mathcal{D},R}  -d_c\mathcal{D}_f^R \Lambda^2 \frac{2\tilde{\lambda}^{f,\mathcal{D},R}(\tilde{\lambda}^{u,\mathcal{D}_f,R} + \tilde{\mu}^{u,\mathcal{D}_f,R} )+\tilde{\mu}^{f,\mathcal{D},R}(2 \tilde{\lambda}^{u,\mathcal{D}_f,R} + \tilde{\mu}^{u,\mathcal{D}_f,R} )- \tilde{K}_{odd}^{f,\mathcal{D},R} \tilde{K}_{odd}^{u,\mathcal{D}_f,R} }{8 \pi [\mathcal{D}_f^R \Lambda^2 +\tilde{\sigma}^{R}]^2} \nonumber \\
&  \beta_{\tilde{K}_{odd}^{u,\mathcal{D},R}} = 2 \tilde{K}_{odd}^{u,\mathcal{D},R} -d_c \mathcal{D}_f^R \Lambda^2 \frac{ \tilde{K}_{odd}^{f,\mathcal{D},R} \tilde{\mu}^{u,\mathcal{D}_f,R} +\tilde{\mu}^{f,\mathcal{D},R} \tilde{K}_{odd}^{u,\mathcal{D}_f,R} }{8 \pi [\mathcal{D}_f^R \Lambda^2 +\tilde{\sigma}^{R}]^2} \nonumber \\
&  \beta_{\tilde{A}_{odd}^{u,\mathcal{D},R}} = 2 \tilde{A}_{odd}^{u,\mathcal{D},R} -d_c\mathcal{D}_f^R \Lambda^2\frac{ \tilde{A}_{odd}^{f,\mathcal{D},R} (\tilde{\mu}^{u,\mathcal{D}_f,R} +\tilde{\lambda}^{u,\mathcal{D}_f,R}) }{4 \pi [\mathcal{D}_f^R \Lambda^2 +\tilde{\sigma}^{R}]^2} \nonumber \\
& \beta_{\tilde{\mu}^{u,\mathcal{D}_f,R}} = 2 \tilde{\mu}^{u,\mathcal{D}_f,R} -d_c \mathcal{D}_f^R \Lambda^2\frac{\tilde{\mu}^{f,\mathcal{D}_f,R} \tilde{\mu}^{u,\mathcal{D}_f,R} - \tilde{K}_{odd}^{f,\mathcal{D}_f,R} \tilde{K}_{odd}^{u,\mathcal{D}_f,R} }{8 \pi [\mathcal{D}_f^R \Lambda^2 +\tilde{\sigma}^{R}]^2}- \frac{\tilde{\mu}^{u,\mathcal{D}_f,R}}{\mathcal{D}_f^R}\frac{\partial log Z_{\mathcal{D}_f}}{\partial b} \nonumber \\
&  \beta_{\tilde{\lambda}^{u,\mathcal{D}_f,R}} = 2 \tilde{\lambda}^{u,\mathcal{D}_f,R} -d_c \mathcal{D}_f^R \Lambda^2 \frac{2\tilde{\lambda}^{f,\mathcal{D}_f,R}(\tilde{\lambda}^{u,\mathcal{D}_f,R} + \tilde{\mu}^{u,\mathcal{D}_f,R} )+\tilde{\mu}^{f,\mathcal{D}_f,R}(2 \tilde{\lambda}^{u,\mathcal{D}_f,R} + \tilde{\mu}^{u,\mathcal{D}_f,R} )}{8 \pi [\mathcal{D}_f^R \Lambda^2+\tilde{\sigma}^{R}]^2} \\ & +d_c \mathcal{D}_f^R \Lambda^2\frac{ \tilde{K}_{odd}^{f,\mathcal{D}_f,R} \tilde{K}_{odd}^{u,\mathcal{D}_f,R} }{8 \pi [\mathcal{D}_f^R \Lambda^2 +\tilde{\sigma}^{R}]^2} - \frac{\tilde{\lambda}^{u,\mathcal{D}_f,R}}{\mathcal{D}_f^R}\frac{\partial log Z_{\mathcal{D}_f}}{\partial b} \nonumber \\
&  \beta_{\tilde{K}_{odd}^{u,\mathcal{D}_f,R}} = 2 \tilde{K}_{odd}^{u,\mathcal{D}_f,R} -d_c\mathcal{D}_f^R \Lambda^2\frac{ \tilde{K}_{odd}^{f,\mathcal{D}_f,R} \tilde{\mu}^{u,\mathcal{D}_f,R} +\tilde{\mu}^{f,\mathcal{D}_f,R} \tilde{K}_{odd}^{u,\mathcal{D}_f,R} }{8 \pi [\mathcal{D}_f^R \Lambda^2 +\tilde{\sigma}^{R}]^2}  - \frac{\tilde{K}_{odd}^{u,\mathcal{D}_f,R}}{\mathcal{D}_f^R}\frac{\partial log Z_{\mathcal{D}_f}}{\partial b} \nonumber \\
&  \beta_{\tilde{A}_{odd}^{u,\mathcal{D}_f,R}} = 2 \tilde{A}_{odd}^{u,\mathcal{D}_f,R} -d_c\mathcal{D}_f^R \Lambda^2\frac{ \tilde{A}_{odd}^{f,\mathcal{D}_f,R} (\tilde{\mu}^{u,\mathcal{D}_f,R} +\tilde{\lambda}^{u,\mathcal{D}_f,R}) }{4 \pi [\mathcal{D}_f^R \Lambda^2 +\tilde{\sigma}^{R}]^2}- \frac{\tilde{A}_{odd}^{u,\mathcal{D}_f,R}}{\mathcal{D}_f^R}\frac{\partial log Z_{\mathcal{D}_f}}{\partial b} \nonumber \\
& \beta_{\tilde{\mu}^{f,\mathcal{D},R}} = 2 \tilde{\mu}^{f,\mathcal{D},R} -d_c\mathcal{D}_f^R \Lambda^2\frac{\tilde{\mu}^{f,\mathcal{D},R} \tilde{\mu}^{f,\mathcal{D}_f,R} - \tilde{K}_{odd}^{f,\mathcal{D},R} \tilde{K}_{odd}^{f,\mathcal{D}_f,R} }{8 \pi [\mathcal{D}_f^R \Lambda^2 +\tilde{\sigma}^{R}]^2} \nonumber \\
&  \beta_{\tilde{\lambda}^{f,\mathcal{D},R}} = 2 \tilde{\lambda}^{f,\mathcal{D},R} -d_c\mathcal{D}_f^R \Lambda^2\frac{2\tilde{\lambda}^{f,\mathcal{D},R}(\tilde{\lambda}^{f,\mathcal{D}_f,R} + \tilde{\mu}^{f,\mathcal{D}_f,R} )+\tilde{\mu}^{f,\mathcal{D},R}(2 \tilde{\lambda}^{f,\mathcal{D}_f,R} + \tilde{\mu}^{f,\mathcal{D}_f,R} )- \tilde{K}_{odd}^{f,\mathcal{D},R} \tilde{K}_{odd}^{f,\mathcal{D}_f,R} }{8 \pi [\mathcal{D}_f^R \Lambda^2 +\tilde{\sigma}^{R}]^2} \nonumber \\
&  \beta_{\tilde{K}_{odd}^{f,\mathcal{D},R}} = 2 \tilde{K}_{odd}^{f,\mathcal{D},R} -d_c \mathcal{D}_f^R \Lambda^2 \frac{ \tilde{K}_{odd}^{f,\mathcal{D},R} \tilde{\mu}^{f,\mathcal{D}_f,R} +\tilde{\mu}^{f,\mathcal{D},R} \tilde{K}_{odd}^{f,\mathcal{D}_f,R} }{8 \pi [\mathcal{D}_f^R \Lambda^2 +\tilde{\sigma}^{R}]^2} \nonumber \\
&  \beta_{\tilde{A}_{odd}^{f,\mathcal{D},R}} = 2 \tilde{A}_{odd}^{f,\mathcal{D},R} -d_c \mathcal{D}_f^R \Lambda^2 \frac{ \tilde{A}_{odd}^{f,\mathcal{D},R} (\tilde{\mu}^{f,\mathcal{D}_f,R} +\tilde{\lambda}^{f,\mathcal{D}_f,R}) }{4 \pi [\mathcal{D}_f^R \Lambda^2 +\tilde{\sigma}^{R}]^2} \nonumber \\
& \beta_{\tilde{\mu}^{f,\mathcal{D}_f,R}} = 2 \tilde{\mu}^{f,\mathcal{D}_f,R} -d_c\mathcal{D}_f^R \Lambda^2\frac{\tilde{\mu}^{f,\mathcal{D}_f,R} \tilde{\mu}^{f,\mathcal{D}_f,R} - \tilde{K}_{odd}^{f,\mathcal{D}_f,R} \tilde{K}_{odd}^{f,\mathcal{D}_f,R} }{8 \pi [\mathcal{D}_f^R \Lambda^2 +\tilde{\sigma}^{R}]^2}- \frac{\tilde{\mu}^{f,\mathcal{D}_f,R}}{\mathcal{D}_f^R}\frac{\partial log Z_{\mathcal{D}_f}}{\partial b} \nonumber \\
&  \beta_{\tilde{\lambda}^{f,\mathcal{D}_f,R}} = 2 \tilde{\lambda}^{f,\mathcal{D}_f,R} -d_c \mathcal{D}_f^R \Lambda^2 \frac{2\tilde{\lambda}^{f,\mathcal{D}_f,R}(\tilde{\lambda}^{f,\mathcal{D}_f,R} + \tilde{\mu}^{f,\mathcal{D}_f,R} )+\tilde{\mu}^{f,\mathcal{D}_f,R}(2 \tilde{\lambda}^{f,\mathcal{D}_f,R} + \tilde{\mu}^{f,\mathcal{D}_f,R} )}{8 \pi [\mathcal{D}_f^R \Lambda^2 +\tilde{\sigma}^{R}]^2} \nonumber \\ &+d_c \mathcal{D}_f^R \Lambda^2 \frac{\tilde{K}_{odd}^{f,\mathcal{D}_f,R} \tilde{K}_{odd}^{f,\mathcal{D}_f,R} }{8 \pi [\mathcal{D}_f^R \Lambda^2 +\tilde{\sigma}^{R}]^2} - \frac{\tilde{\lambda}^{f,\mathcal{D}_f,R}}{\mathcal{D}_f^R}\frac{\partial log Z_{\mathcal{D}_f}}{\partial b}  \nonumber \\
&  \beta_{\tilde{K}_{odd}^{f,\mathcal{D}_f,R}} = 2 \tilde{K}_{odd}^{f,\mathcal{D}_f,R} -d_c \mathcal{D}_f^R \Lambda^2 \frac{ \tilde{K}_{odd}^{f,\mathcal{D}_f,R} \tilde{\mu}^{f,\mathcal{D}_f,R} +\tilde{\mu}^{f,\mathcal{D}_f,R} \tilde{K}_{odd}^{f,\mathcal{D}_f,R} }{8 \pi [\mathcal{D}_f^R \Lambda^2 +\tilde{\sigma}^{R}]^2} - \frac{\tilde{K}_{odd}^{f,\mathcal{D}_f,R}}{\mathcal{D}_f^R}\frac{\partial log Z_{\mathcal{D}_f}}{\partial b} \nonumber \\
&  \beta_{\tilde{A}_{odd}^{f,\mathcal{D}_f,R}} = 2 \tilde{A}_{odd}^{f,\mathcal{D}_f,R} -d_c \mathcal{D}_f^R \Lambda^2 \frac{ \tilde{A}_{odd}^{f,\mathcal{D}_f,R} (\tilde{\mu}^{f,\mathcal{D}_f,R} +\tilde{\lambda}^{f,\mathcal{D}_f,R}) }{4 \pi [\mathcal{D}_f^R \Lambda^2 +\tilde{\sigma}^{R}]^2} - \frac{\tilde{A}_{odd}^{f,\mathcal{D}_f,R}}{\mathcal{D}_f^R}\frac{\partial log Z_{\mathcal{D}_f}}{\partial b} \nonumber \\
&  \beta_{\tilde{\sigma}^{R}} = 2\tilde{\sigma}^{R}+\frac{\partial log Z_{\sigma}}{\partial b} \nonumber \\
&  \beta_{\tilde{\mathcal{L}}^R} = 2 \frac{\tilde{\mathcal{L}}^R}{\mathcal{D}_f^R}\frac{\partial log Z_{\mathcal{D}_f}}{\partial b} \nonumber \\
&  \beta_{\mathcal{D}_f^R} =  \frac{\partial log Z_{\mathcal{D}_f}}{\partial b}, \nonumber \\
\end{align}
\end{subequations}
where we have used that $Z_{\mathcal{L}} =1$ since no Feynman diagrams contribute to zero-th order in the wave vectors to $\Upsilon_j^2$. Furthermore $\Lambda$ is the UV cutoff, in other words the Fourier wave-vector corresponding to the microscopic length scale of the theory (for example, the lattice spacing of the system). The parameter $b = e^s$ where $s$ is the re-scaling parameter \cite{tongstatistical,peskin2018introduction,kardar2007statistical}. Note as well that the stress does in fact renormalize, thus from a zero-stress state, a non-zero stress can be generated in the presence of odd elastic moduli or a mis-match of diffusivities. More on this will be mentioned in Sec.~\ref{sec:NumRG}.

\subsubsection{Stability Analysis}
With Eq.~\eqref{eq:OddElasticRGEquations}, we may perform a stability analysis of the Aronovitz-Lubensky fixed point \cite{aronovitz1988fluctuations}. We are interested in having only physically relevant parameters. Thus we further reduce the equations so that the only parameters considered are: \begin{equation}
\begin{split}
& \{\tilde{\mathcal{C}}_{ijkl}^{u,\mathcal{D},R}/\tilde{\mathcal{L}}^R ,\tilde{\mathcal{C}}_{ijkl}^{f,\mathcal{D},R}/\tilde{\mathcal{L}}^R,\tilde{\mathcal{C}}_{ijkl}^{u,\mathcal{D}_f,R}/\mathcal{D}_f^R,\tilde{\mathcal{C}}_{ijkl}^{f,\mathcal{D}_f,R}/\mathcal{D}_f^R, \tilde{\sigma}^R /\mathcal{D}_f^R\}  = \{ \hat{\tilde{\mathcal{C}}}_{ijkl}^{u,\mathcal{D},R} ,\hat{\tilde{\mathcal{C}}}_{ijkl}^{f,\mathcal{D},R},\hat{\tilde{\mathcal{C}}}_{ijkl}^{u,\mathcal{D}_f,R},\hat{\tilde{\mathcal{C}}}_{ijkl}^{f,\mathcal{D}_f,R},\hat{\tilde{\sigma}}^{R}\}
\end{split}
\end{equation}
The equations derived are then:
\begin{align}
& \beta_{\hat{\tilde{\mu}}^{u,\mathcal{D},R}} = 2 \hat{\tilde{\mu}}^{u,\mathcal{D},R} -d_c(\mathcal{D}_f^R)^2 \Lambda^2\frac{\hat{\tilde{\mu}}^{f,\mathcal{D},R} \hat{\tilde{\mu}}^{u,\mathcal{D}_f,R} - \hat{\tilde{K}}_{odd}^{f,\mathcal{D},R} \hat{\tilde{K}}_{odd}^{u,\mathcal{D}_f,R} }{8 \pi  [\mathcal{D}_f^R \Lambda^2 +\hat{\tilde{\sigma}}^{R}]^2} - 2\frac{\hat{\tilde{\lambda}}^{u,\mathcal{D},R}}{\mathcal{D}_f^R}\frac{\partial log Z_{\mathcal{D}_f}}{\partial b}  \nonumber \\
&  \beta_{\hat{\tilde{\lambda}}^{u,\mathcal{D},R}} = 2 \hat{\tilde{\lambda}}^{u,\mathcal{D},R}-d_c(\mathcal{D}_f^R)^2 \Lambda^2\frac{2\hat{\tilde{\lambda}}^{f,\mathcal{D},R}(\hat{\tilde{\lambda}}^{u,\mathcal{D}_f,R} + \hat{\tilde{\mu}}^{u,\mathcal{D}_f,R} )+\hat{\tilde{\mu}}^{f,\mathcal{D},R}(2 \hat{\tilde{\lambda}}^{u,\mathcal{D}_f,R} + \hat{\tilde{\mu}}^{u,\mathcal{D}_f,R} )}{8 \pi [\mathcal{D}_f^R \Lambda^2 +\hat{\tilde{\sigma}}^{R}]^2} \nonumber  \\
&+ d_c(\mathcal{D}_f^R)^2 \Lambda^2  \frac{ \hat{\tilde{K}}_{odd}^{f,\mathcal{D},R} \hat{\tilde{K}}_{odd}^{u,\mathcal{D}_f,R} }{8 \pi [\mathcal{D}_f^R \Lambda^2 +\hat{\tilde{\sigma}}^{R}]^2}- 2\frac{\hat{\tilde{\lambda}}^{u,\mathcal{D},R}}{\mathcal{D}_f^R}\frac{\partial log Z_{\mathcal{D}_f}}{\partial b} \nonumber  \\
&  \beta_{\hat{\tilde{K}}_{odd}^{u,\mathcal{D},R}} = 2 \hat{\tilde{K}}_{odd}^{u,\mathcal{D},R} -d_c(\mathcal{D}_f^R)^2 \Lambda^2\frac{ \hat{\tilde{K}}_{odd}^{f,\mathcal{D},R} \hat{\tilde{\mu}}^{u,\mathcal{D}_f,R} +\hat{\tilde{\mu}}^{f,\mathcal{D},R} \hat{\tilde{K}}_{odd}^{u,\mathcal{D}_f,R} }{8 \pi [\mathcal{D}_f^R \Lambda^2 +\hat{\tilde{\sigma}}^{R}]^2}- 2\frac{\hat{\tilde{K}}_{odd}^{u,\mathcal{D},R}}{\mathcal{D}_f^R}\frac{\partial log Z_{\mathcal{D}_f}}{\partial b} \nonumber  \\
&  \beta_{\hat{\tilde{A}}_{odd}^{u,\mathcal{D},R}} = 2 \hat{\tilde{A}}_{odd}^{u,\mathcal{D},R} -d_c(\mathcal{D}_f^R)^2 \Lambda^2\frac{ \hat{\tilde{A}}_{odd}^{f,\mathcal{D},R} (\hat{\tilde{\mu}}^{u,\mathcal{D}_f,R} +\hat{\tilde{\lambda}}^{u,\mathcal{D}_f,R}) }{4 \pi [\mathcal{D}_f^R \Lambda^2 +\hat{\tilde{\sigma}}^{R}]^2} - 2\frac{\hat{\tilde{A}}_{odd}^{u,\mathcal{D},R}}{\mathcal{D}_f^R}\frac{\partial log Z_{\mathcal{D}_f}}{\partial b} \nonumber  \\
& \beta_{\hat{\tilde{\mu}}^{u,\mathcal{D}_f,R}} = 2 \hat{\tilde{\mu}}^{u,\mathcal{D}_f,R} -d_c(\mathcal{D}_f^R)^2 \Lambda^2\frac{\hat{\tilde{\mu}}^{f,\mathcal{D}_f,R} \hat{\tilde{\mu}}^{u,\mathcal{D}_f,R} - \hat{\tilde{K}}_{odd}^{f,\mathcal{D}_f,R} \hat{\tilde{K}}_{odd}^{u,\mathcal{D}_f,R} }{8 \pi [\mathcal{D}_f^R \Lambda^2 +\hat{\tilde{\sigma}}^{R}]^2} - 2\frac{\hat{\tilde{\mu}}^{u,\mathcal{D}_f,R}}{\mathcal{D}_f^R}\frac{\partial log Z_{\mathcal{D}_f}}{\partial b} \nonumber  \\
&  \beta_{\hat{\tilde{\lambda}}^{u,\mathcal{D}_f,R}} = 2 \hat{\tilde{\lambda}}^{u,\mathcal{D}_f,R} -d_c(\mathcal{D}_f^R)^2 \Lambda^2\frac{2\hat{\tilde{\lambda}}^{f,\mathcal{D}_f,R}(\hat{\tilde{\lambda}}^{u,\mathcal{D}_f,R} + \hat{\tilde{\mu}}^{u,\mathcal{D}_f,R} )+\hat{\tilde{\mu}}^{f,\mathcal{D}_f,R}(2 \hat{\tilde{\lambda}}^{u,\mathcal{D}_f,R} + \hat{\tilde{\mu}}^{u,\mathcal{D}_f,R} ) }{8 \pi [\mathcal{D}_f^R \Lambda^2 +\hat{\tilde{\sigma}}^{R}]^2} \nonumber  \\
&+\frac{\hat{\tilde{K}}_{odd}^{f,\mathcal{D}_f,R} \hat{\tilde{K}}_{odd}^{u,\mathcal{D}_f,R} }{8 \pi [\mathcal{D}_f^R \Lambda^2 +\hat{\tilde{\sigma}}^{R}]^2}- 2\frac{\hat{\tilde{\lambda}}^{u,\mathcal{D}_f,R}}{\mathcal{D}_f^R}\frac{\partial log Z_{\mathcal{D}_f}}{\partial b} \nonumber  \\
&  \beta_{\hat{\tilde{K}}_{odd}^{u,\mathcal{D}_f,R}} = 2 \hat{\tilde{K}}_{odd}^{u,\mathcal{D}_f,R} -d_c(\mathcal{D}_f^R)^2 \Lambda^2\frac{ \hat{\tilde{K}}_{odd}^{f,\mathcal{D}_f,R} \hat{\tilde{\mu}}^{u,\mathcal{D}_f,R} +\hat{\tilde{\mu}}^{f,\mathcal{D}_f,R} \hat{\tilde{K}}_{odd}^{u,\mathcal{D}_f,R} }{8 \pi [\mathcal{D}_f^R \Lambda^2 +\hat{\tilde{\sigma}}^{R}]^2} - 2\frac{\hat{\tilde{K}}_{odd}^{u,\mathcal{D}_f,R}}{\mathcal{D}_f^R}\frac{\partial log Z_{\mathcal{D}_f}}{\partial b} \nonumber  \\
&  \beta_{\hat{\tilde{A}}_{odd}^{u,\mathcal{D}_f,R}} = 2 \hat{\tilde{A}}_{odd}^{u,\mathcal{D}_f,R} -d_c(\mathcal{D}_f^R)^2 \Lambda^2\frac{ \hat{\tilde{A}}_{odd}^{f,\mathcal{D}_f,R} (\hat{\tilde{\mu}}^{u,\mathcal{D}_f,R} +\hat{\tilde{\lambda}}^{u,\mathcal{D}_f,R}) }{4 \pi  [\mathcal{D}_f^R \Lambda^2 +\hat{\tilde{\sigma}}^{R}]^2}- 2\frac{\hat{\tilde{A}}_{odd}^{u,\mathcal{D}_f,R}}{\mathcal{D}_f^R}\frac{\partial log Z_{\mathcal{D}_f}}{\partial b} \nonumber \\
& \beta_{\hat{\tilde{\mu}}^{f,\mathcal{D},R}} = 2 \hat{\tilde{\mu}}^{f,\mathcal{D},R} -d_c(\mathcal{D}_f^R)^2 \Lambda^2\frac{\hat{\tilde{\mu}}^{f,\mathcal{D},R} \hat{\tilde{\mu}}^{f,\mathcal{D}_f,R} - \hat{\tilde{K}}_{odd}^{f,\mathcal{D},R} \hat{\tilde{K}}_{odd}^{f,\mathcal{D}_f,R} }{8 \pi  [\mathcal{D}_f^R \Lambda^2 +\hat{\tilde{\sigma}}^{R}]^2} - 2\frac{\hat{\tilde{\mu}}^{f,\mathcal{D},R}}{\mathcal{D}_f^R}\frac{\partial log Z_{\mathcal{D}_f}}{\partial b} \nonumber  \\
&  \beta_{\hat{\tilde{\lambda}}^{f,\mathcal{D},R}} = 2 \hat{\tilde{\lambda}}^{f,\mathcal{D},R} -d_c(\mathcal{D}_f^R)^2 \Lambda^2\frac{2\hat{\tilde{\lambda}}^{f,\mathcal{D},R}(\hat{\tilde{\lambda}}^{f,\mathcal{D}_f,R} + \hat{\tilde{\mu}}^{f,\mathcal{D}_f,R} )+\hat{\tilde{\mu}}^{f,\mathcal{D},R}(2 \hat{\tilde{\lambda}}^{f,\mathcal{D}_f,R} + \hat{\tilde{\mu}}^{f,\mathcal{D}_f,R} ) }{8 \pi [\mathcal{D}_f^R \Lambda^2 +\hat{\tilde{\sigma}}^{R}]^2} \nonumber  \\ &+ d_c(\mathcal{D}_f^R)^2 \Lambda^2\frac{\hat{\tilde{K}}_{odd}^{f,\mathcal{D},R} \hat{\tilde{K}}_{odd}^{f,\mathcal{D}_f,R} }{8 \pi [\mathcal{D}_f^R \Lambda^2 +\hat{\tilde{\sigma}}^{R}]^2} - 2\frac{\hat{\tilde{\lambda}}^{f,\mathcal{D},R}}{\mathcal{D}_f^R}\frac{\partial log Z_{\mathcal{D}_f}}{\partial b} \nonumber  \\
&  \beta_{\hat{\tilde{K}}_{odd}^{f,\mathcal{D},R}} = 2 \hat{\tilde{K}}_{odd}^{f,\mathcal{D},R} -d_c(\mathcal{D}_f^R)^2 \Lambda^2\frac{ \hat{\tilde{K}}_{odd}^{f,\mathcal{D},R} \hat{\tilde{\mu}}^{f,\mathcal{D}_f,R} +\hat{\tilde{\mu}}^{f,\mathcal{D},R} \hat{\tilde{K}}_{odd}^{f,\mathcal{D}_f,R} }{8 \pi [\mathcal{D}_f^R \Lambda^2 +\hat{\tilde{\sigma}}^{R}]^2} - 2\frac{\hat{\tilde{K}}_{odd}^{f,\mathcal{D},R}}{\mathcal{D}_f^R}\frac{\partial log Z_{\mathcal{D}_f}}{\partial b} \nonumber  \\
&  \beta_{\hat{\tilde{A}}_{odd}^{f,\mathcal{D},R}} = 2 \hat{\tilde{A}}_{odd}^{f,\mathcal{D},R} -d_c(\mathcal{D}_f^R)^2 \Lambda^2\frac{ \hat{\tilde{A}}_{odd}^{f,\mathcal{D},R} (\hat{\tilde{\mu}}^{f,\mathcal{D}_f,R} +\hat{\tilde{\lambda}}^{f,\mathcal{D}_f,R}) }{4 \pi [\mathcal{D}_f^R \Lambda^2 +\hat{\tilde{\sigma}}^{R}]^2}- 2\frac{\hat{\tilde{A}}_{odd}^{f,\mathcal{D},R}}{\mathcal{D}_f^R}\frac{\partial log Z_{\mathcal{D}_f}}{\partial b} \nonumber \\
& \beta_{\hat{\tilde{\mu}}^{f,\mathcal{D}_f,R}} = 2 \hat{\tilde{\mu}}^{f,\mathcal{D}_f,R} -d_c(\mathcal{D}_f^R)^2 \Lambda^2\frac{\hat{\tilde{\mu}}^{f,\mathcal{D}_f,R} \hat{\tilde{\mu}}^{f,\mathcal{D}_f,R} - \hat{\tilde{K}}_{odd}^{f,\mathcal{D}_f,R} \hat{\tilde{K}}_{odd}^{f,\mathcal{D}_f,R} }{8 \pi [\mathcal{D}_f^R \Lambda^2 +\hat{\tilde{\sigma}}^{R}]^2}   - 2\frac{\hat{\tilde{\mu}}^{f,\mathcal{D}_f,R}}{\mathcal{D}_f^R}\frac{\partial log Z_{\mathcal{D}_f}}{\partial b}  \nonumber  \\
&  \beta_{\hat{\tilde{\lambda}}^{f,\mathcal{D}_f,R}} = 2 \hat{\tilde{\lambda}}^{f,\mathcal{D}_f^R}  -d_c(\mathcal{D}_f^R)^2 \Lambda^2\frac{2\hat{\tilde{\lambda}}^{f,\mathcal{D}_f,R}(\hat{\tilde{\lambda}}^{f,\mathcal{D}_f,R} + \hat{\tilde{\mu}}^{f,\mathcal{D}_f,R} )+\hat{\tilde{\mu}}^{f,\mathcal{D}_f,R}(2 \hat{\tilde{\lambda}}^{f,\mathcal{D}_f,R} + \hat{\tilde{\mu}}^{f,\mathcal{D}_f,R} ) }{8 \pi [\mathcal{D}_f^R \Lambda^2 +\hat{\tilde{\sigma}}^{R}]^2} \nonumber \\ & + d_c(\mathcal{D}_f^R)^2 \Lambda^2 \frac{\hat{\tilde{K}}_{odd}^{f,\mathcal{D}_f,R} \hat{\tilde{K}}_{odd}^{f,\mathcal{D}_f,R} }{8 \pi [\mathcal{D}_f^R \Lambda^2 +\hat{\tilde{\sigma}}^{R}]^2} - 2\frac{\hat{\tilde{\lambda}}^{f,\mathcal{D}_f,R}}{\mathcal{D}_f^R}\frac{\partial log Z_{\mathcal{D}_f}}{\partial b} \nonumber \\
&  \beta_{\hat{\tilde{K}}_{odd}^{f,\mathcal{D}_f,R}} = 2 \hat{\tilde{K}}_{odd}^{f,\mathcal{D}_f,R} -d_c(\mathcal{D}_f^R)^2 \Lambda^2\frac{ \hat{\tilde{K}}_{odd}^{f,\mathcal{D}_f,R} \hat{\tilde{\mu}}^{f,\mathcal{D}_f,R} +\hat{\tilde{\mu}}^{f,\mathcal{D}_f,R} \hat{\tilde{K}}_{odd}^{f,\mathcal{D}_f,R} }{8 \pi  [\mathcal{D}_f^R \Lambda^2 +\hat{\tilde{\sigma}}^{R}]^2} - 2\frac{\hat{\tilde{K}}_{odd}^{f,\mathcal{D}_f,R}}{\mathcal{D}_f^R}\frac{\partial log Z_{\mathcal{D}_f}}{\partial b} \nonumber \\
&  \beta_{\hat{\tilde{A}}_{odd}^{f,\mathcal{D}_f,R}} = 2 \hat{\tilde{A}}_{odd}^{f,\mathcal{D}_f,R} -d_c(\mathcal{D}_f^R)^2 \Lambda^2\frac{ \hat{\tilde{A}}_{odd}^{f,\mathcal{D}_f,R} (\hat{\tilde{\mu}}^{f,\mathcal{D}_f,R} +\hat{\tilde{\lambda}}^{f,\mathcal{D}_f,R}) }{4 \pi  [\mathcal{D}_f^R \Lambda^2 +\hat{\tilde{\sigma}}^{R}]^2} - 2\frac{\hat{\tilde{A}}_{odd}^{f,\mathcal{D}_f,R}}{\mathcal{D}_f^R}\frac{\partial log Z_{\mathcal{D}_f}}{\partial b} \nonumber  \\
&  \beta_{\hat{\tilde{\sigma}}^{R}} =  2\hat{\tilde{\sigma}}^{R}+\frac{\partial log Z_{\sigma}}{\partial b} \nonumber  \\
\end{align}
Formulated in terms of these variables, one in principle should solve for the fixed point/manifold. However, as we know $\frac{\partial log Z_{\mathcal{D}_f}}{\partial b}$ is a complicated expression and thus renders obtainment of these solutions analytically intractable. In addition we note that, $\frac{\partial log Z_{\sigma}}{\partial b}$ is also quite complicated as an expression, however it indicates that stress is generated when fluctuation dissipation is broken or when odd elastic parameters are present. Despite the complexity of these expressions, if we define the Aronovitz-Lubensky fixed point as:
\begin{equation}
    \{\hat{\tilde{\mu}}^{\gamma,\delta,R},\hat{\tilde{\lambda}}^{\gamma,\delta,R},\hat{\tilde{K}}_{odd}^{\gamma,\delta,R},\hat{\tilde{A}}_{odd}^{\gamma,\delta,R},\hat{\tilde{\sigma}}^{R}\} = \{ \frac{16 \pi \Lambda^2}{4+d_c},\frac{-8 \pi \Lambda^2}{4+d_c},0,0,0\}
\end{equation}
where $\gamma \in \{ u,f\},\delta \in \{ \mathcal{D},\mathcal{D}_f\}$, one can check that it is indeed a fixed point of the equations and we can analyze its stability. Performing a stability analysis on these 16 variables gives the following eigen-values:
\begin{equation}
\begin{split}
    & \text{Eigen-values} =  \{ 0,0,0,0,0,0,0,0,\frac{-2d_c}{4+d_c},\frac{-2d_c}{4+d_c},-2,\frac{2d_c}{4+d_c},\frac{2d_c}{4+d_c},\frac{2d_c}{4+d_c},\frac{2d_c}{4+d_c},\frac{2d_c}{4+d_c},\frac{2(2+d_c)}{4+d_c}\}
\end{split}
\end{equation}
The eight zero eigen-values indicate that the Aronovitz-Lubensky fixed point potentially belongs to a higher-dimensional fixed manifold, although perhaps not a stable one. Four of these eight zero eigen-values are perturbations purely in $\{\hat{\tilde{\mu}}^{\gamma,\delta,R},\hat{\tilde{\lambda}}^{\gamma,\delta,R} \}$ and three of them are somewhat analytically complicated, however they indicate that there are potentially a broader set of fixed points where fluctuation-dissipation may not necessarily hold \cite{le2021thermal}. The one simple eigen-vector that is a pure perturbation in $\{\hat{\tilde{\mu}}^{\gamma,\delta,R},\hat{\tilde{\lambda}}^{\gamma,\delta,R} \}$ is:
\begin{equation}
\begin{split}
&[\delta \hat{\tilde{A}}_{odd}^{u,\mathcal{D},R},\delta \hat{\tilde{A}}_{odd}^{f,\mathcal{D},R},\delta \hat{\tilde{A}}_{odd}^{u,\mathcal{D}_f,R},\delta \hat{\tilde{A}}_{odd}^{f,\mathcal{D}_f,R},\delta \hat{\tilde{K}}_{odd}^{u,\mathcal{D},R},\delta \hat{\tilde{K}}_{odd}^{f,\mathcal{D},R},\delta \hat{\tilde{K}}_{odd}^{u,\mathcal{D}_f,R},\delta \hat{\tilde{K}}_{odd}^{f,\mathcal{D}_f,R},\\ & \delta  \hat{\tilde{\lambda}}^{u,\mathcal{D},R},\delta \hat{\tilde{\lambda}}^{f,\mathcal{D},R},\delta \hat{\tilde{\lambda}}^{u,\mathcal{D}_f,R},\delta \hat{\tilde{\lambda}}^{f,\mathcal{D}_f,R},\delta \hat{\tilde{\mu}}^{u,\mathcal{D},R},\delta \hat{\tilde{\mu}}^{f,\mathcal{D},R},\delta \hat{\tilde{\mu}}^{u,\mathcal{D}_f,R},\delta \hat{\tilde{\mu}}^{f,\mathcal{D}_f,R}, \delta \hat{\tilde{\sigma}}^{R}]\\ &
= 
[0,0,
0,
0,
0,
0,
0,
0,
-\frac{-4-3d_c}{16},
-\frac{-3(4+d_c)}{32},
-\frac{-3(4+d_c)}{32},
-1/2,
-1,
0,
0,
1,0
]
\end{split}
\end{equation}
The other 4 zero eigen-values are perturbations in $\{ \hat{\tilde{K}}_{odd}^{\gamma,\delta,R},\hat{\tilde{A}}_{odd}^{\gamma,\delta,R}\}$ and include:
\begin{equation}
\begin{split}
&[\delta \hat{\tilde{A}}_{odd}^{u,\mathcal{D},R},\delta \hat{\tilde{A}}_{odd}^{f,\mathcal{D},R},\delta \hat{\tilde{A}}_{odd}^{u,\mathcal{D}_f,R},\delta \hat{\tilde{A}}_{odd}^{f,\mathcal{D}_f,R},\delta \hat{\tilde{K}}_{odd}^{u,\mathcal{D},R},\delta \hat{\tilde{K}}_{odd}^{f,\mathcal{D},R},\delta \hat{\tilde{K}}_{odd}^{u,\mathcal{D}_f,R},\delta \hat{\tilde{K}}_{odd}^{f,\mathcal{D}_f,R},\\ & \delta  \hat{\tilde{\lambda}}^{u,\mathcal{D},R},\delta \hat{\tilde{\lambda}}^{f,\mathcal{D},R},\delta \hat{\tilde{\lambda}}^{u,\mathcal{D}_f,R},\delta \hat{\tilde{\lambda}}^{f,\mathcal{D}_f,R},\delta \hat{\tilde{\mu}}^{u,\mathcal{D},R},\delta \hat{\tilde{\mu}}^{f,\mathcal{D},R},\delta \hat{\tilde{\mu}}^{u,\mathcal{D}_f,R},\delta \hat{\tilde{\mu}}^{f,\mathcal{D}_f,R}, \delta \hat{\tilde{\sigma}}^{R}] \\ &
= 
[1,
1,
0,
0,
0,
0,
0,
0,
0,
0,
0,
0,
0,
0,
0,
0,0], \\&
[
0,
0,
1,
1,
0,
0,
0,
0,
0,
0,
0,
0,
0,
0,
0,
0,0], \\&
[
0,
0,
0,
0,
1,
0,
1,
0,
0,
0,
0,
0,
0,
0,
0,
0,0], \\ &
[
0,
0,
0,
0,
1,
1,
0,
0,
0,
0,
0,
0,
0,
0,
0,
0,0]
\end{split}
\end{equation}
These four marginal perturbations indicate that perturbations in $\hat{A}_{odd}^{\gamma,\delta,R}$ that preserve fluctuation dissipation are marginal. Furthermore the two eigen-vectors corresponding to the eigen-value $\frac{-2d_c}{4+d_c}$ are:
\begin{equation}
\begin{split}
&[\delta \hat{\tilde{A}}_{odd}^{u,\mathcal{D},R},\delta \hat{\tilde{A}}_{odd}^{f,\mathcal{D},R},\delta \hat{\tilde{A}}_{odd}^{u,\mathcal{D}_f,R},\delta \hat{\tilde{A}}_{odd}^{f,\mathcal{D}_f,R},\delta \hat{\tilde{K}}_{odd}^{u,\mathcal{D},R},\delta \hat{\tilde{K}}_{odd}^{f,\mathcal{D},R},\delta \hat{\tilde{K}}_{odd}^{u,\mathcal{D}_f,R},\delta \hat{\tilde{K}}_{odd}^{f,\mathcal{D}_f,R},\\ & \delta  \hat{\tilde{\lambda}}^{u,\mathcal{D},R},\delta \hat{\tilde{\lambda}}^{f,\mathcal{D},R},\delta \hat{\tilde{\lambda}}^{u,\mathcal{D}_f,R},\delta \hat{\tilde{\lambda}}^{f,\mathcal{D}_f,R},\delta \hat{\tilde{\mu}}^{u,\mathcal{D},R},\delta \hat{\tilde{\mu}}^{f,\mathcal{D},R},\delta \hat{\tilde{\mu}}^{u,\mathcal{D}_f,R},\delta \hat{\tilde{\mu}}^{f,\mathcal{D}_f,R}, \delta \hat{\tilde{\sigma}}^{R}]\\&
= 
[0,
0,
0,
0,
0,
0,
0,
0,
-5/4,
-5/4,
-5/4,
-5/4,
1,
1,
1,
1
,0] , \\ &
[
0,
0,
0,
0,
1,
1,
1,
1,
0,
0,
0,
0,
0,
0,
0,
0,0]
\end{split}
\end{equation}
and the eigen-value $-2$ corresponds to the eigen-vector:
\begin{equation}
\begin{split}
&[\delta \hat{\tilde{A}}_{odd}^{u,\mathcal{D},R},\delta \hat{\tilde{A}}_{odd}^{f,\mathcal{D},R},\delta \hat{\tilde{A}}_{odd}^{u,\mathcal{D}_f,R},\delta \hat{\tilde{A}}_{odd}^{f,\mathcal{D}_f,R},\delta \hat{\tilde{K}}_{odd}^{u,\mathcal{D},R},\delta \hat{\tilde{K}}_{odd}^{f,\mathcal{D},R},\delta \hat{\tilde{K}}_{odd}^{u,\mathcal{D}_f,R},\delta \hat{\tilde{K}}_{odd}^{f,\mathcal{D}_f,R},\\ & \delta  \hat{\tilde{\lambda}}^{u,\mathcal{D},R},\delta \hat{\tilde{\lambda}}^{f,\mathcal{D},R},\delta \hat{\tilde{\lambda}}^{u,\mathcal{D}_f,R},\delta \hat{\tilde{\lambda}}^{f,\mathcal{D}_f,R},\delta \hat{\tilde{\mu}}^{u,\mathcal{D},R},\delta \hat{\tilde{\mu}}^{f,\mathcal{D},R},\delta \hat{\tilde{\mu}}^{u,\mathcal{D}_f,R},\delta \hat{\tilde{\mu}}^{f,\mathcal{D}_f,R}, \delta \hat{\tilde{\sigma}}^{R}]\\ &
= 
[0,0,
0,
0,
0,
0,
0,
0,
-1/2,
-1/2,
-1/2,
-1/2,
1,
1,
1,
1,0
]
\end{split}
\end{equation}
One of these negative eigen-values and their respective eigen-vectors tells us that perturbations in $\hat{K}_{odd}^{\gamma,\delta,R}$ that preserve fluctuation-dissipation are irrelevant. In addition it gives us the associated eigen-value $\frac{-2d_c}{4+d_c}$ and thus we expect $\hat{K}_{odd}^R/\hat{\mu}^R$ to re-scale to zero with this exponent. Meanwhile, the other two correspond to eigen-vectors found in a fixed point analysis using Boltzmann weights for equilibrium elastic membranes (using Boltzmann weights \cite{aronovitz1988fluctuations}), indeed they indicate that if we restrict ourselves to the domain of phase space where fluctuation-dissipation holds and $\hat{A}_{odd}^{\gamma,\delta,R},\hat{K}_{odd}^{\gamma,\delta,R}$ are odd, then the Aronovitz-Lubensky fixed point is the globally stable fixed point. The eigen-vectors corresponding to the eigen-value $2d_c/(2+d_c)$ correspond to:
\begin{equation}
\begin{split}
&[\delta \hat{\tilde{A}}_{odd}^{u,\mathcal{D},R},\delta \hat{\tilde{A}}_{odd}^{f,\mathcal{D},R},\delta \hat{\tilde{A}}_{odd}^{u,\mathcal{D}_f,R},\delta \hat{\tilde{A}}_{odd}^{f,\mathcal{D}_f,R},\delta \hat{\tilde{K}}_{odd}^{u,\mathcal{D},R},\delta \hat{\tilde{K}}_{odd}^{f,\mathcal{D},R},\delta \hat{\tilde{K}}_{odd}^{u,\mathcal{D}_f,R},\delta \hat{\tilde{K}}_{odd}^{f,\mathcal{D}_f,R},\\ & \delta  \hat{\tilde{\lambda}}^{u,\mathcal{D},R},\delta \hat{\tilde{\lambda}}^{f,\mathcal{D},R},\delta \hat{\tilde{\lambda}}^{u,\mathcal{D}_f,R},\delta \hat{\tilde{\lambda}}^{f,\mathcal{D}_f,R},\delta \hat{\tilde{\mu}}^{u,\mathcal{D},R},\delta \hat{\tilde{\mu}}^{f,\mathcal{D},R},\delta \hat{\tilde{\mu}}^{u,\mathcal{D}_f,R},\delta \hat{\tilde{\mu}}^{f,\mathcal{D}_f,R}, \delta \hat{\tilde{\sigma}}^{R}]
\\ &  = 
[
0,
0,
0,
0,
0,
0,
0,
0,
\frac{-31 - 9 dc}{26 },
-1/2,
-1/2,
-1/2,
 \frac{32 + 3 dc}{26},
1,
1,
1
,0], \\ &
[
0,
0,
0,
0,
0,
0,
0,
0,
\frac{8 \pi(34+9d_c)}{13(4+d_c)},
0,
0,
0,
-\frac{8 \pi(20+3d_c)}{13(4+d_c)},
0,
0,
0,1], \\ &
[
0,
0,
0,
0,
1,
0,
0,
0,
0,
0,
0,
0,
0,
0,
0,
0,0], \\ &
[
0,
0,
1,
0,
0,
0,
0,
0,
0,
0,
0,
0,
0,
0,
0,
0
,0], \\
& 
[
1,
0,
0,
0,
0,
0,
0,
0,
0,
0,
0,
0,
0,
0,
0,
0,0]
\end{split}
\end{equation}
The last eigen-vector corresponding to eigen-value $2(2+d_c)/(4+d_c)$ is:
\begin{equation}
\begin{split}
&[\delta \hat{\tilde{A}}_{odd}^{u,\mathcal{D},R},\delta \hat{\tilde{A}}_{odd}^{f,\mathcal{D},R},\delta \hat{\tilde{A}}_{odd}^{u,\mathcal{D}_f,R},\delta \hat{\tilde{A}}_{odd}^{f,\mathcal{D}_f,R},\delta \hat{\tilde{K}}_{odd}^{u,\mathcal{D},R},\delta \hat{\tilde{K}}_{odd}^{f,\mathcal{D},R},\delta \hat{\tilde{K}}_{odd}^{u,\mathcal{D}_f,R},\delta \hat{\tilde{K}}_{odd}^{f,\mathcal{D}_f,R},\\ & \delta  \hat{\tilde{\lambda}}^{u,\mathcal{D},R},\delta \hat{\tilde{\lambda}}^{f,\mathcal{D},R},\delta \hat{\tilde{\lambda}}^{u,\mathcal{D}_f,R},\delta \hat{\tilde{\lambda}}^{f,\mathcal{D}_f,R},\delta \hat{\tilde{\mu}}^{u,\mathcal{D},R},\delta \hat{\tilde{\mu}}^{f,\mathcal{D},R},\delta \hat{\tilde{\mu}}^{u,\mathcal{D}_f,R},\delta \hat{\tilde{\mu}}^{f,\mathcal{D}_f,R}, \delta \hat{\tilde{\sigma}}^{R}]\\ &
= 
[0,0,
0,
0,
0,
0,
0,
0,
-\frac{8(2+d_c) \pi}{(3+d_c)(4+d_c)},
-\frac{8(2+d_c) \pi}{(3+d_c)(4+d_c)},
\\ &
-\frac{8(2+d_c) \pi}{(3+d_c)(4+d_c)},
-\frac{8(2+d_c) \pi}{(3+d_c)(4+d_c)},
\frac{16(2+d_c) \pi}{(3+d_c)(4+d_c)},
\frac{16(2+d_c) \pi}{(3+d_c)(4+d_c)}, \\ &
\frac{16(2+d_c) \pi}{(3+d_c)(4+d_c)},
\frac{16(2+d_c) \pi}{(3+d_c)(4+d_c)},1
]
\end{split}
\end{equation}

All in all, we can then summarize our findings via the following statements:
\paragraph{} If we enforce fluctuation dissipation and insist upon the odd elastic constants being zero, then indeed the Aronovitz-Lubensky fixed point is the globally stable fixed point.
\paragraph{} If we enforce fluctuation dissipation but do not insist upon the odd elastic constants being zero, then indeed the Aronovitz-Lubensky fixed point is stable to perturbations in $\hat{\tilde{K}}_{odd}^{\gamma,\delta,R}$ but is marginally stable with respect to perturbations in $\hat{\tilde{A}}_{odd}^{\gamma,\delta,R}$. Stress however is always an unstable direction.
\paragraph{} If we do not enforce fluctuation dissipation and insist upon the odd elastic constants being zero, then the Aronovitz-Lubensky fixed point potentially belongs to a larger manifold and the fixed point is no longer globally stable. Relevant unstable directions could potentially take us to a new fixed point to be obtained. Indeed this has been better explored in \cite{le2021thermal}. 
\paragraph{} If we do not enforce fluctuation dissipation and do not insist upon the odd elastic constants being zero, then the Aronovitz-Lubensky fixed point is not globally stable. A globally stable fixed point has not been established.
\paragraph{} All the above analysis has been only done to 1-loop order and thus a 2-loop expansion is in principle necessary to establish whether the zero eigen-value directions stay zero or become non-zero (in the absence of a symmetry that enforces the eigen-values to be zero). This would be important for comprehending how these other perturbative eigen-vectors impact the stability of the Aronovitz-Lubensky fixed point.

\subsubsection{Derivation of Thermal Length Scale Assuming Fluctuation-Dissipation Relations \label{sec:ThermalLengthOdd}}
As can be seen from the Fig.~2 and  Fig.~3 of the main text, we have non-dimensionalized the linear system size by a thermal length scale which we have yet to define. Given that the odd elastic parameters are now present, there is no reason that the thermal length scale should take the value given in the equilibrium Hamiltonian theory \cite{kovsmrlj2016response}:
\begin{equation}
\label{eq:LthEquilibriumOddReference}
    \ell_{\text{th}} = \sqrt{\frac{16 \pi^3  (\kappa^o)^2}{3k_B T  Y^o}}
\end{equation}
where we use un-scaled variables and $Y^o$ is the bare Young's modulus. The superscript $o$ marks the fact that the bare parameters are being used. The manner in which one may obtain this length scale in the equilibrium Hamiltonian case is by comparing the an-harmonic $\partial log Z_{\mathcal{D}_f} / \partial log b$ with $\mathcal{D}_f^o$ and obtain for what value of UV-cutoff, $\Lambda$, for which the two are comparable. The procedure is no different in the odd elastic case, however, $\partial log Z_{\mathcal{D}_f} / \partial log b$ is now significantly more complicated. Without assuming fluctuation-dissipation to hold or even making some simplifications explained in the following sentence, it is difficult to obtain a thermal length scale. Thus we resort to obtaining a provisional form done in the Mathematica code whereby one assumes fluctuation-dissipation to hold and obtains the lowest powers of $\Lambda$ in $\partial log Z_{\mathcal{D}_f} / \partial log b$ (in order to reduce the power of the polynomial in $\Lambda$ that one has to solve for) and then comparing them to $\mathcal{D}_f^o$. By doing this, we obtain that in terms of the bare elastic moduli of the theory:
\begin{equation}
\begin{split}
\label{eq:LTHODD}
    & \ell_{\text{th}}=\bigg( \pi^2 \kappa^o  [6 k_BT \mathcal{D}_f^o (K_{odd}^o)^2(\lambda^o+\mu^o)^2+4 \pi \kappa^o \mathcal{D}^o ((K_{odd}^o)^2 + \mu^o(\lambda^o+2\mu^o))^2 \\ &  +(A_{odd}^o)^2((K_{odd}^o)^2(k_BT \mathcal{D}_f^o +4 \pi \mathcal{D}^o \kappa^o ) +k_BT \mathcal{D}_f^o (\mu^o)^2) +8 \pi \mathcal{D}^o \kappa^o A_{odd}^o  ((K_{odd}^o)^2 \\ & + \mu_{odd}^o(\lambda^o +2 \mu^o)) -6 k_BT\mathcal{D}_f^oA_{odd}^oK_{odd}^o]/ (3 k_BT \mathcal{D}^o  [(K_{odd}^o)^2 + (\mu^o)^2] \\ & [A_{odd}^o K_{odd}^o+(K_{odd}^o)^2+\mu^o(\lambda^o+2\mu^o)]  [\lambda^o+\mu^o] ) \bigg)^{1/2}
\end{split}
\end{equation}
where we have re-inserted all variables, including temperature and bending rigidity, that we scaled out in Eq.~(3) and Eq.~(4) of the main text. This equation also reduces to the form Eq.~\eqref{eq:LthEquilibriumOddReference} when $A_{odd}$ and $K_{odd}$ are both zero and the fluctuation-dissipation relations are satisfied. We use this definition of $\ell_{\text{th}}$ to plot Fig.~2 and Fig.~3 of the main text.

\subsubsection{Numerical Renormalization With Fluctuation-Dissipation Relations \label{sec:NumRG}}
\begin{figure}[h!]
\includegraphics[width=\textwidth]{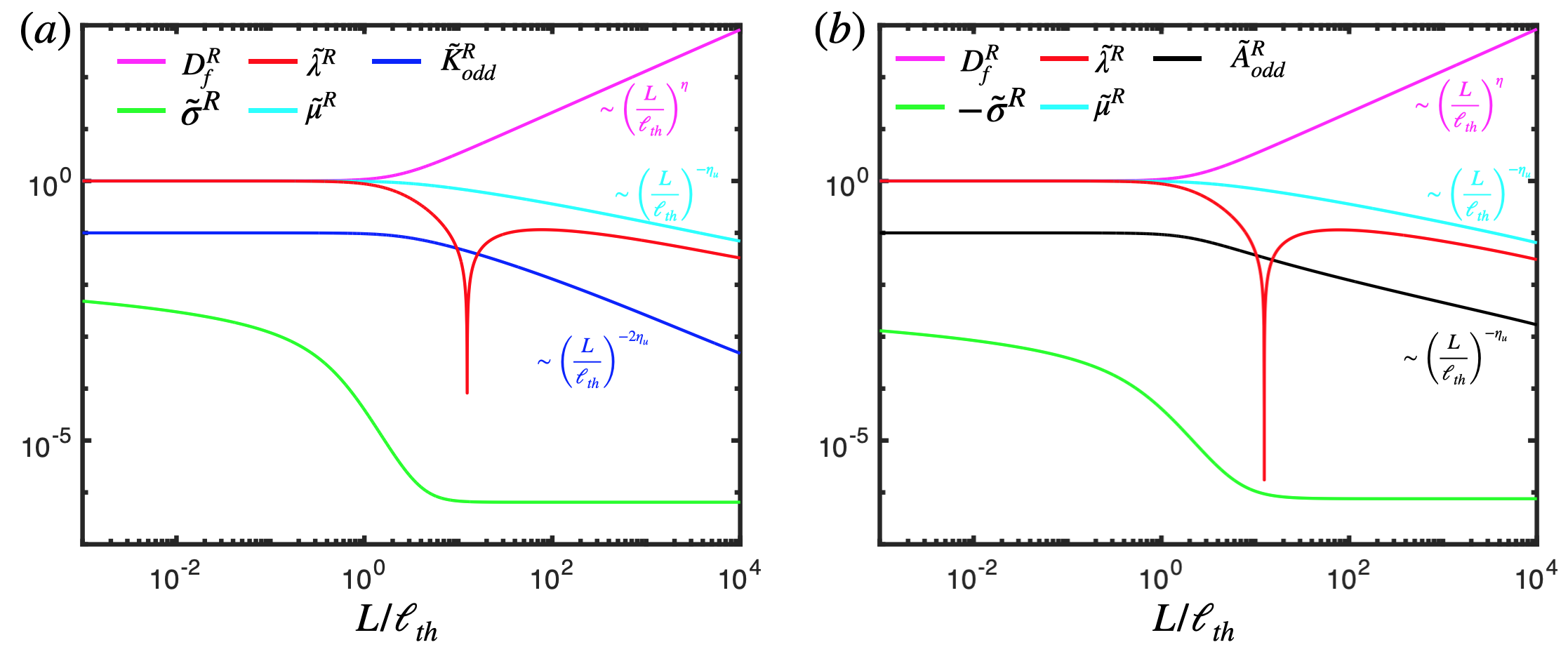}
\caption{In the above plots, we observe the elastic moduli plotted against the system size normalized by our new definition of $\ell_{\text{th}}$ given in Eq.~\eqref{eq:LTHODD}. For reference, $\eta \approx .8$ and $\eta_u \approx .4$. In addition, the microscopic odd elastic moduli ($A_{odd}^o, K_{odd}^o$) are a tenth of $\lambda^o, \mu^o$. In case (a) we plot the elastic moduli in the present of $K_{odd}$. As one can see, we tune the microscopic stress to a positive value ($\approx 5.94 \cdot 10^{-3}$ at $L/\ell_{\text{th}} = 10^{-4}$) in an attempt to tune the stress at $L/\ell_{\text{th}} \approx 10^4$ to be as close to zero as possible. In case (b) we plot the elastic moduli in the present of $A_{odd}$. As one can see, we tune the microscopic stress to a negative value ($\approx -1.58 \cdot 10^{-3}$ at $L/\ell_{\text{th}} = 10^{-4}$) in an attempt to tune the stress at $L/\ell_{\text{th}} \approx 10^4$ to be as close to zero as possible.}
\label{fig:TheoryPlot}
\end{figure}

We now aim to show how we obtain theoretical plots which indicate the scaling of our moduli when fluctuation-dissipation relations hold. For a 1-loop $1/d_c$ analysis, the diagrams we . With the assumption that fluctuation dissipation holds ($\mathcal{L} = \mathcal{D}$ and $\mathcal{L}_f = \mathcal{D}_f$), we may rewrite Eq.~\eqref{eq:OddElasticRGEquations} and drop the hats, $\hat{}$, from above our moduli to instead write down:
 \begin{align}
 \label{eq:FDRG}
& \beta_{\tilde{\mu}^{R}} = 2 \tilde{\mu}^{R} -d_c(\mathcal{D}_f^R)^2 \Lambda^2\frac{(\tilde{\mu}^{R})^2  - (\tilde{K}_{odd}^{R})^2 }{8 \pi [\mathcal{D}_f^R \Lambda^2 +\tilde{\sigma}^{R}]^2}- 2\frac{\tilde{\mu}^{R}}{\mathcal{D}_f^R}\frac{\partial log Z_{\mathcal{D}_f}}{\partial b} \nonumber  \\
&  \beta_{\tilde{\lambda}^{R}} = 2 \tilde{\lambda}^{R} \nonumber   -d_c(\mathcal{D}_f^R)^2 \Lambda^2\frac{2\tilde{\lambda}^{R}(\tilde{\lambda}^{R} + \tilde{\mu}^{R} )+\tilde{\mu}^{R}(2 \tilde{\lambda}^{R} + \tilde{\mu}^{R} )- (\tilde{K}_{odd}^{R})^2 }{8 \pi [\mathcal{D}_f^R \Lambda^2 +\tilde{\sigma}^{R}]^2} - 2\frac{\tilde{\lambda}^{R}}{\mathcal{D}_f^R}\frac{\partial log Z_{\mathcal{D}_f}}{\partial b} \nonumber  \\
&  \beta_{\tilde{K}_{odd}^{R}} = 2 \tilde{K}_{odd}^{R} -d_c(\mathcal{D}_f^R)^2 \Lambda^2\frac{ 2\tilde{\mu}^{R} \tilde{K}_{odd}^{R} }{8 \pi [\mathcal{D}_f^R \Lambda^2 +\tilde{\sigma}^{R}]^2}- 2\frac{\tilde{K}_{odd}^{R}}{\mathcal{D}_f^R}\frac{\partial log Z_{\mathcal{D}_f}}{\partial b} \nonumber  \\
&  \beta_{\tilde{A}_{odd}^{R}} = 2 \tilde{A}_{odd}^{R} -d_c(\mathcal{D}_f^R)^2 \Lambda^2\frac{ \tilde{A}_{odd}^{R} (\tilde{\mu}^{R} +\tilde{\lambda}^{R}) }{4 \pi  [\mathcal{D}_f^R \Lambda^2 +\tilde{\sigma}^{R}]^2}- 2\frac{\tilde{A}_{odd}^{R}}{\mathcal{D}_f^R}\frac{\partial log Z_{\mathcal{D}_f}}{\partial b} \nonumber \\
&  \beta_{\tilde{\sigma}^{R}} = 2 \tilde{\sigma}^{R}  + (\mathcal{D}_f^R)^2\Lambda^2 [\tilde{A}_{odd}^R+2\tilde{K}_{odd}^R][(\tilde{A}_{odd}^R-2\tilde{K}_{odd}^R)(\mathcal{D}_f^R \Lambda^2 + \tilde{\sigma}^R) +2 \tilde{\mathcal{L}}^R \tilde{A}_{odd}^R \tilde{\mu}^R]/ \nonumber \\ &  \{4 \pi [\tilde{\lambda}^R+3 \tilde{\mu}^R][(\tilde{\mathcal{L}}^R)^2(\tilde{K}_{odd}^R\tilde{A}_{odd}^R + (\tilde{K}_{odd}^R)^2+\tilde{\mu}^R(\tilde{\lambda}^R+2 \tilde{\mu}^R)) \nonumber +\tilde{\mathcal{L}}^R (\mathcal{D}_f^R \Lambda^2 +\tilde{\sigma}^R)(\tilde{\lambda}^R+3 \tilde{\mu}^R) +  (\mathcal{D}_f^R \Lambda^2+\tilde{\sigma}^R)^2 ]\} \nonumber  \\
&  \beta_{\tilde{\mathcal{L}}^R} = 2 \frac{\tilde{\mathcal{L}}^R}{\mathcal{D}_f^R}\frac{\partial log Z_{\mathcal{D}_f}}{\partial b} \nonumber \\
&  \beta_{\mathcal{D}_f^R} =  \frac{\partial log Z_{\mathcal{D}_f}}{\partial b} \nonumber \\
\end{align}
As one can see, the diagrammatic contribution for the stress greatly simplifies and can be written down under the assumption of fluctuation-dissipation relations. Furthermore, this contribution is generically non-zero in the presence of odd elastic parameters. Keeping in mind that simulations are done using a barostat that tunes the system to zero pressure for the box, we must numerically integrate these equations with some microscopic stress in order to tune, as best we can, the stress at the system size to zero. This is done in both Fig.~\ref{fig:TheoryPlot}. We tune the microscopic stresses such that a target system size (a specific value of $L/\ell_{\text{th}}$) satisfies $\sigma^R(L/\ell_{\text{th}}) \approx 0$ in order to compare the theory to simulation results. This is equivalent to tuning the stress such that we are at the critical point in the thermodynamic limit. In the case of $K_{odd}$ in Fig.~\ref{fig:TheoryPlot}(a), we need to tune the microscopic stress to a positive value because negative stress is generated by $K_{odd}$. This can be seen from the sign of the $K_{odd}$ portion of the diagrammatic contribution in the stress equation, $\beta_{\tilde{\sigma}^R}$, in Eq.~\eqref{eq:FDRG}. On the other hand, in the case of $A_{odd}$ in Fig.~\ref{fig:TheoryPlot}(a), we need to tune the microscopic stress to a negative value because positive stress is generated by $A_{odd}$. This can be seen from the sign of the $A_{odd}$ portion of the diagrammatic contribution in the stress equation, $\beta_{\tilde{\sigma}^R}$, in Eq.~\eqref{eq:FDRG}.

We finish by mentioning that the results for both odd moduli are in agreement with the stability analysis, taking $d_c=1$.

\end{document}